\documentclass[times,twocolumn]{aastex62}

\usepackage{cases}
\usepackage{graphicx}
\usepackage{subfigure}
\usepackage{natbib}
\usepackage{color}
\usepackage{tabularx}
\usepackage{bm}
\usepackage{threeparttable}
\newcommand{\thetae}{\theta_{\rm E}}
\newcommand{\teff}{t_{\rm eff}}

\newcommand{\pie}{{\pi}_{\rm E}}

\newcommand{\te}{t_{\rm E}}
\newcommand{\eventa}{KMT-2017-BLG-1194}
\newcommand{\eventb}{KMT-2017-BLG-0428}
\newcommand{\eventc}{KMT-2019-BLG-1806}
\newcommand{\eventd}{KMT-2017-BLG-1003}
\newcommand{\evente}{KMT-2019-BLG-1367}
\newcommand{\eventf}{OGLE-2017-BLG-1806}
\newcommand{\eventg}{KMT-2016-BLG-1105}

\newcommand{\Sp}{{\it Spitzer}}
\newcommand{\hjd}{{\rm HJD}^{\prime}}
\DeclareGraphicsExtensions{.pdf,.png,.jpg}

\shorttitle{}
\shortauthors{Zang et al.}

\begin{document}
\title{{\large Systematic KMTNet Planetary Anomaly Search, Paper VII: Complete Sample of $q < 10^{-4}$ Planets from the First Four-Year Survey}}

\correspondingauthor{Weicheng Zang}
\email{3130102785@zju.edu.cn}

\author[0000-0001-6000-3463]{Weicheng Zang}
\affiliation{Department of Astronomy, Tsinghua University, Beijing 100084, China}
\affiliation{Center for Astrophysics $|$ Harvard \& Smithsonian, 60 Garden St.,Cambridge, MA 02138, USA}

\author{Youn Kil Jung}
\affiliation{Korea Astronomy and Space Science Institute, Daejon 34055, Republic of Korea}
\affiliation{University of Science and Technology, Korea, (UST), 217 Gajeong-ro Yuseong-gu, Daejeon 34113, Republic of Korea}
%ykjung21@kasi.re.kr

\author[0000-0003-0626-8465]{Hongjing Yang}
\affiliation{Department of Astronomy, Tsinghua University, Beijing 100084, China}

\author{Xiangyu Zhang}
\affiliation{Max-Planck-Institute for Astronomy, K\"onigstuhl 17, 69117 Heidelberg, Germany}
%zhangxia17@mails.tsinghua.edu.cn

\author[0000-0001-5207-5619]{Andrzej Udalski}
\affiliation{Astronomical Observatory, University of Warsaw, Al. Ujazdowskie 4, 00-478 Warszawa, Poland}
%udalski@astrouw.edu.pl

\author{Jennifer C. Yee}
\affiliation{Center for Astrophysics $|$ Harvard \& Smithsonian, 60 Garden St.,Cambridge, MA 02138, USA}
%jyee@cfa.harvard.edu

\author{Andrew Gould}
\affiliation{Max-Planck-Institute for Astronomy, K\"onigstuhl 17, 69117 Heidelberg, Germany}
\affiliation{Department of Astronomy, Ohio State University, 140 W. 18th Ave., Columbus, OH 43210, USA}
%andygould47@yahoo.com

\author{Shude Mao}
\affiliation{Department of Astronomy, Tsinghua University, Beijing 100084, China}
\affiliation{National Astronomical Observatories, Chinese Academy of Sciences, Beijing 100101, China}

\collaboration{(Leading Authors)}

\author{Michael D. Albrow}
\affiliation{University of Canterbury, Department of Physics and Astronomy, Private Bag 4800, Christchurch 8020, New Zealand}
%michael.albrow@canterbury.ac.nz

\author{Sun-Ju Chung}
\affiliation{Korea Astronomy and Space Science Institute, Daejon 34055, Republic of Korea}
\affiliation{University of Science and Technology, Korea, (UST), 217 Gajeong-ro Yuseong-gu, Daejeon 34113, Republic of Korea}
%sjchung@kasi.re.kr

\author{Cheongho Han}
\affiliation{Department of Physics, Chungbuk National University, Cheongju 28644, Republic of Korea}
%cheongho@astroph.chungbuk.ac.kr

\author{Kyu-Ha Hwang}
\affiliation{Korea Astronomy and Space Science Institute, Daejon 34055, Republic of Korea}
%kyuha@kasi.re.kr

\author{Yoon-Hyun Ryu}
\affiliation{Korea Astronomy and Space Science Institute, Daejon 34055, Republic of Korea}
%yoonhyunryu@gmail.com

\author{In-Gu Shin}
\affiliation{Center for Astrophysics $|$ Harvard \& Smithsonian, 60 Garden St.,Cambridge, MA 02138, USA}
%ingushin@gmail.com

\author{Yossi Shvartzvald}
\affiliation{Department of Particle Physics and Astrophysics, Weizmann Institute of Science, Rehovot 76100, Israel}
%yossishv@gmail.com

\author{Sang-Mok Cha}
\affiliation{Korea Astronomy and Space Science Institute, Daejon 34055, Republic of Korea}
\affiliation{School of Space Research, Kyung Hee University, Yongin, Kyeonggi 17104, Republic of Korea} 
%chasm@kasi.re.kr

\author{Dong-Jin Kim}
\affiliation{Korea Astronomy and Space Science Institute, Daejon 34055, Republic of Korea}
%keaton03@kasi.re.kr

\author{Hyoun-Woo Kim}
\affiliation{Korea Astronomy and Space Science Institute, Daejon 34055, Republic of Korea}
%hwkim@kasi.re.kr

\author{Seung-Lee Kim}
\affiliation{Korea Astronomy and Space Science Institute, Daejon 34055, Republic of Korea}
\affiliation{University of Science and Technology, Korea, (UST), 217 Gajeong-ro Yuseong-gu, Daejeon 34113, Republic of Korea}
%slkim@kasi.re.kr

\author{Chung-Uk Lee}
\affiliation{Korea Astronomy and Space Science Institute, Daejon 34055, Republic of Korea}
%leecu@kasi.re.kr

\author{Dong-Joo Lee}
\affiliation{Korea Astronomy and Space Science Institute, Daejon 34055, Republic of Korea}
%marin678@kasi.re.kr

\author{Yongseok Lee}
\affiliation{Korea Astronomy and Space Science Institute, Daejon 34055, Republic of Korea}
\affiliation{School of Space Research, Kyung Hee University, Yongin, Kyeonggi 17104, Republic of Korea}
%yslee@kasi.re.kr

\author{Byeong-Gon Park}
\affiliation{Korea Astronomy and Space Science Institute, Daejon 34055, Republic of Korea}
\affiliation{University of Science and Technology, Korea, (UST), 217 Gajeong-ro Yuseong-gu, Daejeon 34113, Republic of Korea}
%bgpark@kasi.re.kr

\author{Richard W. Pogge}
\affiliation{Department of Astronomy, Ohio State University, 140 W. 18th Ave., Columbus, OH  43210, USA}
%pogge.1@osu.edu

\collaboration{(The KMTNet Collaboration)}

\author[0000-0001-7016-1692]{Przemek Mr\'{o}z}
\affiliation{Astronomical Observatory, University of Warsaw, Al. Ujazdowskie 4, 00-478 Warszawa, Poland}
%pmroz@caltech.edu

\author[0000-0002-2335-1730]{Jan Skowron}
\affiliation{Astronomical Observatory, University of Warsaw, Al. Ujazdowskie 4, 00-478 Warszawa, Poland}
%jskowron@astrouw.edu.pl

\author[0000-0002-9245-6368]{Radoslaw Poleski}
\affiliation{Astronomical Observatory, University of Warsaw, Al. Ujazdowskie 4, 00-478 Warszawa, Poland}
%rpoleski@astrouw.edu.pl

\author[0000-0002-0548-8995]{Micha{\l}~K. Szyma\'{n}ski}
\affiliation{Astronomical Observatory, University of Warsaw, Al. Ujazdowskie 4, 00-478 Warszawa, Poland}
%msz@astrouw.edu.pl

\author[0000-0002-7777-0842]{Igor Soszy\'{n}ski}
\affiliation{Astronomical Observatory, University of Warsaw, Al. Ujazdowskie 4, 00-478 Warszawa, Poland}
%soszynsk@astrouw.edu.pl

\author[0000-0002-2339-5899]{Pawe{\l} Pietrukowicz}
\affiliation{Astronomical Observatory, University of Warsaw, Al. Ujazdowskie 4, 00-478 Warszawa, Poland}
%pietruk@astrouw.edu.pl

\author[0000-0003-4084-880X]{Szymon Koz{\l}owski}
\affiliation{Astronomical Observatory, University of Warsaw, Al. Ujazdowskie 4, 00-478 Warszawa, Poland}
%simkoz@astrouw.edu.pl

\author[0000-0001-6364-408X]{Krzysztof Ulaczyk}
\affiliation{Department of Physics, University of Warwick, Gibbet Hill Road, Coventry, CV4~7AL,~UK}
%kulaczyk@astrouw.edu.pl

\author[0000-0002-9326-9329]{Krzysztof A. Rybicki}
\affiliation{Astronomical Observatory, University of Warsaw, Al. Ujazdowskie 4, 00-478 Warszawa, Poland}
%krybicki@astrouw.edu.pl

\author[0000-0002-6212-7221]{Patryk Iwanek}
\affiliation{Astronomical Observatory, University of Warsaw, Al. Ujazdowskie 4, 00-478 Warszawa, Poland}
%piwanek@astrouw.edu.pl

\author[0000-0002-3051-274X]{Marcin Wrona}
\affiliation{Astronomical Observatory, University of Warsaw, Al. Ujazdowskie 4, 00-478 Warszawa, Poland}
%mwrona@astrouw.edu.pl

\author[0000-0002-1650-1518]{Mariusz Gromadzki}
\affiliation{Astronomical Observatory, University of Warsaw, Al. Ujazdowskie 4, 00-478 Warszawa, Poland}
%Mazki@astrouw.edu.pl

\collaboration{(The OGLE Collaboration)}

%shude.mao@gmail.com

\author{Hanyue Wang}
\affiliation{Center for Astrophysics $|$ Harvard \& Smithsonian, 60 Garden St.,Cambridge, MA 02138, USA}
%\affiliation{Harvard College, Harvard University, MA 02138, USA}

\author[0000-0002-1279-0666]{Jiyuan Zhang}
\affiliation{Department of Astronomy, Tsinghua University, Beijing 100084, China}

\author[0000-0003-4027-4711]{Wei Zhu}
\affiliation{Department of Astronomy, Tsinghua University, Beijing 100084, China}
%weizhu@mail.tsinghua.edu.cn

\collaboration{(The MAP Collaboration)}

\begin{abstract}

We present the analysis of seven microlensing planetary events with planet/host mass ratios $q < 10^{-4}$: KMT-2017-BLG-1194, KMT-2017-BLG-0428, KMT-2019-BLG-1806, KMT-2017-BLG-1003, KMT-2019-BLG-1367, OGLE-2017-BLG-1806, and KMT-2016-BLG-1105. They were identified by applying the Korea Microlensing Telescope Network (KMTNet) AnomalyFinder algorithm to 2016--2019 KMTNet events. A Bayesian analysis indicates that all the lens systems consist of a cold super-Earth orbiting an M or K dwarf. Together with 17 previously published and three that will be published elsewhere, AnomalyFinder has found a total of 27 planets that have solutions with $q < 10^{-4}$ from 2016--2019 KMTNet events, which lays the foundation for the first statistical analysis of the planetary mass-ratio function based on KMTNet data. By reviewing the 27 planets, we find that the missing planetary caustics problem in the KMTNet planetary sample has been solved by AnomalyFinder. We also find a desert of high-magnification planetary signals ($A \gtrsim 65$), and a follow-up project for KMTNet high-magnification events could detect at least two more $q < 10^{-4}$ planets per year and form an independent statistical sample. 

\end{abstract}

\section{Introduction}\label{intro}

Among current exoplanet detection methods, a unique capability of the gravitational microlensing technique \citep{Shude1991,Andy1992} is to detect low-mass ($M_{\rm planet} \lesssim 20 M_{\oplus}$) cold planets beyond the snow line \citep{Hayashi1981, Min2011}, including Neptune-mass cold planets, which are common (Uranus and Neptune) in our Solar System and cold terrestrial planets, which are absent in our Solar System. Because the typical host stars of the microlensing planetary systems are M and K dwarfs, detections of $q < 10^{-4}$ planets (where $q$ is the planet/host mass ratio) can reveal the abundance of low-mass cold planets and answer how common the outer solar system is. 

However, since the first microlensing planet, which was detected in 2003 \citep{OB03235}, the first 13 years of microlensing planetary detections only discovered six $q < 10^{-4}$ planets\footnote{They are OGLE-2005-BLG-169Lb \citep{OB05169}, OGLE-2005-BLG-390Lb \citep{OB05390}, OGLE-2007-BLG-368Lb \citep{OB07368}, MOA-2009-BLG-266Lb \citep{MB09266}, OGLE-2013-BLG-0341Lb \citep{OB130341}, OGLE-2015-BLG-1670 \citep{OB151670}.} and none of them had mass ratios below $4.4 \times 10^{-5}$. The paucity of detected $q < 10^{-4}$ planets led to important statistical implications for cold planets. \cite{Suzuki2016} analyzed 1474 microlensing events discovered by the Microlensing Observations in Astrophysics (MOA) survey \citep{Sako2008} and formed a homogeneously selected sample including 22 planets. They found that the mass-ratio function of microlensing planets increases as $q$ decreases until a break at $q \sim 1.7 \times 10^{-4}$, below which the planetary occurrence rate likely drops. This break suggests that the Neptune-mass planets are likely to be the most common of cold planets. However, the \cite{Suzuki2016} sample only contains two $q < 10^{-4}$ and thus may be affected by small number statistics. To examine the existence of the break, a larger $q < 10^{-4}$ sample is needed.  

After its commissioning season in 2015, the new-generation microlensing survey, the Korea Microlensing Telescope Network (KMTNet, \citealt{KMT2016}), has been conducting near-continuous, wide-area, high-cadence surveys for $\sim 96~{\rm deg}^2$. The fields with cadences of $\Gamma \geq 2~{\rm hr}^{-1}$ are the KMTNet prime fields ($\sim 12~{\rm deg}^2$) and the other fields are the KMTNet sub-prime fields ($\sim 84~{\rm deg}^2$). Since 2016, the detections of $q < 10^{-4}$ planets have been greatly increased in two ways, and the KMTNet data played a major or decisive role in all detections. First, more than ten $q < 10^{-4}$ planets have been detected from by-eye searches, including three with $q < 2 \times 10^{-5}$ \citep{KB180029,OB190960,KB200414}. Second, \cite{OB191053,2019_prime} developed the KMTNet AnomalyFinder algorithm to systematically search for planetary signals. This algorithm has been applied to the 2018 and 2019 KMTNet prime fields ($\Gamma \geq 2~{\rm hr}^{-1}$) and uncovered five new $q < 10^{-4}$ planets \citep{OB191053,KB190253,2018_prime}. Moreover, the systematic search opens a window for a homogeneous large-scale KMTNet planetary sample. According to the experience from 2018 and 2019 KMTNet prime fields, we expect to detect $\gtrsim 20$ planets with $q < 10^{-4}$ from 2016--2019 seasons. This will be an order of magnitude larger than the \cite{Suzuki2016} sample at $q < 10^{-4}$.

%and which combined with the planetary detection efficiency calculator based on AnomalyFinder (Zang, Jung, et al. in prep), will be the first to measure the occurrence rate of $q < 10^{-4}$ planets with 

To build the first KMTNet $q < 10^{-4}$ statistical sample, we applied the KMTNet AnomalyFinder algorithm to the 2016--2019 KMTNet microlensing events. In this paper, we introduce seven new $q < 10^{-4}$ events from this search. They are \eventa, \eventb, \eventc/OGLE-2019-BLG-1250, \eventd, \evente, \eventf/KMT-2017-BLG-1021, and \eventg. Together with 17 already published and three that will be published elsewhere, the KMTNet AnomalyFinder algorithm found 27 events that can be fit by $q < 10^{-4}$ models from 2016--2019 KMTNet data. However, whether a planet can be used for statistical studies requires further investigations, which is beyond the scope of this paper.

The paper is structured as follows. In Section \ref{anomaly}, we briefly introduce the KMTNet AnomalyFinder algorithm and the procedure to form the $q < 10^{-4}$ sample. In Sections \ref{obser}, \ref{model} and \ref{lens}, we present the observations and the analysis of seven $q < 10^{-4}$ events. Finally, we discuss the implications from the 2016--2019 KMTNet $q < 10^{-4}$ planetary sample in Section \ref{dis}.

%one event (OGLE-2017-BLG-0448) has $q \sim 10^{-3}$. The remaining event, KMT-2016-BLG-0625, has three degenerate solutions ($q = 3.9 \times 10^{-5}$, $q = 2.2 \times 10^{-4}$ and $q = 2.4 \times 10^{-4}$) and will be published together with other 2016 KMTNet prime-field planets from AnomalyFinder (Shin et al. in prep). 
 
 %KMT-2019-BLG-1806/OGLE-2019-BLG-1250, which has the lowest mass ratio among all 2016--2019 KMTNet planets, will be published together with the AnomalyFinder planetary detection efficiency calculator (Zang, Jung et al. in prep).

\section{The Basic of AnomalyFinder and the Procedure}\label{anomaly}

Section 2 of \cite{OB191053} and Section 2 of \cite{2019_prime} together introduced the KMTNet AnomalyFinder algorithm. The AnomalyFinder uses a \cite{Gould2D} 2-dimensional grid of $(t_0, \teff)$ to search for and fit anomalies from the residuals to a point-source point-lens (PSPL, \citealt{Paczynski1986}) model. Here $t_0$ is the time of maximum magnification, and $\teff$ is the effective timescale. For our search, the shortest $\teff$ is 0.05 days and the longest $\teff$ is 6.65 days. The parameters that evaluate the significance of a candidate anomaly are $\Delta\chi^2_0$ and $\Delta\chi^2_{\rm flat}$. See Equation (4) of \cite{OB191053} for their definitions. The criteria of $\Delta\chi^2_0$ and $\Delta\chi^2_{\rm flat}$ are the same as the criteria used in \citet{2019_prime,2018_prime,2018_subprime}, with $\Delta\chi^2_0 > 200$, or $\Delta\chi^2_0 > 120$ and $\Delta\chi^2_{\rm flat} > 60$ for the KMTNet prime-field events and $\Delta\chi^2_0 > 100$, or $\Delta\chi^2_0 > 60$ and $\Delta\chi^2_{\rm flat} > 30$ for the KMTNet sub-prime-field events. Future statistical studies should use the same criteria. In addition, an anomaly is required to contain at least three successive points $\geq 2\sigma$ away from a PSPL model.

As a result, we found 464 and 608 candidate anomalies from 2016--2019 KMTNet prime-field and sub-prime-field events, respectively. We checked whether the data from other surveys are consistent with the KMTNet-based anomalies and cross-checked with C. Han's modeling. We fitted all the $q < 10^{-3}$ candidates with online data and found 13 new candidates with $q < 2 \times 10^{-4}$. Then, we conducted tender-loving care (TLC) re-reductions and re-fitted the 13 events. Of these, eight events unambiguously have $q < 10^{-4}$, three events, KMT-2016-BLG-1307, KMT-2017-BLG-0849, and KMT-2017-BLG-1057, have $10^{-4} < q < 2 \times 10^{-4}$, and two events, KMT-2016-BLG-0625 (Shin et al. in prep) and OGLE-2017-BLG-0448/KMT-2017-BLG-0090 (Zhai et al. in prep), have ambiguous mass ratios at $10^{-5} \lesssim q \lesssim 10^{-3}$ and will be published elsewhere.

Among the eight unambiguous $q < 10^{-4}$ events, one event, OGLE-2016-BLG-0007/MOA-2016-BLG-088/KMT-2016-BLG-1991, will be published elsewhere because it has the lowest-$q$ of this sample. We analyze and publish the remaining seven events in this paper. We note that the planetary signals of the seven events are not strong, although they are confirmed by at least two data sets. We thus further check whether the light curves have other similar anomalies, to exclude the possibility of unknown systematic errors. We applied the AnomalyFinder algorithm to the re-reduction data. For all of the seven events, besides the known planetary signals no anomaly with $\Delta\chi^2_0 > 20$ was detected. Therefore, the light curves of the seven events are stable and planetary signals are reliable.

\section{Observations and Data Reductions}\label{obser}

\begin{table*}
    \renewcommand\arraystretch{1.5}
    \centering
    \caption{Event Names, Alerts, Locations, and Cadences for the six planetary events}
    \begin{tabular}{c c c c c c c}
    \hline
    \hline
    Event Name & Alert Date & ${\rm RA}_{\rm J2000}$ & ${\rm Decl.}_{\rm J2000}$ & $\ell$ & $b$ & $\Gamma ({\rm hr}^{-1})$ \\
    \eventa & Post Season & 18:17:17.31 & $-$25:19:26.18 & +6.63 & $-$4.34 & 0.4 \\
    \hline
    \eventb & Post Season & 18:05:32.46 & $-$28:29:25.01 & +2.59 & $-$3.55 & 4.0 \\
    \hline
    \eventc & 26 Jul 2019 & 18:02:09.01 & $-$29:24:53.60 & +1.41 & $-$3.35 & 1.0 \\
    OGLE-2019-BLG-1250 & & & & & & 0.3 \\
    \hline 
    \eventd & Post Season & 17:41:38.76 & $-$24:22:26.18 & +3.42 & +3.15 & 1.0 \\
    \hline
    \evente & 27 Jun 2019 & 18:09:53.12 & $-$29:45:43.96 & +1.93 & $-$4.99 & 0.4 \\
    \hline
    \eventf & 14 Oct 2017 & 17:46:29.58 & $-$24:16:20.17 & +4.09 & +2.26 & 0.3 \\
    KMT-2017-BLG-1021 & & & & & & 1.0 \\
    \hline
    \eventg & Post Season & 17:45:47.34 & $-$26:15:58.93 & +2.30 & +1.16 & 1.0 \\
    \hline
    \hline
    \end{tabular}
    \label{event_info}
\end{table*}

Table \ref{event_info} lists the basic observational information for the seven events, including event names, the first discovery date, the coordinates in the equatorial and galactic systems, and the nominal cadences ($\Gamma$). The seven planetary events were all identified by the KMTNet post-season EventFinder algorithm \citep{KMTeventfinder}. Of them, \eventc/OGLE-2019-BLG-1250 and \eventf/KMT-2017-BLG-1021 were discovered by the KMTNet alert-finder system \citep{KMTAF} and the Early Warning System \citep{Udalski1994,Udalski2003} of the Optical Gravitational Lensing Experiment (OGLE, \citealt{OGLEIV}),  respectively, during their observational seasons. Hereafter, we designate \eventc/OGLE-2019-BLG-1250 and \eventf/KMT-2017-BLG-1021 by their first-discovery name, \eventc\ and \eventf. During the 2019 observational season, the KMTNet alert-finder system also discovered \evente. In addition, OGLE observed the locations of \evente\ and \eventg\ but did not alert them. We also include the OGLE data for these two events into the light-curve analysis, for which the OGLE data confirm the planetary signals found by the KMTNet. MOA did not issue alerts for any of the seven events, and there were no follow-up data to the best of our knowledge. 

KMTNet conducted observations from three identical 1.6 m telescopes equipped with $4~{\rm deg}^2$ cameras in Chile (KMTC), South Africa (KMTS), and Australia (KMTA). OGLE took data using an 1.3m telescope with 1.4 ${\rm deg}^2$ field of view in Chile. For both surveys, most of the images were taken in the $I$ band, and a fraction of $V$-band images were acquired for source color measurements. Each KMTNet V-band data point was taken one minute before or after one KMTNet I-band data point of the same field.

The KMTNet and OGLE data used in the light-curve analysis were reduced using the custom photometry pipelines based on the difference imaging technique \citep{Tomaney1996,Alard1998}: pySIS (\citealt{pysis}, Yang et al. in prep) for the KMTNet data, and \cite{Wozniak2000} for the OGLE data. For each event, the KMTC data were additionally reduced using the pyDIA photometry pipeline \citep{pyDIA} to measure the source color. Except for \eventf\ and \eventg\ whose sources are not located in any OGLE star catalog, the $I$-band magnitudes of the other five events reported in this paper have been calibrated to the standard $I$-band magnitude using the OGLE-III star catalog \citep{OGLEIII}.

\section{Light-curve Analysis}\label{model}

\subsection{Preamble}\label{preamble}
Because all seven events contain short-lived deviations from a PSPL model, we first introduce the common methods for the light-curve analysis. The PSPL model is described by three parameters, $t_0$, $u_0$, and $\te$, which respectively represent the time of lens-source closest approach, the closest lens-source projected separation normalized to the angular Einstein radius $\thetae$, and the Einstein timescale,
\begin{equation}\label{eqn:1}
\te = \frac{\thetae}{\mu_{\rm rel}}; \qquad \thetae = \sqrt{\kappa M_{\rm L} \pi_{\rm rel}},
\end{equation}  
where $\kappa \equiv \frac{4G}{c^2\mathrm{au}} \simeq 8.144 \frac{{\rm mas}}{M_{\odot}}$, $M_{\rm L}$ is the lens mass, and $(\pi_{\rm rel}, \mu_{\rm rel})$ are the lens-source relative (parallax, proper motion). In addition, for each data set $i$, we introduce two linear parameters, ($f_{{\rm S},i}$, $f_{{\rm B},i}$), to fit the flux of the source and any blend flux, respectively. 

We search for binary-lens single-source (2L1S) models for each event. A 2L1S model requires four parameters in addition to the PSPL parameters, $(s, q, \alpha, \rho)$, which respectively denote the planet-host projected separation in units of $\thetae$, the planet/host mass ratio, the angle between the source trajectory and the binary axis, and the angular source radius $\theta_*$ scaled to $\thetae$, i.e., $\rho = \theta_*/\thetae$. 

%Because all of the planetary signals of the seven events occurred in the low-magnification regions, the sources probably cross or come close to the planetary caustic and some of the 2L1S parameters can be estimated before detailed numerical analysis. 

Although the final results need detailed numerical analysis, some of the 2L1S parameters can be estimated by heuristic analysis. A PSPL fit excluding the data points around the anomaly can yield the three PSPL parameters, $t_0$, $u_0$, and $\te$. If an anomaly occurred at $t_{\rm anom}$, the corresponding lens-source offset, $u_{\rm anom}$, and $\alpha$ can be estimated by 
\begin{equation}\label{alpha}
    u_{\rm anom} = \sqrt{u_0^2 + \left(\frac{t_{\rm anom} - t_0}{\te}\right)^2}; \quad \left|\alpha\right|=\left|\sin^{-1} \frac{u_{0}}{u_{\rm anom}}\right|.
\end{equation}
Because the planetary caustics are located at the position of $|s - s^{-1}| \sim u_{\rm anom}$, we obtain
\begin{equation}\label{s}
    s_{\pm}\sim\frac{\sqrt{u_{\rm anom}^{2}+4} \pm u_{\rm anom}}{2},
\end{equation}
where $s = s_{+}$ and $s = s_{-}$ correspond to the major-image (quadrilateral) and the minor-image (triangular) planetary caustics, respectively. For two degenerate solutions with similar $q$ but different $s$, \cite{KMT2021_mass1} suggested that the geometric mean of two solutions satisfies
\begin{equation}\label{equ:offset}
    s_{\rm mean} = s_{\pm}.
\end{equation}
In addition, \cite{Zhang2022} suggested a slightly different formalism, and \cite{ZhangGaudi2022} provided a theoretical treatment of it. For a dip-type planetary signal, \cite{KB190253} pointed out that the mass ratio can be estimated by
\begin{equation}\label{tdip}
    q = \left(\frac{\Delta t_{\rm dip}}{4\te} \right)^2 \frac{s}{u_0} |\sin^3\alpha|,
\end{equation}
where $\Delta t_{\rm dip}$ is the duration of the dip, and the accuracy of Equation (\ref{tdip}) should be at a factor of $\sim 2$ level. 

To find all the possible 2L1S models, we conduct two-phase grid searches for the parameters, ($\log s$, $\log q$, $\alpha$, $\rho$). In the first phase, we conduct a sparse grid, which consists of 21 values equally spaced between $-1.0 \leq \log s \leq 1.0$, 20 values equally spaced between $0^{\circ}\leq \alpha < 360^{\circ}$, 61 values equally spaced between $-6.0 \leq \log q \leq 0.0$ and five values equally spaced between $-3.5 \leq \log \rho \leq -1.5$. We use a code based on the advanced contour integration code \citep{Bozza2010,Bozza2018}, \texttt{VBBinaryLensing}\footnote{\url{http://www.fisica.unisa.it/GravitationAstrophysics/VBBinaryLensing.htm}} to compute the 2L1S magnification. For each grid point, we search for the minimum $\chi^2$ by Markov chain Monte Carlo (MCMC) $\chi^2$ minimization using the \texttt{emcee} ensemble sampler \citep{emcee}, with fixed ($\log q$, $\log s$) and free ($t_0, u_0, \te, \rho, \alpha$). In the second phase, we conduct a denser ($\log s$, $\log q$, $\alpha$, $\rho$) grid search around each local minimum (e.g., \citealt{OB180799}). Finally, we refine the best-fit models by MCMC with all parameters free. 

For degenerate solutions, \cite{KB210171} suggested that the phase-space factors can be used to weight the probability of each solution. We follow the procedures of \cite{KB210171} and first calculate the covariance matrix, $\bm{C}$, of ($\log s, \log q, \alpha$) from the MCMC chain. Then, the phase-space factor is
\begin{equation}
    p = \sqrt{{\rm det}(\bm{C})}.
\end{equation}
Because whether a planet and its individual solutions can be used for statistical studies requires further investigations, we provide the phase-space factors for the event with multiple solutions but do not use them to weight or reject solutions.

We also investigate whether the inclusion of two high-order effects can improve the fit. The first is the microlensing parallax effect \citep{Gould1992,Gould2000,Gouldpies2004}, which is due to the Earth's orbital acceleration around the Sun. We fit it by two parameters, $\pi_{\rm E,N}$ and $\pi_{\rm E,E}$, which are the north and east components of the microlensing parallax vector $\bm{\pi}_{\rm E}$ in equatorial coordinates, 
\begin{equation}\label{equ:pie}
    \bm{\pi}_{\rm E} \equiv \frac{\pi_{\rm rel}}{\thetae} \frac{\bm{\mu}_{\rm rel}}{\mu_{\rm rel}}.
\end{equation}
We also fit the $u_0 > 0$ and $u_0 < 0$ solutions to consider the ``ecliptic degeneracy'' \citep{Jiang2004, Poindexter2005}. For four cases in this paper, the parallax contours take the form of elongated ellipses, so we report the constraints on the minor axes of the error ellipse, ($\pi_{\rm E, \parallel}$), which is approximately parallel with the direction of the Earth’s acceleration. For the major axes of the parallax contours, $\pi_{\rm E, \bot} \sim \pi_{\rm E, N}$, we only report it when the constraint is useful. 

The second effect is the lens orbital motion effect \citep{MB09387, OB09020}, and we fit it by the parameter $\vec{\gamma} =\left(\frac{ds/dt}{s}, \frac{d\alpha}{dt}\right)$, where $ds/dt$ and $d\alpha/dt$ represent the instantaneous changes in the separation and orientation of the two components defined at $t_0$, respectively. To exclude unbound systems, we restrict the MCMC trials to $\beta < 1.0$. Here $\beta$ is the absolute value of the ratio of projected kinetic to potential energy \citep{An2002,OB050071D},
\begin{equation}
    \beta \equiv \left| \frac{{\rm KE}_{\perp}}{{\rm PE}_{\perp}} \right| = \frac{\kappa M_{\odot} {\rm yr}^2}{8\pi^2}\frac{\pie}{\thetae}\gamma^2\left(\frac{s}{\pie + \pi_{\rm S}/\thetae}\right)^3,
\end{equation}
and where $\pi_{\rm S}$ is the source parallax estimated by the mean distance to red clump stars in the direction of each event \citep{Nataf2013}. 

\subsection{``Dip'' Anomalies}

\subsubsection{\eventa}

\begin{figure}
    \centering
    \includegraphics[width=0.95\columnwidth]{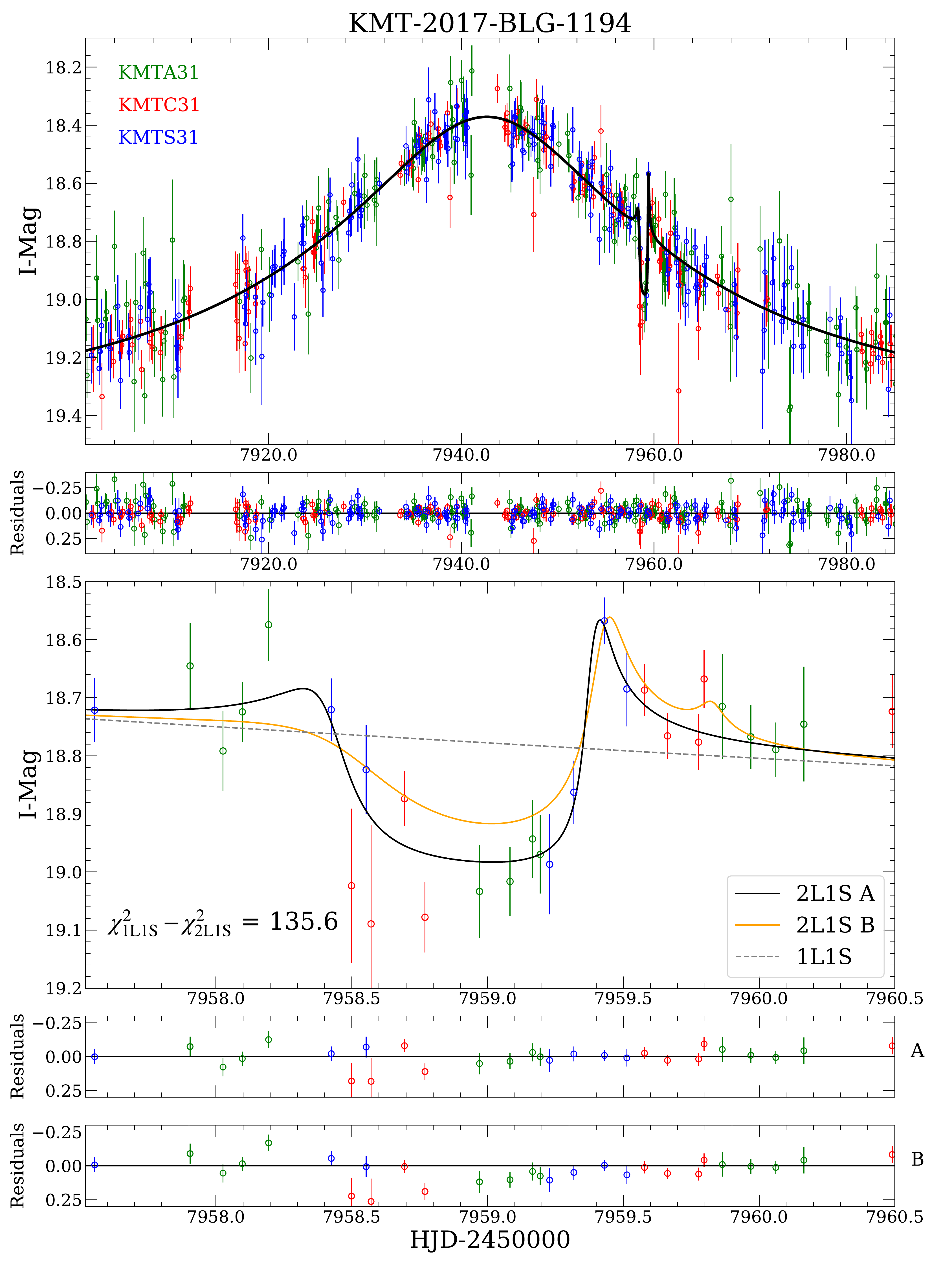}
    \caption{The observed data and the 2L1S (the black and orange solid lines) and 1L1S models (the grey dashed line) for \eventa. The data taken from different data sets are shown with different colors. The bottom panels show a close-up of the dip-type planetary signal and the residuals to the 2L1S models.}
    \label{lc1}
\end{figure}

\begin{table}
    \renewcommand\arraystretch{1.25}
    \centering
    \caption{2L1S Parameters for \eventa}
    \begin{tabular}{c c c}
     \hline
    Parameter & A & B \\
    \hline
    $\chi^2$/dof & 928.0/928 & 950.6/928 \\
    \hline
    $t_{0}$ (${\rm HJD}^{\prime}$) & $7942.66 \pm 0.13$ & $7942.59 \pm 0.13$ \\
    $u_{0}$  & $0.256 \pm 0.018$ & $0.246 \pm 0.011$  \\
    $\te$ (days)  & $47.0 \pm 2.5$ & $47.9 \pm 1.7$ \\
    $\rho (10^{-3})$ & $<2.6$ & $<1.4$ \\
    $\alpha$ (rad) & $2.505 \pm 0.013$ & $2.515 \pm 0.011$ \\
    $s$ & $0.8063 \pm 0.0103$ & $0.8055 \pm 0.0065$ \\
    $\log q$ & $-4.582 \pm 0.058$ & $-4.585 \pm 0.074$ \\
    $I_{\rm S, OGLE}$ & $20.28 \pm 0.08$ & $20.34 \pm 0.06$ \\
    \hline
    \end{tabular}
    \tablecomments{The upper limit on $\rho$ is $3\sigma$.}
    \label{parm1}
\end{table}

Figure \ref{lc1} shows the observed data together with the best-fit PSPL and 2L1S models for \eventa. There is a dip centered on $\hjd \sim 7958.9~(\hjd = {\rm HJD} - 2450000)$ , i.e., $t_{\rm anom} \sim 7958.9$, with a duration of $\Delta t_{\rm dip} \sim 1.05$ days. The dip and the ridge around the dip are covered by three KMTNet sites, so the anomaly is secure. A PSPL fit yields ($t_0$, $u_0$, $\te$) = (7942.7, 0.26, 46), and using the heuristic formalism of Section \ref{preamble}, we obtain
\begin{equation}\label{{Equ:parm1}}
    \alpha \sim 143.5^\circ;\quad s = s_{-} \sim 0.807;\quad \log q \sim -4.68. 
\end{equation}

\begin{figure}[htb] 
    \centering
    \includegraphics[width=0.98\columnwidth]{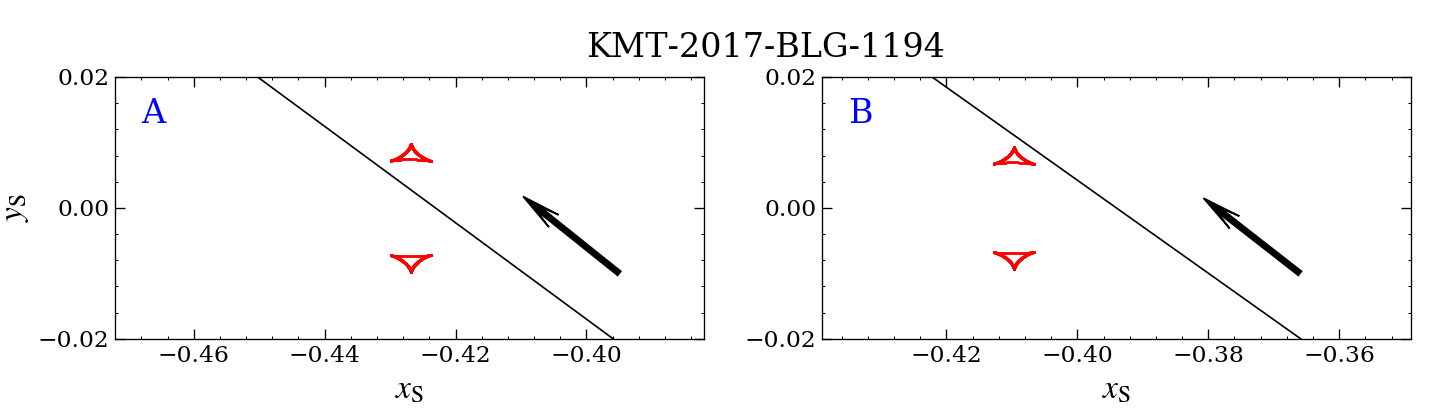}
    \includegraphics[width=0.98\columnwidth]{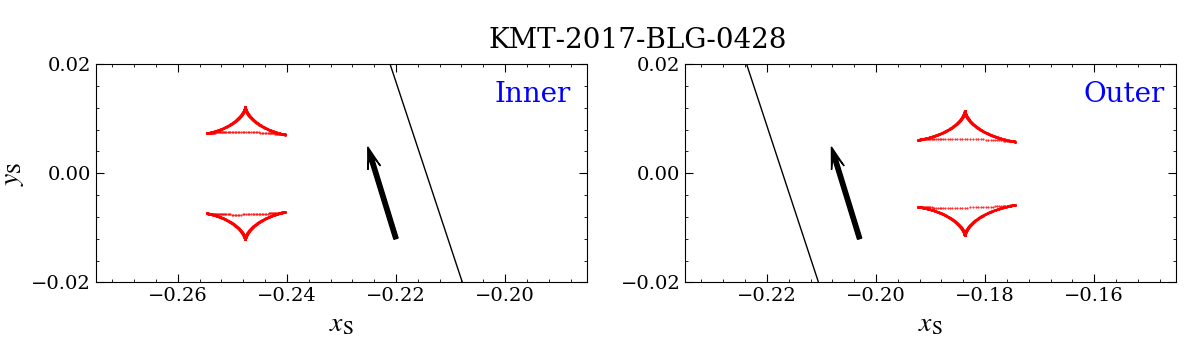}
    \includegraphics[width=0.98\columnwidth]{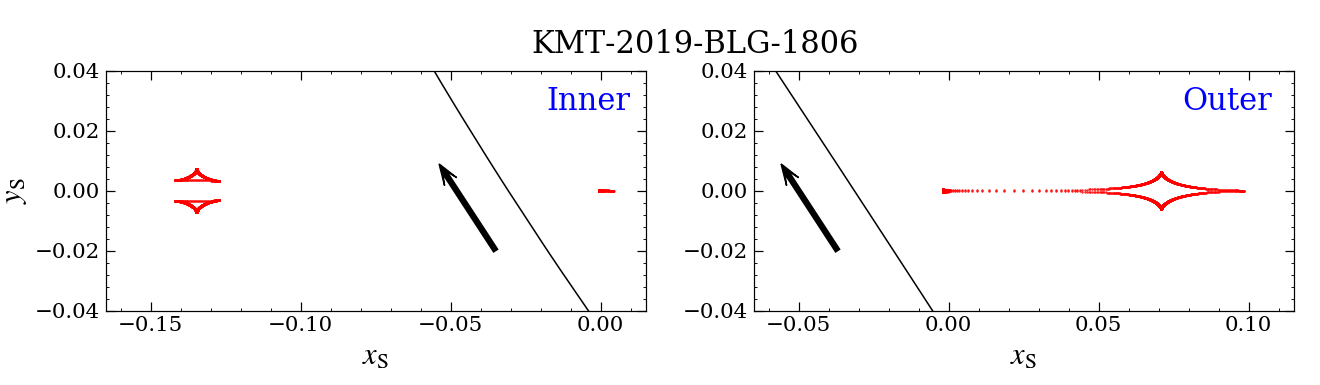}
    \includegraphics[width=0.98\columnwidth]{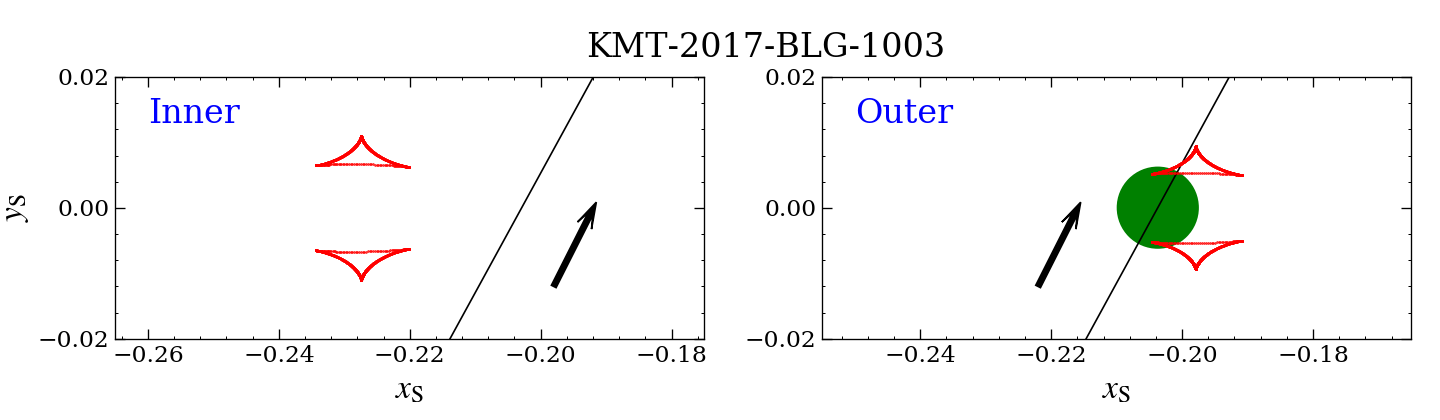}
    \includegraphics[width=0.98\columnwidth]{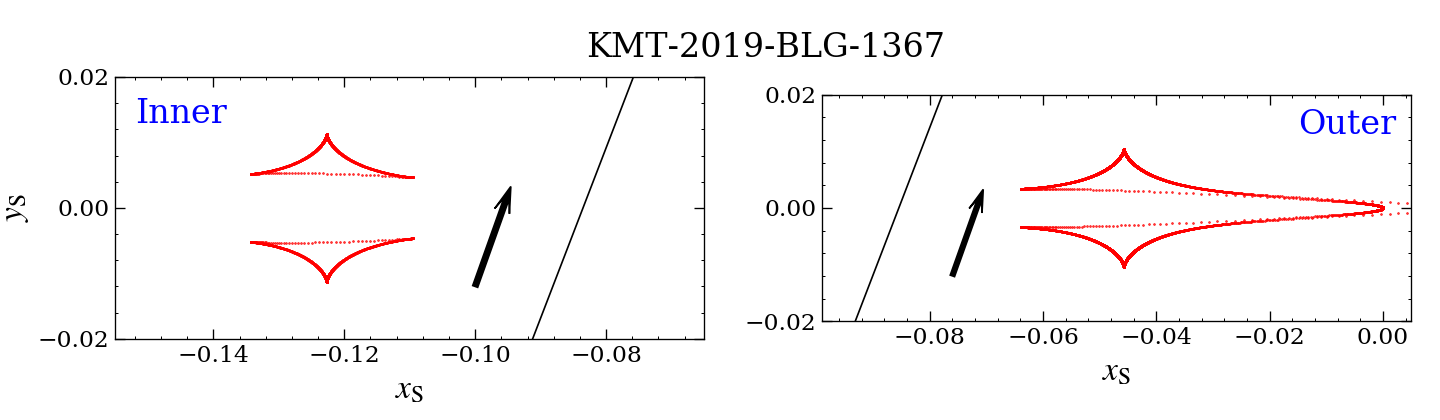}
    \caption{Geometries of the five ``dip'' planetary events. In each panel, the red lines represent the caustic, the black solid line represents the source trajectory, and the line with an arrow indicates the direction of the source motion. For the outer solution of \eventd, $\rho$ is constrained at the $>3\sigma$ level, so the radius of the green dot represents the source radius. For other solutions, $\rho$ only has weak constraints with $<3\sigma$, so their source radii are not shown.}
    \label{cau1}
\end{figure}

The grid search yields one solution. Its parameters are presented in Table \ref{parm1} and are in good agreement with the heuristic estimates. The top left panel of Figure \ref{cau1} displays the caustic structure and the source trajectory, for which the two minor-image planetary caustics are located on both sides of the source trajectory. We label the solution as the solution ``A''. To further investigate the parameter space and check whether the event has the inner/outer solutions \citep{GG1997}, for which the source passes inside (the ``Inner'' solution) the two planetary caustics (closer to the central caustic) or outside (the ``Outer'' solution), we follow the procedures of \cite{OB170173}. First, we conduct a ``hotter'' MCMC with the error bar inflated by a factor of $\sqrt{3.0}$. Second, we make a scatter plot of $\log q$ versus $\Delta\xi$ from the ``hotter'' MCMC chain. Here $\Delta\xi$ represents the offset between the source and the planetary caustic as the source crosses the binary axis,
\begin{equation}
    \Delta\xi = u_0\csc(\alpha) - (s - s^{-1}).
\end{equation}

\begin{figure}[htb] 
    \centering
    \includegraphics[width=\columnwidth]{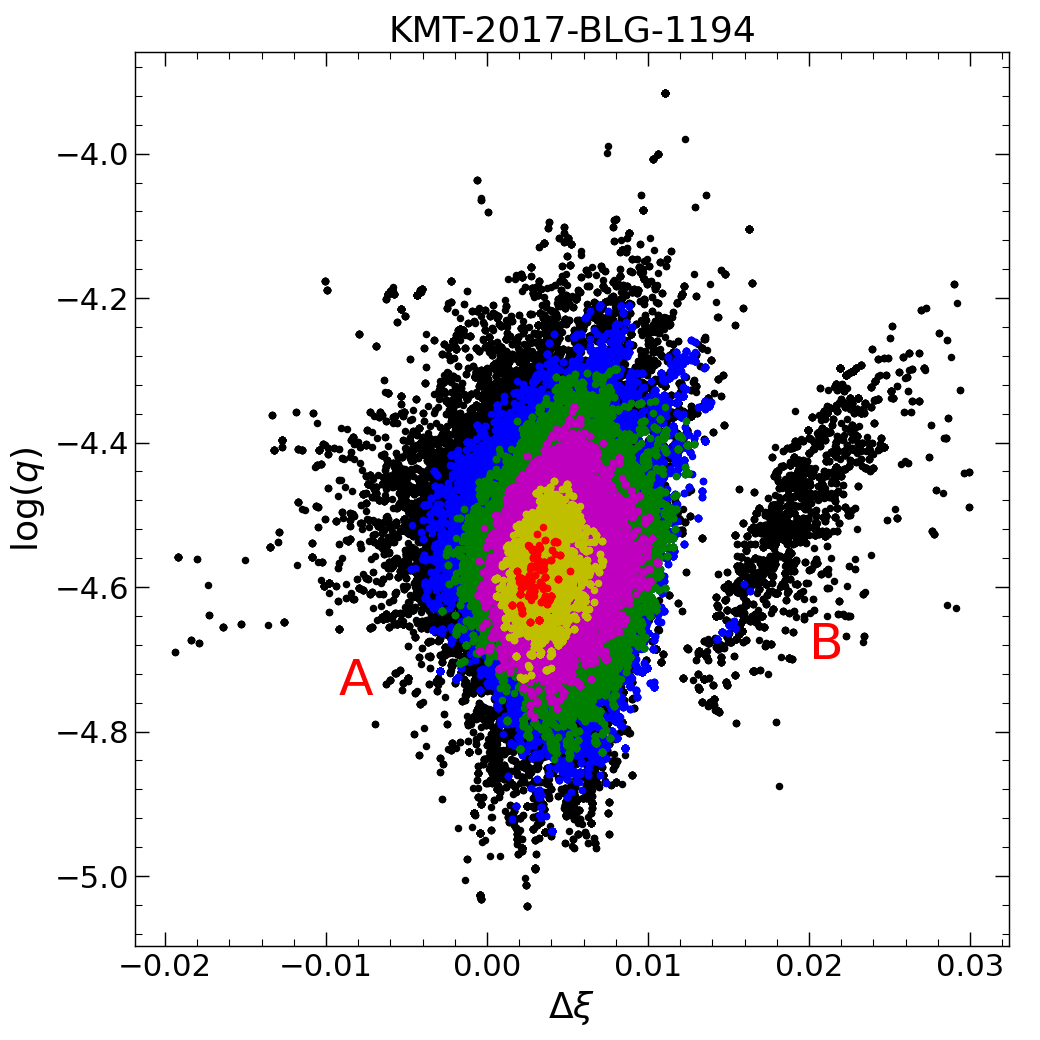}
    \caption{Scatter plot of $\log q$ vs. $\Delta\xi$ for \eventa, where $\Delta\xi = u_0\csc(\alpha) - (s - s^{-1})$ represents the offset between the source and the center of the planetary caustic at the moment that the source crosses the binary axis. The distribution is derived by inflating the error bars by a factor of $\sqrt{3}$ and then multiplying the resulting $\chi^2$ by 3 for the plot. Red, yellow, magenta, green, blue and black colors represent $\Delta\chi^2 < 2 \times (1, 4, 9, 16, 25, \infty)$. ``A'' and ``B'' represent two local minima and the corresponding parameters are given in Table \ref{parm1}.}
    \label{xi_kb171194}
\end{figure}

The resulting scatter plot is shown in Figure \ref{xi_kb171194}, from which we find another local minimum at $\Delta\xi \sim 0.02$. We label this solution as the ``B'' solution. As shown in the top right panel of Figure \ref{cau1}, the ``B'' solution corresponds to the ``Inner'' solution. Its parameters from MCMC are given in Table \ref{parm1} and it is disfavored by $\Delta\chi^2 = 22.6$ compared to the ``A'' solution. In Figure \ref{lc1}, the ``B'' solution cannot fit the anomaly well and all three KMTNet data sets contribute to the $\Delta\chi^2$. The ratio of the phase-space factors is $p_{\rm A} : p_{\rm B} = 1 : 0.54$, which also prefers the ``A'' solution. Thus, we exclude the ``B'' solution. In addition, the models, which have the geometry of the ``Outer'' solution, do not form a local minimum and are disfavored by $\Delta\chi^2 > 60$ compared to the ``A'' solution.

%We find that the MCMC chain contains the models for which the two planetary caustics are both on the left side or right side of the source trajectory, 

For the ``A'' solution a point-source model is consistent within $1\sigma$ and the $3\sigma$ upper limit is $\rho < 0.0026$. The inclusion of higher-order effects yields a constraint on $\pi_{\rm E, \parallel}$, and with the other 2L1S parameters being almost unchanged. We obtain $\pi_{\rm E, \parallel} = -0.18 \pm 0.35$ and adopt the constraints on $\pie$ and $\rho$ in the Bayesian analysis of Section \ref{lens}. This is a new microlensing planet with $q \sim 2.62 \times 10^{-5}$; i.e., about nine times the Earth/Sun mass ratio. 

%where $\pi_{\rm E, \parallel}$ 

\subsubsection{\eventb}

\begin{figure}[htb] 
    \centering
    \includegraphics[width=0.95\columnwidth]{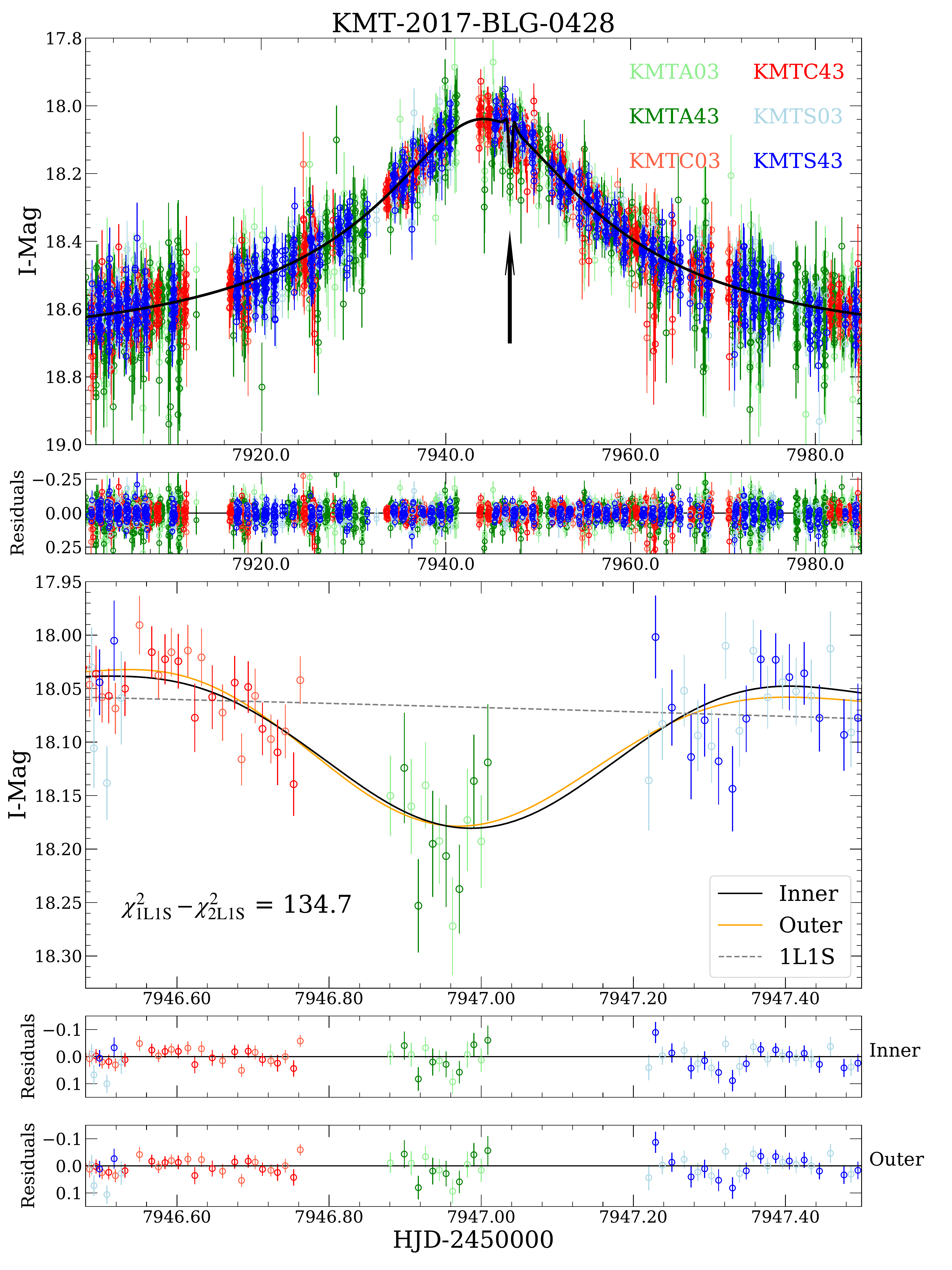}
    \caption{The observed data and models for \eventb. The symbols are similar to those in Figure \ref{lc1}. In the top panel, the black arrow indicates the position of the planetary signal.}
    \label{lc2}
\end{figure}

\begin{table}[htb]
    \renewcommand\arraystretch{1.25}
    \centering
    \caption{2L1S Parameters for \eventb}
    \begin{tabular}{c c c}
     \hline
    Parameter & Inner & Outer \\
    \hline
    $\chi^2$/dof & 9952.0/9952 & 9952.1/9952 \\
    \hline
    $t_{0}$ (${\rm HJD}^{\prime}$) & $7943.976 \pm 0.030$ & $7943.978 \pm 0.031$ \\
    $u_{0}$  & $0.205 \pm 0.009$ & $0.205 \pm 0.009$ \\
    $\te$ (days)  & $44.4 \pm 1.5$ & $44.3 \pm 1.5$ \\
    $\rho (10^{-3})$ & $<6.4$ & $<6.1$ \\
    $\alpha$ (rad) & $1.890 \pm 0.005$ & $1.889 \pm 0.005$ \\
    $s$ & $0.8819 \pm 0.0044$ & $0.9146 \pm 0.0050$ \\
    $\log q$ & $-4.295 \pm 0.072$ & $-4.302 \pm 0.075$ \\
    $I_{\rm S, OGLE}$ & $20.43 \pm 0.05$ & $20.43 \pm 0.05$ \\
    \hline
    \end{tabular}
    %\tablecomments{$\sigma(\pi_{\rm E, \parallel}) \sim 0.2$}
    \label{parm2}
\end{table}

Figure \ref{lc2} shows a $\Delta I \sim 0.12$ mag dip at $t_{\rm anom} \sim 7947.00$, with a duration of $\Delta t_{\rm dip} \sim 0.74$ days. The dip is defined by the KMTA and KMTC data, and the subtle ridges are supported by both the KMTC and KMTS data. These data were taken in good seeing ($1.''4$--$2.''5$) and the anomaly does not correlate with seeing, sky background or airmass. In addition, \cite{MB20135} found that the KMTA data show systematic errors and excluded them from the analysis. In that case, the KMTA data exhibit similar residuals from one-night data in many places of the light curves. For the present case, the anomaly is mainly covered by the KMTA data, but as presented in Section \ref{anomaly}, there is no similar deviation in other places of the light curves. We also carefully checked the KMTA data but did not find any similar residuals. Hence, the anomaly is secure. Applying the heuristic formalism of Section \ref{preamble}, we obtain
\begin{equation}\label{Equ:parm2}
    \alpha \sim 108.4^\circ;\quad s = s_{-} \sim 0.898;\quad \log q \sim -4.19. 
\end{equation}

The 2L1S modeling yields two degenerate solutions with $\Delta \chi^2 = 0.1$. As shown in Figure \ref{cau1}, the two solutions are subjected to the inner/outer degeneracy. Their parameters are given in Table \ref{parm2}, for which $\alpha$ and $q$ are consistent with Equation (\ref{Equ:parm2}). For $s$, we note that the geometric mean of the two solutions, $s_{\rm mean} = 0.898 \pm 0.005$, is in good agreement with Equation (\ref{Equ:parm2}) and thus the formalism of \cite{KMT2021_mass1}. In addition, the observed data only provide a $3\sigma$ upper limit on $\rho$, and a point-source model is consistent within $1\sigma$. The ratio of the phase-space factors is $p_{\rm inner} : p_{\rm outer} = 0.78 : 1$.

With high-order effects, we find that the $\chi^2$ improvement is $\sim 3$ and other parameters are almost the same. The constraint of $\pie$, $\pi_{\rm E, \parallel} = -0.35 \pm 0.26$, will be used in the Bayesian analysis. This is a microlensing planet with a Neptune/Sun mass ratio.

\subsubsection{\eventc}

\begin{figure}[htb] 
    \centering
    \includegraphics[width=0.95\columnwidth]{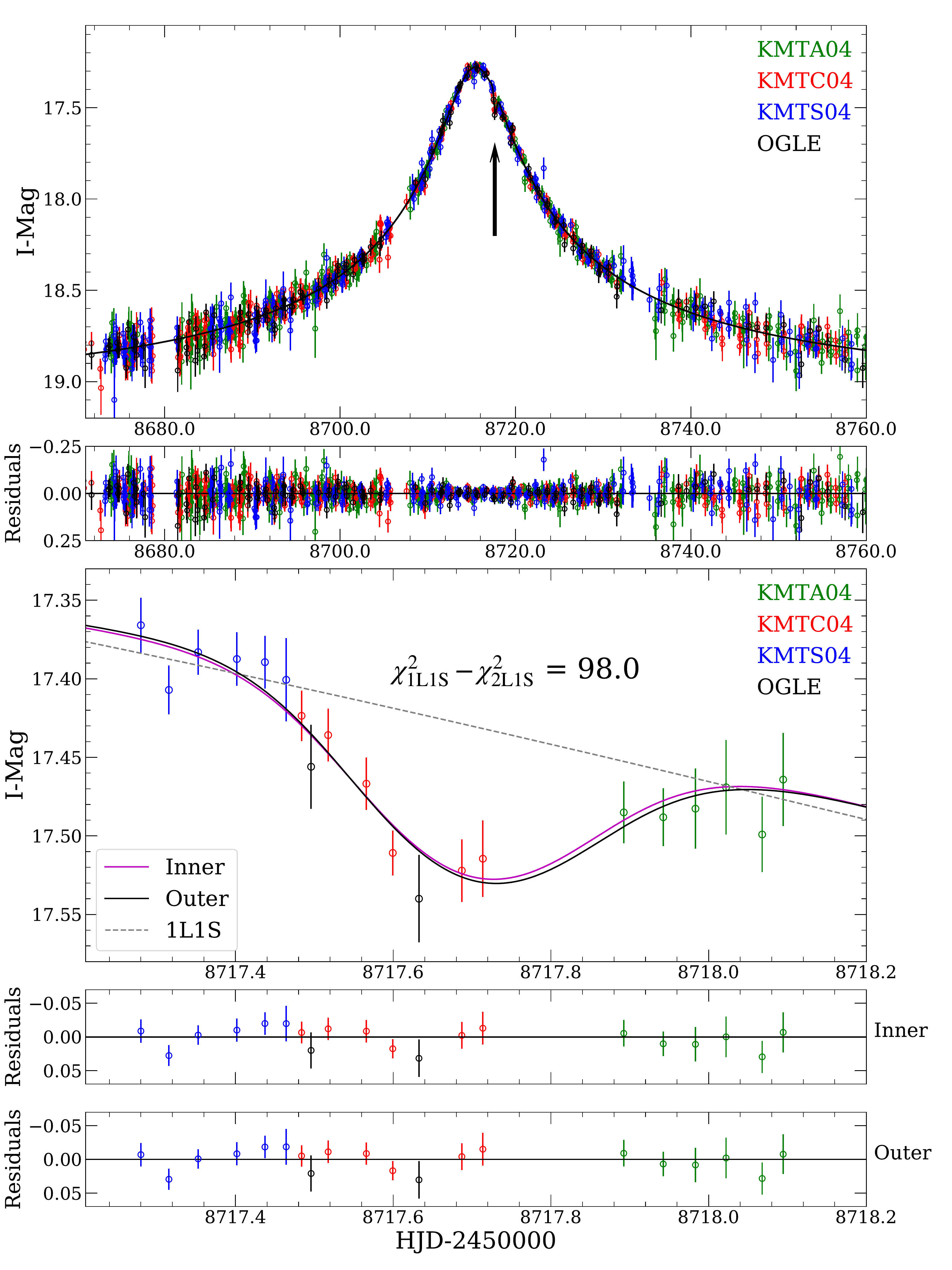}
    \caption{The observed data and models for \eventc. The symbols are similar to those in Figure \ref{lc1}. In the top panel, the black arrow indicates the position of the planetary signal.}
    \label{lc3}
\end{figure}

\begin{table*}[htb]
    \renewcommand\arraystretch{1.25}
    \centering
    \caption{2L1S Parameters \eventc}
    \begin{tabular}{c c c c c}
     \hline
    Parameter & \multicolumn{2}{c}{Inner} & \multicolumn{2}{c}{Outer} \\
    \hline
      & $u_0 > 0$ & $u_0 < 0$ & $u_0 > 0$ & $u_0 < 0$ \\
    $\chi^2$/dof & 3132.5/3132 & 3132.9/3132 & 3132.2/3132 & 3131.8/3132 \\
    \hline
    $t_{0}$ (${\rm HJD}^{\prime}$) & $8715.452 \pm 0.015$ & $8715.451 \pm 0.015$ & $8715.453 \pm 0.014$ & $8715.453 \pm 0.015$  \\
    $u_{0}$  & $0.0260 \pm 0.0017$ & $-0.0251 \pm 0.0020$ & $0.0257 \pm 0.0016$ & $-0.0255 \pm 0.0015$ \\
    $\te$ (days)  & $132.8 \pm 8.1$ & $138.5 \pm 10.8$ & $134.1 \pm 7.9$ & $135.6 \pm 7.9$ \\
    $\rho (10^{-3})$ & $<1.8$ & $<1.8$ & $ <1.9$ & $<1.7$ \\
    $\alpha$ (rad) & $2.150 \pm 0.008$ & $-2.147 \pm 0.008$ & $2.151 \pm 0.009$ & $-2.148 \pm 0.008$ \\
    $s$ & $0.9377 \pm 0.0069$ & $0.9383 \pm 0.0073$ & $1.0339 \pm 0.0084$ & $1.0352 \pm 0.0085$ \\
    $\log q$ & $-4.724 \pm 0.117$ & $-4.734 \pm 0.109$ & $-4.717 \pm 0.117$ & $-4.714 \pm 0.116$ \\
    $\pi_{\rm E, N}$ & $-0.055 \pm 0.150$ & $-0.066 \pm 0.161$ & $-0.060 \pm 0.156$ & $-0.019 \pm 0.160$ \\
    $\pi_{\rm E, E}$ & $-0.058 \pm 0.017$ & $-0.059 \pm 0.014$ & $-0.057 \pm 0.017$ & $-0.060 \pm 0.013$ \\
    %$\pi_{\rm E, \parallel}$ & $-0.065 \pm 0.013$ & $-0.067 \pm 0.016$ & & \\
    %$\pi_{\rm E, \perp}$ & $-0.046 \pm 0.150$ & & & \\
    %$\psi$ (deg) & $81.98 \pm 0.02$ & $82.55 \pm 0.02$ & & \\
    $I_{\rm S}$ & $21.33 \pm 0.07$ & $21.37 \pm 0.09$ & $21.34 \pm 0.07$ & $21.35 \pm 0.07$ \\
    %$I_{\rm B}$  & & & $19.16 \pm 0.01$ & $19.16 \pm 0.01$ \\
    \hline
    \end{tabular}
    %\tablecomments{The upper limits on $\rho$ are $3\sigma$.}
    \label{parm3}
\end{table*}

The anomaly of \eventc\ is also a dip, as shown in Figure \ref{lc3}. The dip has $\Delta t_{\rm dip} \sim 0.6$ days and centers on $t_{\rm anom} \sim 8717.72$. The dip is defined by the KMTC data and the two contemporaneous OGLE points, which were taken in good seeing ($1.''1$--$2.''4$) and low sky background. Hence, the anomaly is secure. Applying the heuristic formalism of Section \ref{preamble}, we obtain
\begin{equation}\label{Equ:parm3}
    \alpha \sim 123.3^\circ;\quad s = s_{-} \sim 0.985;\quad \log q \sim -4.56. 
\end{equation}
In addition, given the Einstein timescale ($\te \sim 135$ days), we expect that $\pie$ should be either measured or strongly constrained.

The 2L1S modeling also finds a pair of inner/outer solutions and combined the $u_0 > 0$ and $u_0 < 0$ degeneracy there are four solutions in total. See Table \ref{parm3} for their parameters. The inclusion of $\pie$ improves the fits by $\Delta\chi^2 = 31$, and all four data sets contribute to the improvement, so the parallax signal is reliable. The angle of the minor axis of the parallax ellipse (north through east) is $\psi = 82.0^\circ$ and $\psi = 82.5^\circ$ for the $u_0 > 0$ and $u_0 < 0$ solutions, respectively. $\pi_{\rm E, \parallel} = 0.06 \pm 0.01$, and $\pi_{\rm E, \bot}$ is constrained to be $\sigma(\pi_{\rm E, \bot}) \sim 0.2$. We obtain $s_{\rm mean} = 0.985 \pm 0.008$, $\alpha = 123.1 \pm 0.5$, and $\log q \sim -4.72$, in good agreement with Equation (\ref{Equ:parm3}). The ratio of the phase-space factors is $p_{\rm inner} : p_{\rm outer} = 0.80 : 1$.

We find that the inclusion of the lens orbital motion effect only improves the fit by $\Delta\chi^2 < 1$ for 2 degree-of-freedom and is not correlated with $\pie$, so we exclude the lens orbital motion effect. With $q \sim 1.9 \times 10^{-5}$, this new planet is the fifth robust $q < 2 \times 10^{-5}$ microlensing planet.

\subsubsection{\eventd}

\begin{figure}[htb] 
    \centering
    \includegraphics[width=0.95\columnwidth]{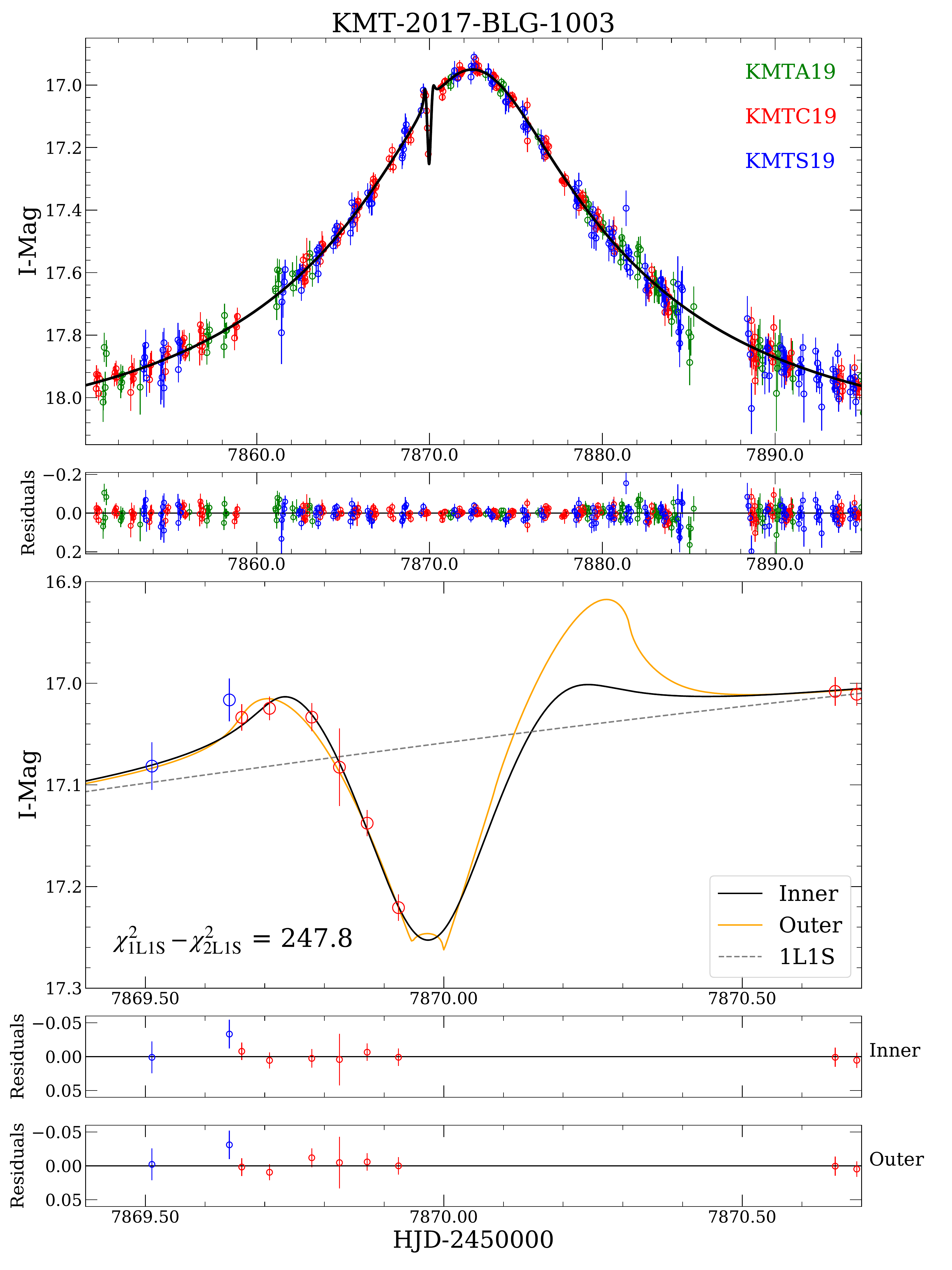}
    \caption{Light curve and models for \eventd. The symbols are similar to those in Figure \ref{lc1}.}
    \label{lc4}
\end{figure}

\begin{table}[htb]
    \renewcommand\arraystretch{1.25}
    \centering
    \caption{2L1S Parameters for \eventd}
    \begin{tabular}{c c c}
     \hline
    Parameter & Inner & Outer \\
    \hline
    $\chi^2$/dof & 2433.2/2433 & 2433.0/2433 \\
    \hline
    $t_{0}$ (${\rm HJD}^{\prime}$) & $7872.484 \pm 0.020$ & $7872.482 \pm 0.020$ \\
    $u_{0}$  & $0.179 \pm 0.005$ & $0.179 \pm 0.005$ \\
    $\te$ (days)  & $25.65 \pm 0.57$ & $25.66 \pm 0.59$ \\
    $\rho (10^{-3})$ & $< 6.7$ & $5.22 \pm 1.16$ \\
    $\alpha$ (rad) & $1.073 \pm 0.006$ & $1.072 \pm 0.006$ \\
    $s$ & $0.8889 \pm 0.0043$ & $0.9096 \pm 0.0045$ \\
    $\log q$ & $-4.260 \pm 0.152$ & $-4.373 \pm 0.144$ \\
    $I_{\rm S, OGLE}$ & $19.30 \pm 0.04$ & $19.30 \pm 0.04$ \\
    \hline
    \end{tabular}
    %\tablecomments{$\sigma(\pi_{\rm E, \parallel}) \sim 0.1$}
    \label{parm4}
\end{table}

Figure \ref{lc4} shows the light curve and the best-fit models for \eventd. The KMTC data show a sudden dip and the ridge is confirmed by the KMTC and KMTS data. These data were taken in good seeing ($1.''2$--$2.''2$) and low sky background, so the anomaly is of astrophysical origin. Although the end of the dip is not covered, the KMTC point at $\hjd = 7870.66$ indicates $\Delta t_{\rm dip} < 0.85$ days, which yields 
\begin{equation}\label{Equ:parm4}
    \alpha \sim 61.3^\circ;\quad s = s_{-} \sim 0.903;\quad \log q < -3.6. 
\end{equation}

The numerical analysis yields a pair of inner/outer solutions, and Table \ref{parm4} lists their parameters. As shown in Figure \ref{cau1}, the ``Outer'' solution has caustic crossings, so its $\rho$ is measured at the $4.5\sigma$ level. For the ``Inner'' solution, a point-source model is consistent within $2\sigma$. We note that the geometric mean of $s$, $s_{\rm mean} = 0.899 \pm 0.004$, which is slightly different from $s_{-}$ by $1\sigma$. This indicates that the prediction of \cite{KMT2021_mass1} might be imperfect for minor-image anomalies with finite-source effects or incomplete coverage. The ratio of the phase-space factors is $p_{\rm inner} : p_{\rm outer} = 0.80 : 1$.

With high-order effects, the $\chi^2$ improvement is 1.7. Although this event is shorter than the first two events, $\pie$ is better constrained due to the about one magnitude brighter data, with $\pi_{\rm E, \parallel} = -0.11 \pm 0.15$. This is another Neptune/Sun mass-ratio planet.

\subsubsection{\evente}

\begin{figure}[htb] 
    \centering
    \includegraphics[width=0.95\columnwidth]{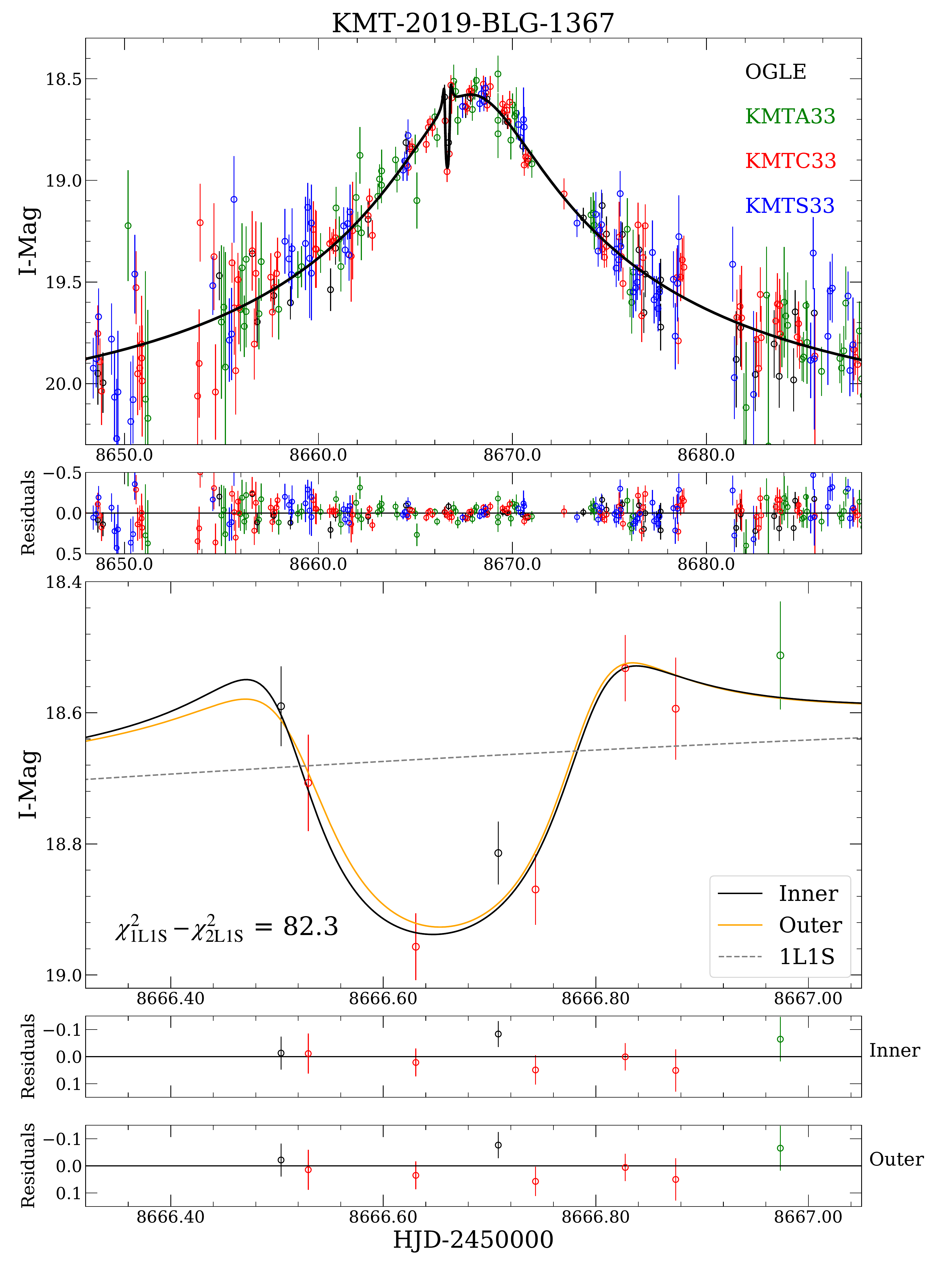}
    \caption{Light curve and models for \evente. The symbols are similar to those in Figure \ref{lc1}.}
    \label{lc5}
\end{figure}

\begin{table}[htb]
    \renewcommand\arraystretch{1.25}
    \centering
    \caption{2L1S Parameters for \evente}
    \begin{tabular}{c c c}
     \hline
    Parameter & Inner & Outer \\
    \hline
    $\chi^2$/dof & 1404.0/1404 & 1404.2/1404 \\
    \hline
    $t_{0}$ (${\rm HJD}^{\prime}$) & $8667.883 \pm 0.051$ & $8667.884 \pm 0.048$ \\
    $u_{0}$  & $0.083 \pm 0.009$ & $0.082 \pm 0.009$ \\
    $\te$ (days)  & $39.3 \pm 3.8$ & $39.8 \pm 4.0$ \\
    $\rho (10^{-3})$ & $<5.3$ & $<5.6$ \\
    $\alpha$ (rad) & $1.208 \pm 0.016$ & $1.207 \pm 0.016$ \\
    $s$ & $0.9389 \pm 0.0066$ & $0.9763 \pm 0.0070$ \\
    $\log q$ & $-4.303 \pm 0.118$ & $-4.298 \pm 0.103$ \\
    $I_{\rm S, OGLE}$ & $21.46 \pm 0.13$ & $21.48 \pm 0.13$ \\
    \hline
    \end{tabular}
    %\tablecomments{$\sigma(\pi_{\rm E, \parallel}) > 1.5$}
    \label{parm5}
\end{table}

Figure \ref{lc5} shows a dip 1.2 days before the peak of an otherwise normal PSPL event, with a duration of $\Delta t_{\rm dip} \sim 0.35$ days. The dip-type anomaly is covered by the KMTC data and one contemporaneous OGLE point, and these data were taken in good seeing ($<2.''0$) and low sky background. Therefore, the anomaly is secure. Applying the heuristic formalism of Section \ref{preamble}, we obtain
\begin{equation}\label{Equ:parm5}
    \alpha \sim 69.5^\circ;\quad s = s_{-} \sim 0.957;\quad \log q \sim -4.33. 
\end{equation}

The 2L1S modeling also yields a pair of inner/outer solutions, with $\Delta\chi^2 = 0.2$. The resulting solutions are given in Table \ref{parm5} and Figure \ref{cau1}. A point-source model is consistent within $1\sigma$ and the $3\sigma$ upper limit is $\rho < 0.0056$, so we expect that the \cite{KMT2021_mass1} formula is applicable. We obtain $s_{\rm mean} = 0.957 \pm 0.007$, in good agreement with $s_{-}$. The ratio of the phase-space factors is $p_{\rm inner} : p_{\rm outer} = 0.82 : 1$. We find that the inclusion of higher-order effects only improves the fitting by $\Delta\chi^2 < 1$ and the $1\sigma$ uncertainty of parallax is $> 0.9$ at all directions, so the constraint on $\pie$ is not useful for the Bayesian analysis. This is another planet with a Neptune/Sun mass ratio.

\subsection{``Bump'' Anomalies}

For bump-type planetary signals, we also check whether the observed data can be fitted by a single-lens binary-source (1L2S) model \citep{Gaudi1998} because it can also produce such anomalies (e.g., \citealt{MB12486, OB160733, MB06074}). For a 1L2S model, its magnification, $A_\lambda$, is the superposition of magnifications for two single-lens single-source (1L1S) models, 
\begin{equation}
    A_\lambda = \frac{A_1f_{1,\lambda}+A_2f_{2,\lambda}}{f_{1,\lambda}+f_{2,\lambda}} = \frac{A_{1}+q_{f,\lambda}A_{2}}{1+q_{f,\lambda}}; \quad     q_{f,\lambda} \equiv \frac{f_{2,\lambda}}{f_{1,\lambda}},
\end{equation}
where $f_{{\rm i},\lambda}$ is the source flux at wavelength $\lambda$, and $i=1$ and $i=2$ correspond to the primary and the secondary sources, respectively.

\subsubsection{\eventf}

\begin{figure}[htb] 
    \centering
    \includegraphics[width=0.95\columnwidth]{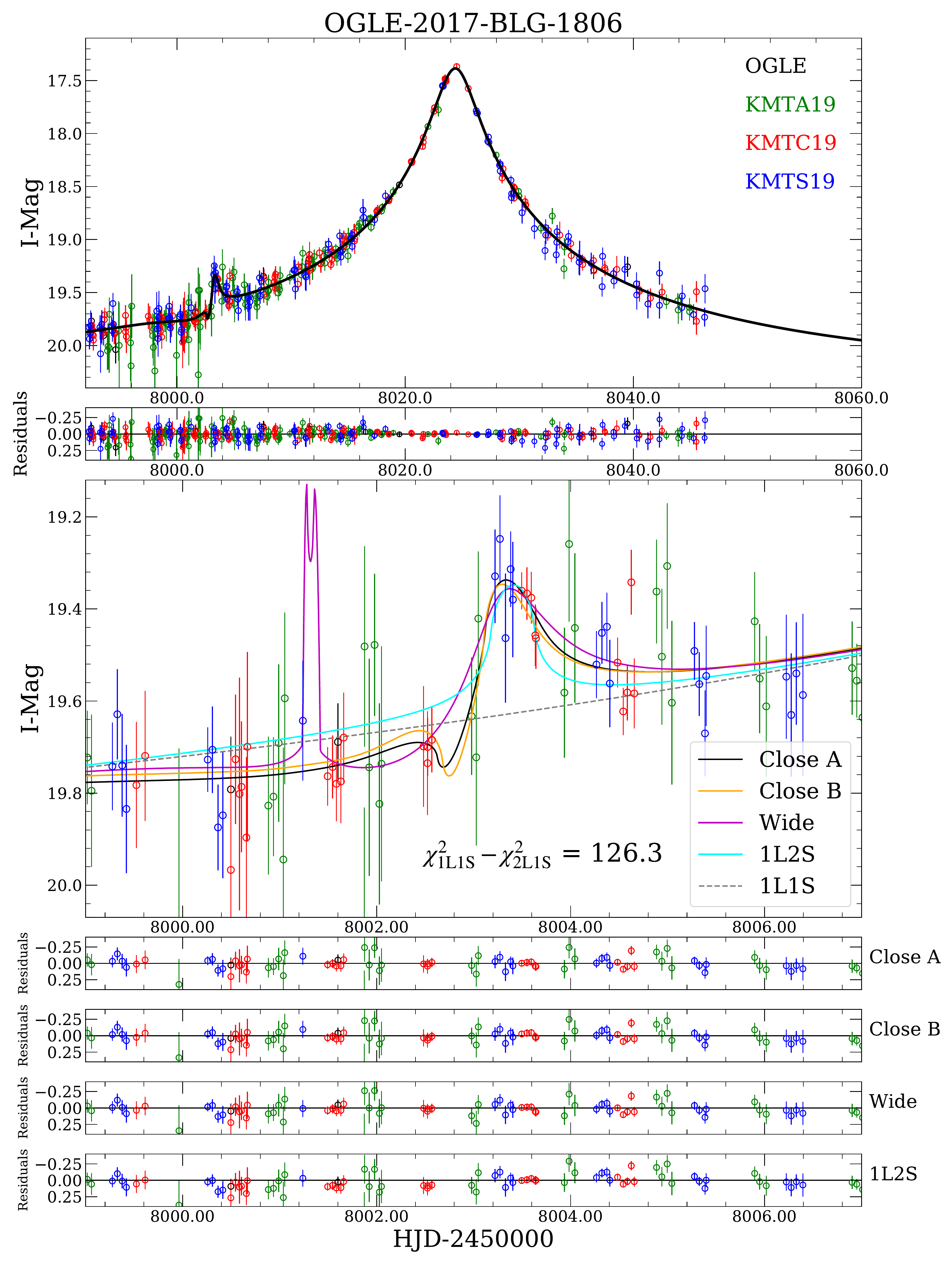}
    \caption{Light curve and models for \eventf. The symbols are similar to those in Figure \ref{lc1}. Different with the previous four events, the anomaly is bump-type, so the best-fit 1L2S model is provided.}
    \label{lc6}
\end{figure}

\begin{figure}[htb] 
    \centering
    \includegraphics[width=0.95\columnwidth]{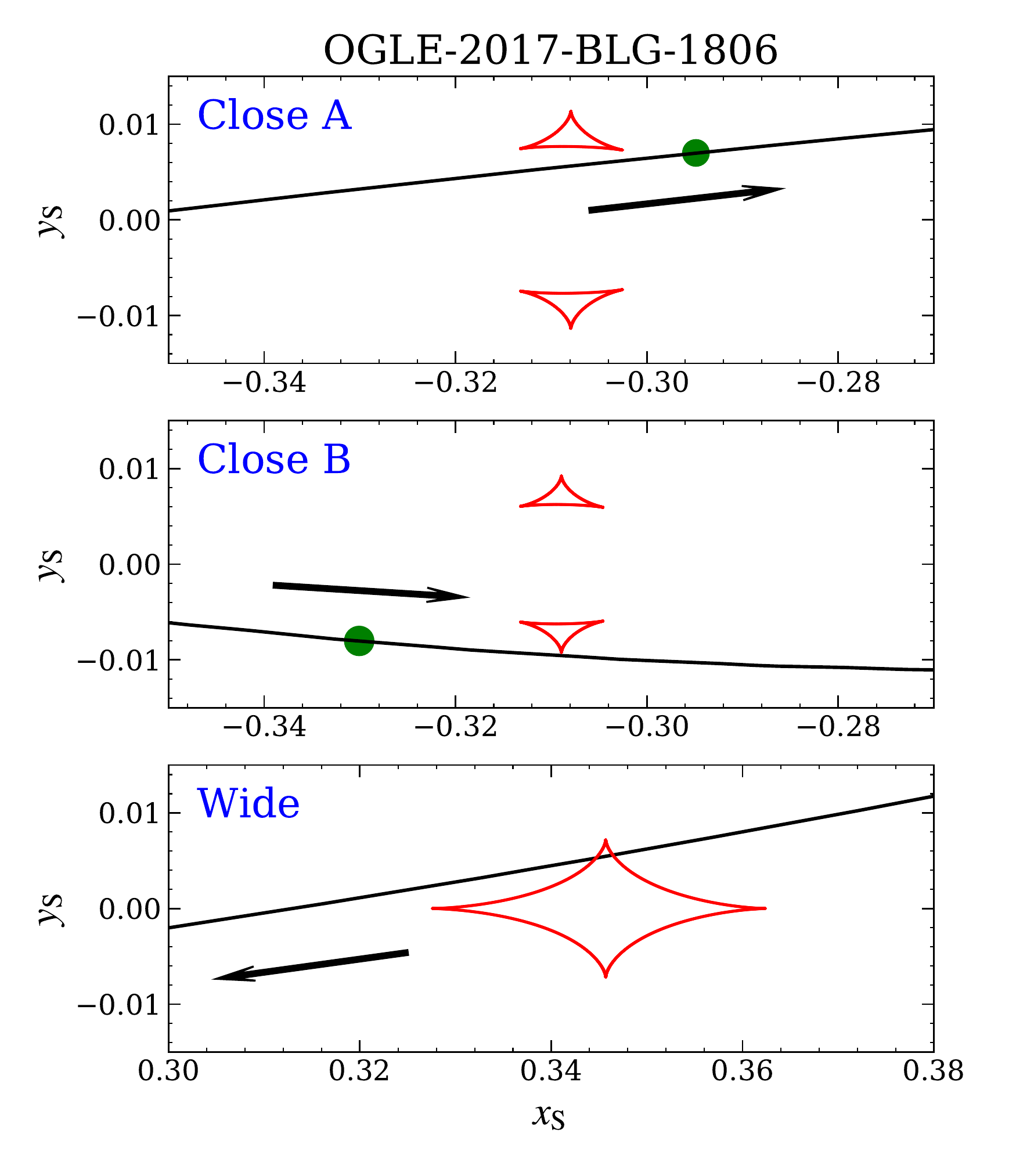}
    \caption{Geometries of \eventf. The symbols are similar to those in Figure \ref{cau1}. For the two ``Close'' solutions, $\rho$ is constrained at the $>3\sigma$ level, so the radius of the two green dots represent the source radius. For the ``Wide'' solution, $\rho$ only has weak constraints with $<3\sigma$, so its source radius is not shown.}
    \label{cau2}
\end{figure}

\begin{table*}[htb]
    \renewcommand\arraystretch{1.25}
    \centering
    \caption{2L1S Parameters for \eventf}
    \begin{tabular}{c c c c c c c}
     \hline
    Parameter & \multicolumn{2}{c}{Close A} & \multicolumn{2}{c}{Close B} & \multicolumn{2}{c}{Wide} \\
    \hline
      & $u_0 > 0$ & $u_0 < 0$ & $u_0 > 0$ & $u_0 < 0$ & $u_0 > 0$ & $u_0 < 0$ \\
    $\chi^2$/dof & 1650.9/1651 & 1650.7/1651 & 1664.8/1651 & 1665.5/1651 & 1659.1/1651 & 1659.0/1651 \\
    \hline
    $t_{0}$ (${\rm HJD}^{\prime}$) & $8024.392 \pm 0.020$ & $8024.393 \pm 0.019$ & $8024.388 \pm 0.020$ & $8024.388 \pm 0.020$ & $8024.388 \pm 0.020$ & $8024.379 \pm 0.020$  \\
    $u_{0}$  & $0.0249 \pm 0.0016$ & $-0.0260 \pm 0.0016$ & $0.0256 \pm 0.0020$ & $-0.0253 \pm 0.0019$ & $0.0269 \pm 0.0018$ & $-0.0257 \pm 0.0017$ \\
    $\te$ (days)  & $69.4 \pm 4.0$ & $66.8 \pm 3.9$ & $69.4 \pm 4.8$ & $69.6 \pm 4.6$ & $64.5 \pm 3.9$ & $67.0 \pm 3.9$ \\
    $\rho (10^{-3})$ & $1.74_{-0.44}^{+0.78}$ & $1.83_{-0.50}^{+0.80}$ & $1.50_{-0.47}^{+0.62}$ & $1.65_{-0.50}^{+0.67}$ & $ <2.8$ & $<2.4$ \\
    $\alpha$ (rad) & $0.001 \pm 0.034$ & $-0.002 \pm 0.037$ & $0.267 \pm 0.066$ & $-0.263 \pm 0.068$ & $3.121 \pm 0.034$ & $-3.121 \pm 0.036$ \\
    $s$ & $0.8609 \pm 0.0069$ & $0.8566 \pm 0.0075$ & $0.8592 \pm 0.0085$ & $0.8601 \pm 0.0080$ & $1.1900 \pm 0.0117$ & $1.1806 \pm 0.0108$ \\
    $\log q$ & $-4.392 \pm 0.180$ & $-4.352 \pm 0.171$ & $-4.766 \pm 0.220$ & $-4.768 \pm 0.209$ & $-4.317 \pm 0.126$ & $-4.441 \pm 0.168$ \\
    $\pi_{\rm E, N}$ & $-0.278 \pm 0.148$ & $0.292 \pm 0.170$ & $0.774 \pm 0.315$ & $-0.756 \pm 0.326$ & $-0.535 \pm 0.175$ & $0.504 \pm 0.170$ \\
    $\pi_{\rm E, E}$ & $0.105 \pm 0.056$ & $0.144 \pm 0.058$ & $0.157 \pm 0.070$ & $0.124 \pm 0.059$ & $0.120 \pm 0.065$ & $0.133 \pm 0.056$ \\
    %$\pi_{\rm E, \parallel}$ & $-0.065 \pm 0.013$ & $-0.067 \pm 0.016$ & & \\
    %$\pi_{\rm E, \perp}$ & $-0.046 \pm 0.150$ & & & \\
    %$\psi$ (deg) & $81.98 \pm 0.02$ & $82.55 \pm 0.02$ & & \\
    $I_{\rm S, KMTC}$ & $21.12 \pm 0.07$ & $21.07 \pm 0.07$ & $21.10 \pm 0.08$ & $21.10 \pm 0.08$ & $21.03 \pm 0.07$ & $21.08 \pm 0.07$ \\
    %$I_{\rm B}$  & & & $19.16 \pm 0.01$ & $19.16 \pm 0.01$ \\
    \hline
    \end{tabular}
    %\tablecomments{The upper limits on $\rho$ are $3\sigma$.}
    \label{parm6}
\end{table*}

\begin{table*}[htb]
    \renewcommand\arraystretch{1.25}
    \centering
    \caption{1L2S Parameters for \eventf\ and \eventg}
    \begin{tabular}{c|c c|c}
    \hline
    \hline
    Parameters & \multicolumn{2}{c|}{\eventf} & \eventg \\
    \hline
     & $u_0 > 0$ & $u_0 < 0$ & \\
    $\chi^2$/dof & $1682.0/1651$ & $1681.4/1651$ & $2298.7/2288$ \\
    \hline
    $t_{0,1}$ (${\rm HJD}^{\prime}$)  & $8024.383 \pm 0.020$ & $8024.381 \pm 0.020$ & $7555.972 \pm 0.094$ \\
    $t_{0,2}$ (${\rm HJD}^{\prime}$) & $8003.876 \pm 0.274$ & $8003.913 \pm 0.253$ & $7547.890 \pm 0.021$  \\
    $u_{0,1}$  & $0.0288 \pm 0.0023$ & $-0.0282 \pm 0.0019$ & $0.143 \pm 0.022$  \\
    $u_{0,2}$ & $0.003 \pm 0.025$ & $-0.004 \pm 0.023$ & $0.0001 \pm 0.0007$ \\
    $\te$ (days)  & $61.2 \pm 4.3$ & $62.2 \pm 3.5$ & $44.9 \pm 5.8$ \\
    $\rho_2$ ($10^{-3}$) & $<7.3$ & $<7.0$ & $<3.3$ \\
    $q_{f,I} (10^{-3})$ & $2.76 \pm 0.76$ & $2.63 \pm 0.74$ & $1.98 \pm 0.48$ \\
    $\pi_{\rm E, N}$ & $0.041 \pm 0.388$ & $0.059 \pm 0.355$ & ... \\
    $\pi_{\rm E, E}$ & $0.111 \pm 0.072$ & $0.117 \pm 0.063$ & ...  \\
    $I_{\rm S, KMTC}$ & $20.96 \pm 0.09$ & $20.98 \pm 0.07$ & $21.31 \pm 0.18$ \\
    \hline
    \hline
    \end{tabular}
    %\tablecomments{$\sigma(\pi_{\rm E, \parallel}) \sim 0.05$}
    \label{1L2S}
\end{table*}

As shown in Figure \ref{lc6}, the light curve of \eventf\ exhibits a bump centered on $t_{\rm anom} \sim 8003.5$, defined by the KMTC and KMTS data. Except for two KMTS points, all the KMTC and KMTS data during $8003 < \hjd < 8005$ were taken in good seeing ($<2.''2$) and low sky background. In addition, most of the data before the bump ($8000 < \hjd < 8003$) are fainter than the 1L1S model. Hence, the signal is secure. Because both the major-image and the two minor-image planetary caustics can produce a bump-type anomaly (e.g., \citealt{OB180383}), we obtain
\begin{equation}\label{Equ:parm6}
    \alpha \sim 184.6^\circ~{\rm for}~s_{+} \sim 1.15;\quad\alpha \sim 4.6^\circ~{\rm for}~s_{-} \sim 0.86. 
\end{equation}

The grid search returns three local minima, and their caustic structures are given in Figure \ref{cau2}. As expected, the three solutions respectively correspond to sources crossing a major-image (quadrilateral) planetary caustic and two minor-image (triangular) planetary caustics. We label the three solutions as ``Close A'', ``Close B'', and ``Wide'', respectively, and their parameters are presented in Table \ref{parm6}. 

The ``Close A'' solution provides the best fit to the observed data, and the ``Close B'' and ``Wide'' solutions are disfavored by $\Delta\chi^2 = 14.1$ and 8.3, respectively. We find that the inclusion of the parallax effect improves the fit by $\Delta\chi^2 = $ 7.8, 20.4, and 11.1 for the ``Close A'', ``Close B'', and ``Wide'' solutions, respectively, and during the anomaly region ($7998 < \hjd < 8008$), $\Delta\chi^2 = $2.2, 22.3, and 6.8. With the anomaly removed, fitting the data by a 1L1S model yields a similar constraint on $\pi_{\rm E, \parallel}$ and a weaker constraint on $\pi_{\rm E, \perp}$, with $\sigma(\pi_{\rm E, \bot}) \sim 0.5$. Thus, the long planetary signal plays an important role in improving the constraint on $\pi_{\rm E, \bot}$ and reduces the $\chi^2$ differences between the three solutions.

The ratio of the phase-space factors is $p_{\rm Close A} : p_{\rm Close B} : p_{\rm Wide} = 1 : 0.95 : 0.61$. For the ``Close A'', and ``Close B'' solutions, the bump was produced by a caustic crossing, so $\rho$ is constrained at the $>3\sigma$ level. For the ``Wide'' solution, the bump was a result of a cusp approach. Although the ``Wide'' solution has caustic crossing features, due to the lack of data during the crossing, a point-source model is consistent within $1\sigma$. 

The 1L2S model is disfavored by $\Delta\chi^2 = 30.7$ compared to the ``Close A'' solution, and the 1L2S parameters are shown in Table \ref{1L2S}. Although the 1L2S model fits the bump well, it provides a worse fit to the observed data before the bump, during which most of the data from the three KMTNet sites are fainter than the 1L2S model. Hence, the 1L2S model is rejected. We find that the lens orbital motion effect is not detectable ($\Delta\chi^2 < 0.5$), so we adopt the parameters with the microlensing parallax effect as our final results.

\subsubsection{\eventg}

\begin{figure}[htb] 
    \centering
    \includegraphics[width=0.95\columnwidth]{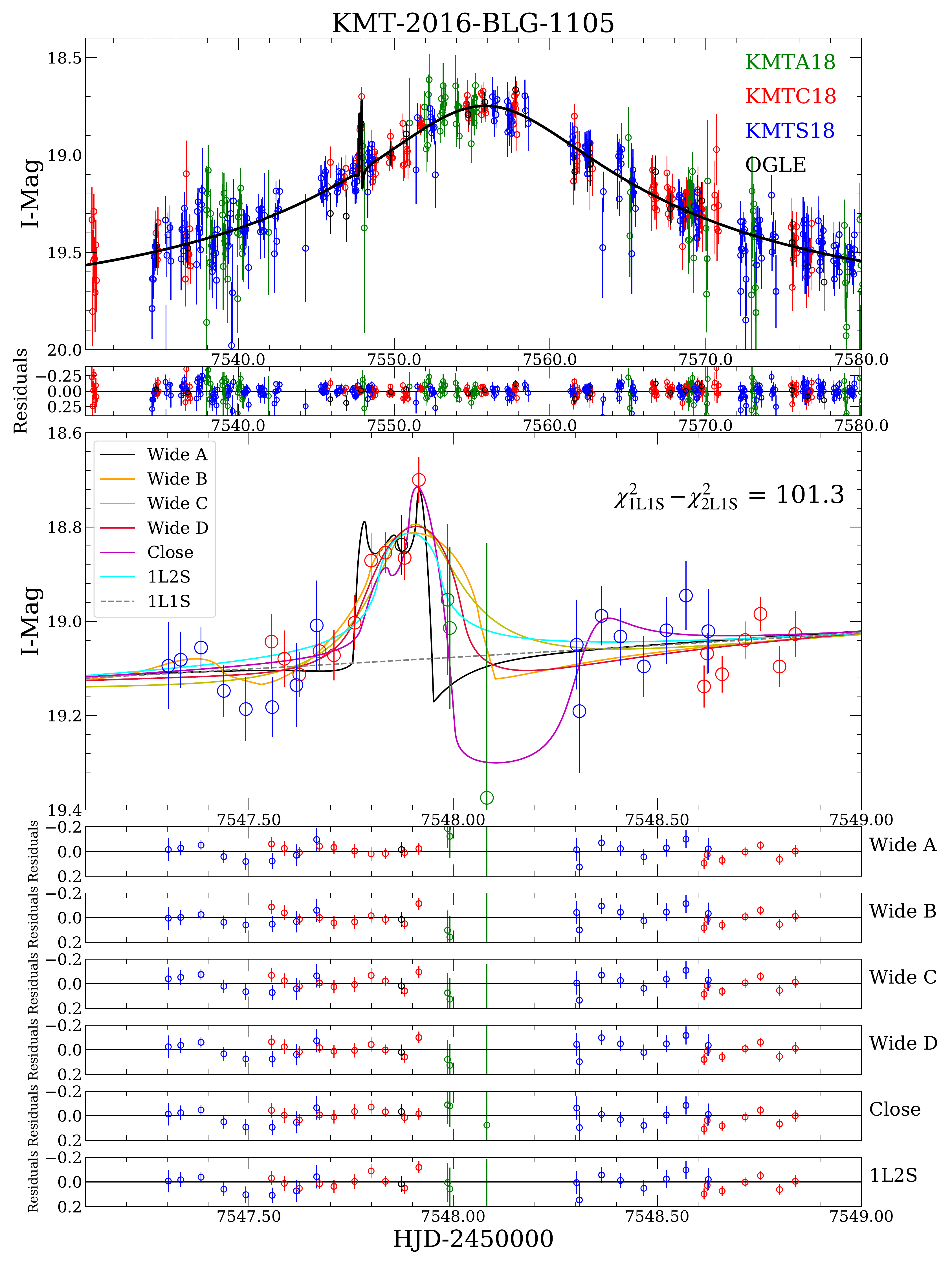}
    \caption{Light curve and models for \eventg. The symbols are similar to those in Figure \ref{lc1}. Because a 1L2S model can produce a short-lived bump, the best-fit 1L2S model is also shown.}
    \label{lc7}
\end{figure}

\begin{figure}[htb] 
    \centering
    \includegraphics[width=0.95\columnwidth]{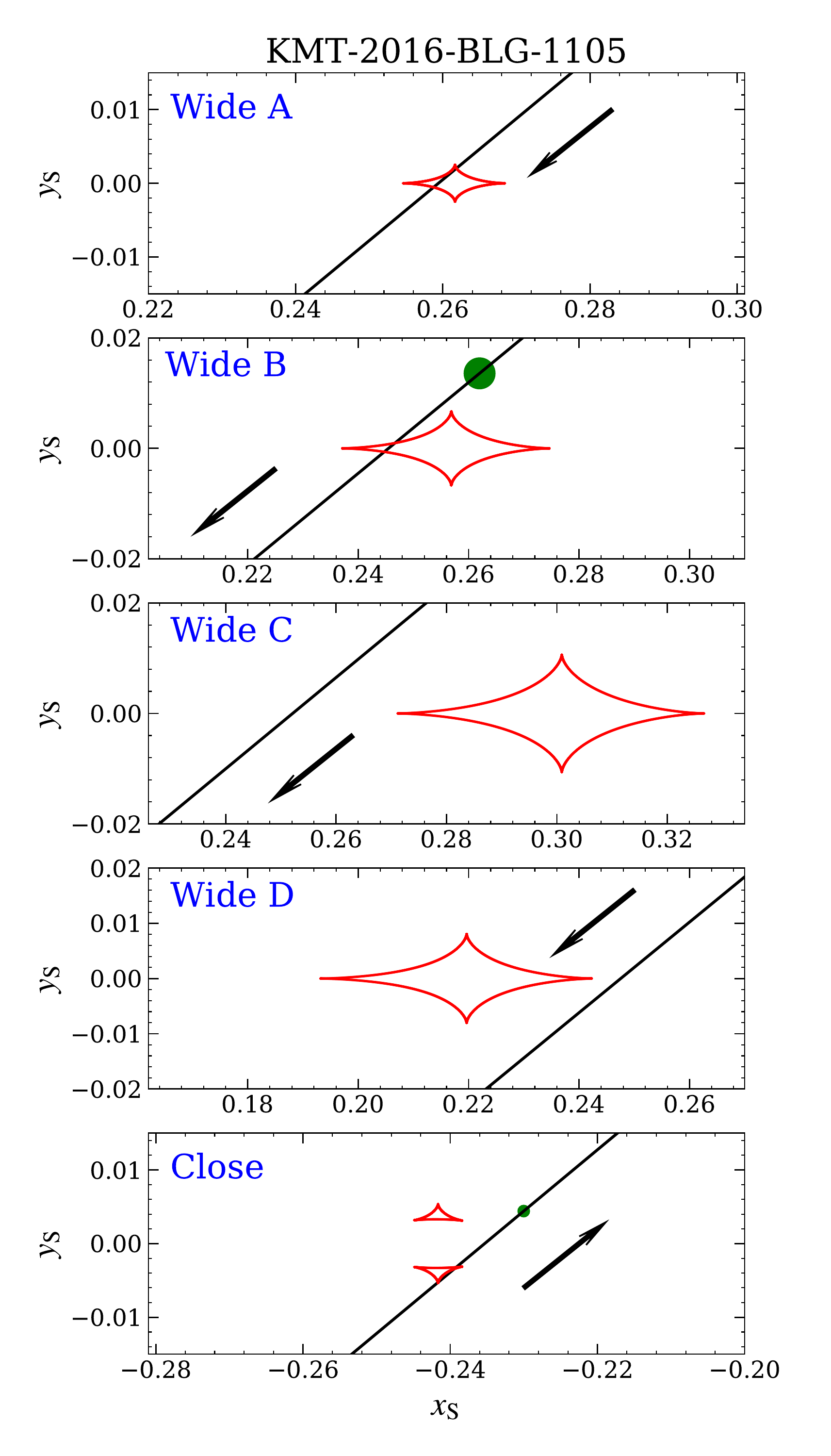}
    \caption{Geometries of \eventg. The symbols are similar to those in Figure \ref{cau1}.}
    \label{cau3}
\end{figure}

\begin{table*}[htb]
    \renewcommand\arraystretch{1.25}
    \centering
    \caption{2L1S Parameters for \eventg}
    \begin{tabular}{c|c c c c c}
    \hline
    \hline
    Parameters & Wide A & Wide B & Wide C & Wide D & Close \\
    \hline
    $\chi^2$/dof & $2286.7/2288$ & $2289.0/2288$ & $2291.1/2288$ & $2289.4/2288$ & $2290.2/2288$ \\
    \hline
    $t_{0}$ (${\rm HJD}^{\prime}$)  & $7555.834 \pm 0.096$ & $7555.789 \pm 0.102$ & $7555.772 \pm 0.093$ & $7555.781 \pm 0.099$ & $7555.896 \pm 0.093$ \\
    $u_{0}$  & $0.171 \pm 0.012$ & $0.153 \pm 0.013$ & $0.154 \pm 0.014$ & $0.154 \pm 0.014$ & $0.148 \pm 0.008$ \\
    $\te$ (days)  & $38.8 \pm 2.0$ & $42.4 \pm 2.9$ & $42.5 \pm 3.1$ & $42.4 \pm 3.1$ & $43.3 \pm 1.8$ \\
    $\rho_1$ ($10^{-3}$)  & $<2.4$ & $2.92 \pm 0.82$ & $<4.6$ & $<5.5$ & $0.75 \pm 0.14$ \\
    $\alpha$ (rad)  & $3.836 \pm 0.014$ & $3.830 \pm 0.016$ & $3.832 \pm 0.014$ & $3.831 \pm 0.014$ & $0.691 \pm 0.021 $ \\
    $s$ & $1.143 \pm 0.009$ & $1.136 \pm 0.011$ & $1.155 \pm 0.012$ & $1.106 \pm 0.013$ & $0.888 \pm 0.007$\\
    $\log q$ & $-5.194 \pm 0.248$ & $-4.423 \pm 0.197$ & $-4.069 \pm 0.182$ & $-4.184 \pm 0.206$ & $-5.027 \pm 0.080$ \\
    $I_{\rm S, KMTC}$ & $21.09 \pm 0.08$ & $21.20 \pm 0.05$ & $21.22 \pm 0.11$ & $21.22 \pm 0.11$ & $21.27 \pm 0.06$\\
    \hline
    \hline
    \end{tabular}
    %\tablecomments{$\sigma(\pi_{\rm E, \parallel}) > 0.9$}
    \label{parm7}
\end{table*}

The anomaly in Figure \ref{lc7} is a short-lived bump centered on $t_{\rm anom} \sim 7547.85$, which is defined by four KMTC data points and supported by one OGLE data point. These data were taken in good seeing ($<2.''0$) and low sky background, so the anomaly is secure. Similar to \eventf, we expect both the major-image and the minor-image planetary caustics can produce the bump and obtain 
\begin{equation}\label{Equ:parm7}
    \alpha \sim 219.3^\circ~{\rm for}~s_{+} \sim 1.13;\quad\alpha \sim 39.3^\circ~{\rm for}~s_{-} \sim 0.89. 
\end{equation}

The 2L1S modeling yields five solutions, including one with the minor-image planetary caustics and four with the major-image planetary caustics. We label them as ``Close'', ``Wide A'', ``Wide B'', ``Wide C'' and ``Wide D'', respectively, and their parameters are given in Table \ref{parm7}. Figure \ref{cau3} displays the caustic structures and source trajectories. The ``Wide A'', ``Wide B'' and ``Close'' solutions exhibit caustic crossings, but only for the ``Wide B'' and ``Close'' solutions $\rho$ are constrained at the $>3\sigma$ level. For the ``Wide A'', ``Wide C'' and ``Wide D'' solutions,  a point-source model is consistent within $\Delta\chi^2 = $ 3, 1, and 1, respectively, and thus we only report their $3\sigma$ upper limit on $\rho$ in Table \ref{parm7}. The ratio of the phase-space factors is $p_{\rm Wide A} : p_{\rm Wide B} : p_{\rm Wide C} : p_{\rm Wide D} : p_{\rm Close} = 0.82 : 0.76 : 0.74 : 1 : 0.41$, so the wide solutions are slightly favored in the phase-space factors.

For the ``Close'' solution, the bump was produced by a cusp approach with the lower triangular planetary caustic, followed by a dip that occurred in the data gap between $\hjd = 7548.0$ and $\hjd = 7548.3$. If the bump were produced by a cusp approach with the upper triangular planetary caustic, there would be a dip before the bump, but the region before the bump is well covered by the KMTS and the KMTC data, which are consistent with the 1L1S model. Thus, the minor-image perturbation only has one solution. 

\begin{figure}[htb] 
    \centering
    \includegraphics[width=\columnwidth]{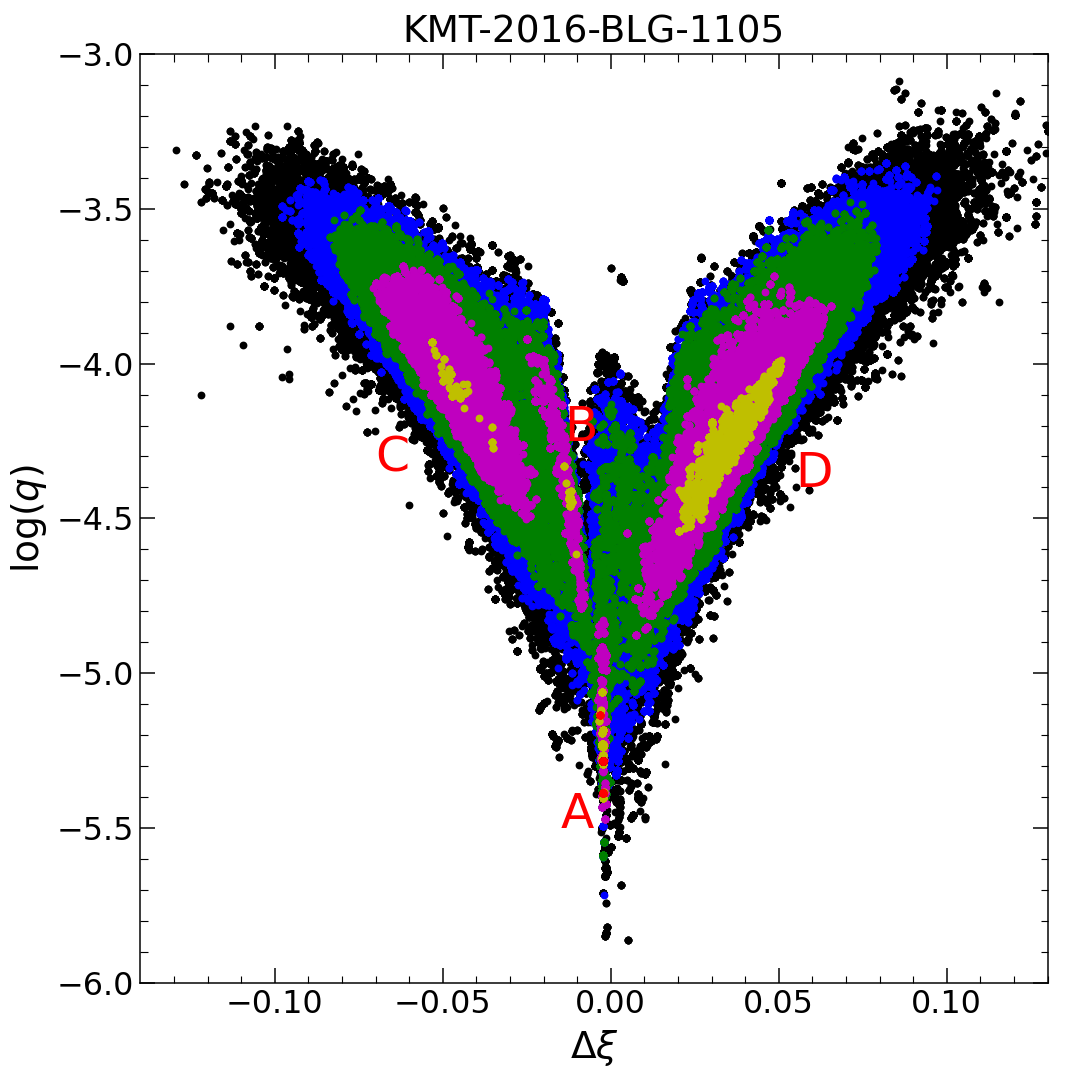}
    \caption{Scatter plot of $\log q$ vs. $\Delta\xi$ for \eventg. The distribution is derived by inflating the error bars by a factor of $\sqrt{2.5}$ and then multiplying the resulting $\chi^2$ by 2.5 for the plot. The colors are the same as those in Figure \ref{xi_kb171194}. ``A'', ``B'', ``C'', and ``D'' represent four local minima and the corresponding parameters are given in Table \ref{parm7}.}
    \label{xi_kb161105}
\end{figure}

For the four ``Wide'' solutions, the ``Wide A'' and ``Wide B'' solutions have a source crossing the planetary caustic, and the ``Wide C'' and ``Wide D'' solutions that contain a source that passes to one side or the other of the planetary caustic. This topology is qualitatively similar to the topology of OGLE-2017-BLG-0173 \citep{OB170173}. We thus also investigate the parameter space by a ``hotter'' MCMC with the error bar inflated by a factor of $\sqrt{2.5}$. The resulting scatter plot is shown in Figure \ref{xi_kb161105}, from which we find that the topology of \eventg\ has differences in three aspects from the topology of OGLE-2017-BLG-0173. First, for the two solutions in which the source passes to one side or the other of the planetary caustic, OGLE-2017-BLG-0173 has caustic crossings and the source is comparable to the size of the planetary caustic, but in the present case, the source does not cross the caustic. Second, for the solution in which the source passes directly over the planetary caustic, the source is much larger than the planetary caustics in the case of OGLE-2017-BLG-0173, while the source of \eventg\ is smaller than the caustic. Third, OGLE-2017-BLG-0173 exhibits a bimodal minimum when the source passes directly over the caustic, and the mass-ratio difference between the two local minima is $\Delta \log q < 0.1$. The corresponding solutions for \eventg, the ``Wide A'' and ``Wide B'' solutions, have $\Delta \log q \sim 1$. We note that the ``Wide A'' and ``Wide B'' solutions have $\Delta\xi \sim$ 0.00 and $-$0.01, respectively. Considering the approximate symmetry with respect to $\Delta\xi$, one might expect an additional minimum that has $\Delta\xi \sim 0.01$ and a similar $\log q$ as the $\log q$ of the ``Wide B'' solution. However, such a potential solution ``disappeared'' from the numerical analysis. Because the trajectories of the ``Wide'' B solution and the putative minimum at $\Delta\xi \sim 0.01$ should be almost symmetric with respect to the center of the caustics, their corresponding planetary signals should also be almost symmetric. As shown in Figure \ref{lc7}, the ``Wide B'' solution drops rapidly during the caustic exit, followed by a dip, so the putative minimum at $\Delta\xi \sim 0.01$ should contain a dip followed by a sudden rise during the caustic entry, which is not supported by the KMTC and KMTS data. Thus, in Figure \ref{xi_kb161105} this topology is absorbed into the MCMC chain of the ``Wide D'' solution and there is no new discrete solution. 

We also check whether the bump-type anomaly can be fitted by a 1L2S model. Table \ref{1L2S} lists the 1L2S parameters. We find that the best-fit 1L2S model is disfavored by $\Delta\chi^2 = 12.0$ compared to the best-fit 2L1S model. The best-fit 1L2S model has $\rho_2 = 0.0018$. We note that the flux ratio is $q_{f,I} \sim 2 \times 10^{-3}$, corresponding to a magnitude difference of 6.7 mag. According to Section \ref{lens}, the primary source lies 4.1 mag below the red giant clump, so the putative source companion would have an absolute magnitude of $M_{I,2} \sim 10.7$ mag, corresponding to an angular source radius of $\theta_{*,2} \sim 0.1~\mu$as. This yields the lens-source relative proper motion of $\mu_{\rm rel} = \theta_{*,2}/\rho_2/\te \sim 0.5~{\rm mas\,yr^{-1}}$, which is lower than the typical $\mu_{\rm rel}$ of bulge microlensing events (See Figure 2 of \cite{Zhu2017spitzer} for examples). However, a model with $\rho_2 = 0$ is only disfavored by $\Delta\chi^2 = 1$, so any reasonable $\mu_{\rm rel}$ is only disfavored by $\Delta\chi^2 < 1$. Thus, while the planetary model is strongly favored, there is a possibility that the anomaly is caused by a second source.

With high-order effects, we find that $\Delta\chi^2 < 1$ and the $1\sigma$ uncertainty of parallax is $> 0.9$ at all directions, so the constraint on $\pie$ is not useful.

%We note that the flux ratio is $q_{f,I} \sim 1.9 \times 10^{-3}$, corresponding to a magnitude difference of 6.8 mag. According to Section \ref{lens}, the primary source lies 4.1 mag below the red giant clump, so the putative source companion would have an absolute magnitude of $M_{I,2} \sim 11$ mag, corresponding to an angular source radius of $\theta_{*,2} \sim 0.1~\mu$as. This yields the lens-source relative proper motion of $\mu_{\rm rel} = \theta_{*,2}/\rho_2/\te \sim 0.5~{\rm mas\,yr^{-1}}$. According to the Appendix of \cite{CMST}, this value have probability of $p = \frac{(\mu_{\rm rel}/\sigma_{\mu})^3}{6\sqrt{\pi}} \sim 5 \times 10^{-4}$, where $\sigma_{\mu} = 2.9~{\rm mas\,yr^{-1}}$ is the proper motion dispersion of bulge lenses and bulge sources\footnote{For disk lenses and bulge sources, the probability is even lower. See Figure 2 of \cite{Zhu2017spitzer} for an example.}. Thus, the best-fit 1L2S model can be excluded by the kinematic argument. However, for a kinematic allowed 1L2S model, which has $\mu_{\rm rel} = \sigma_{\mu}$, and $\rho_2 = 0.28 \times 10^{-3}$, we find an additional $\Delta\chi^2$ of 11.6. Thus, a strict comparison of the 2L1S and the 1L2S models favors the former by $\Delta\chi^2 \sim 7.6 + 11.6 = 19.2$.

\section{Source and Lens Properties}\label{lens}

\subsection{Preamble}

\begin{figure*}
    \centering
    \includegraphics[width=0.33\textwidth]{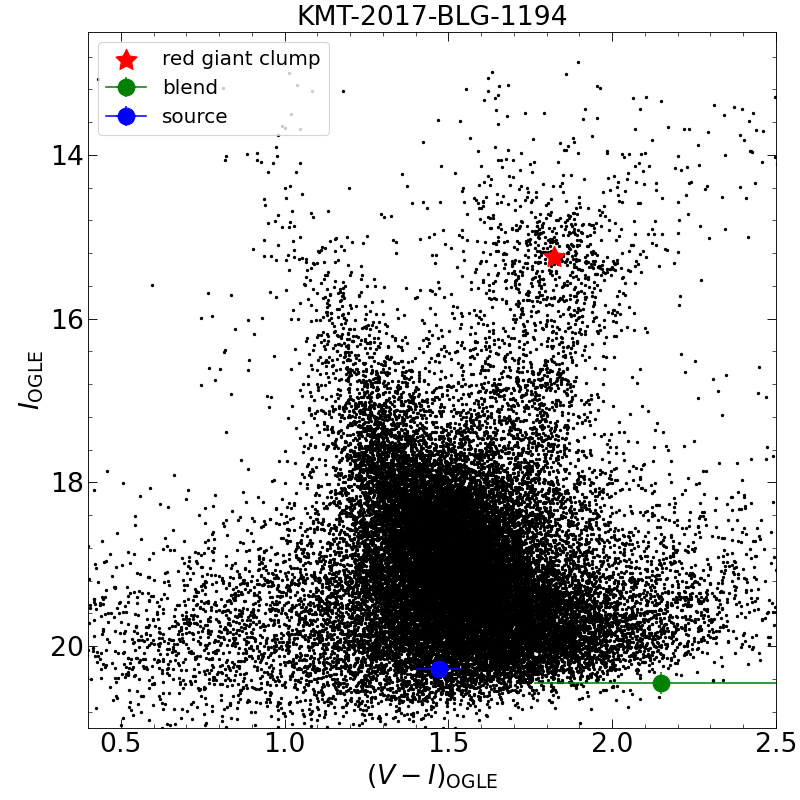}
    \includegraphics[width=0.33\textwidth]{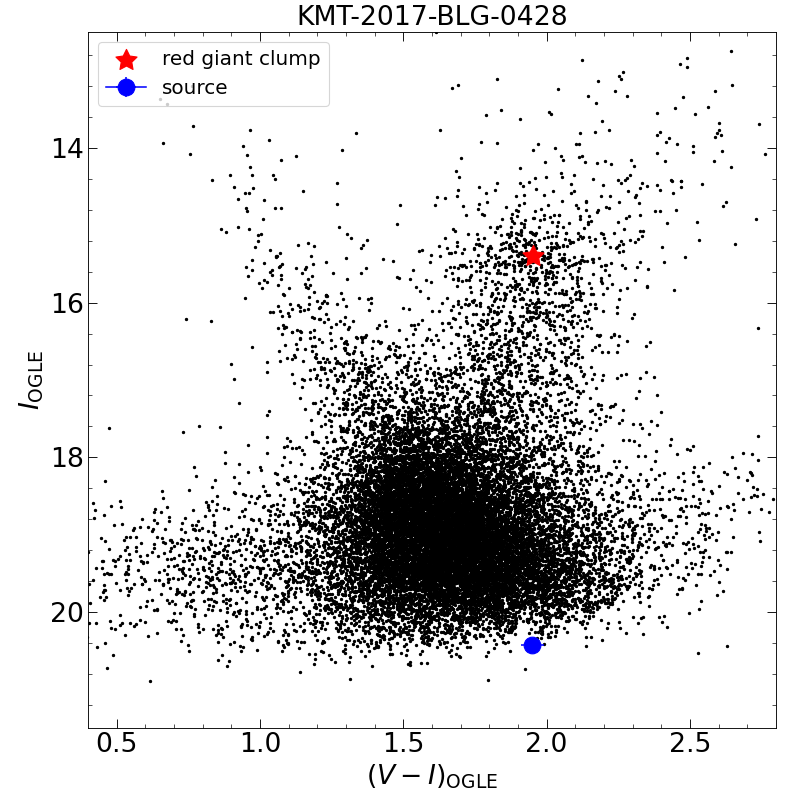}
    \includegraphics[width=0.33\textwidth]{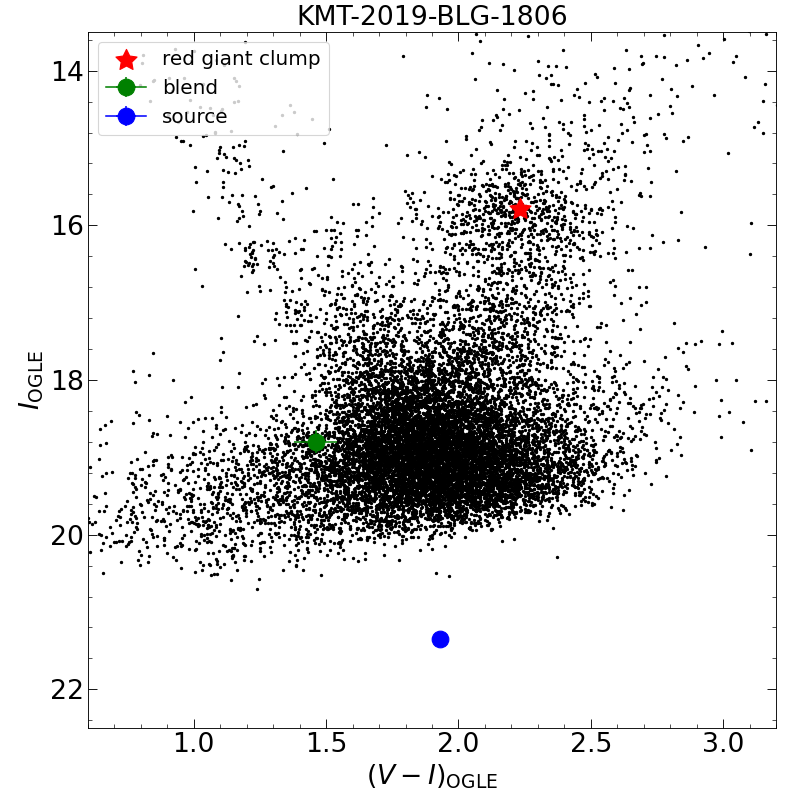}
    \includegraphics[width=0.33\textwidth]{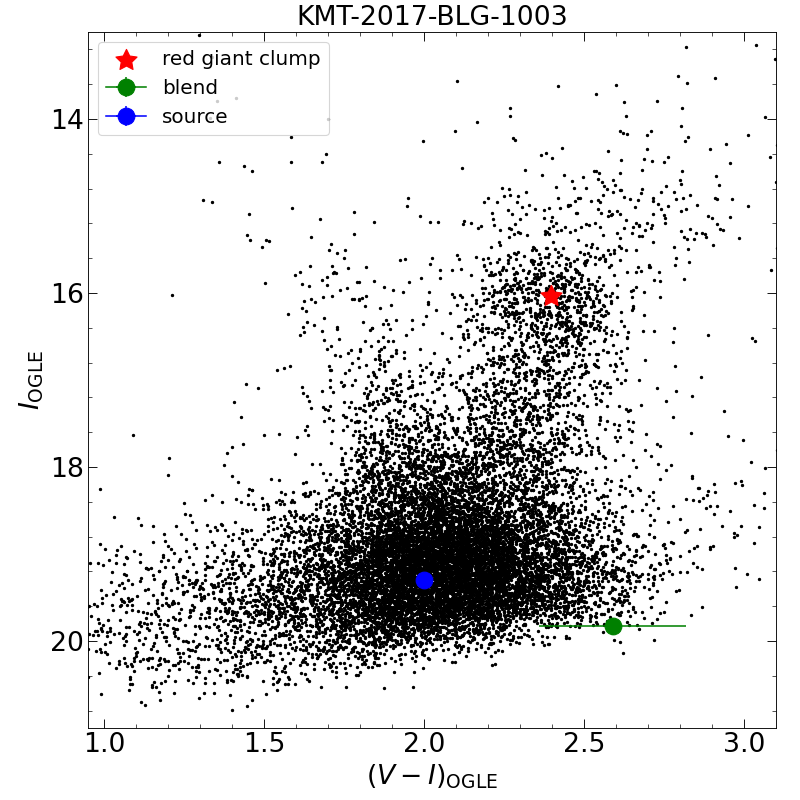}
    \includegraphics[width=0.33\textwidth]{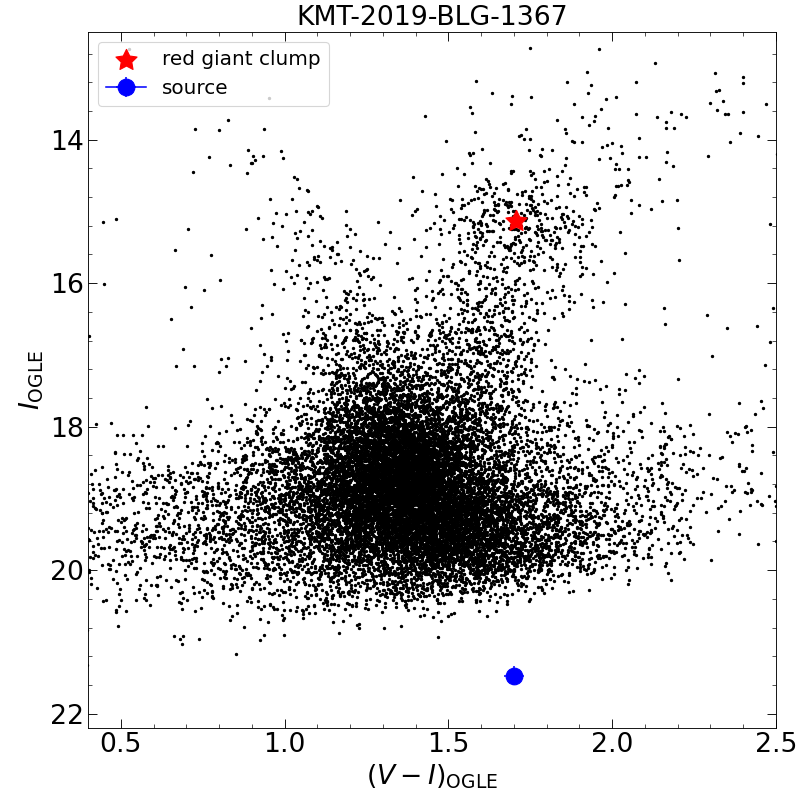}
    \includegraphics[width=0.33\textwidth]{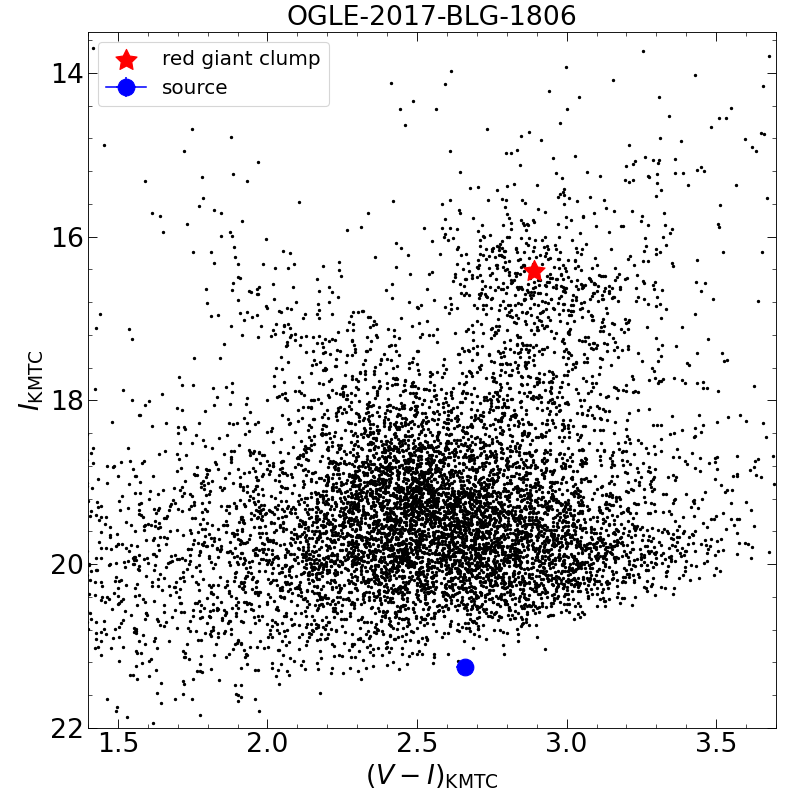}
    \includegraphics[width=0.33\textwidth]{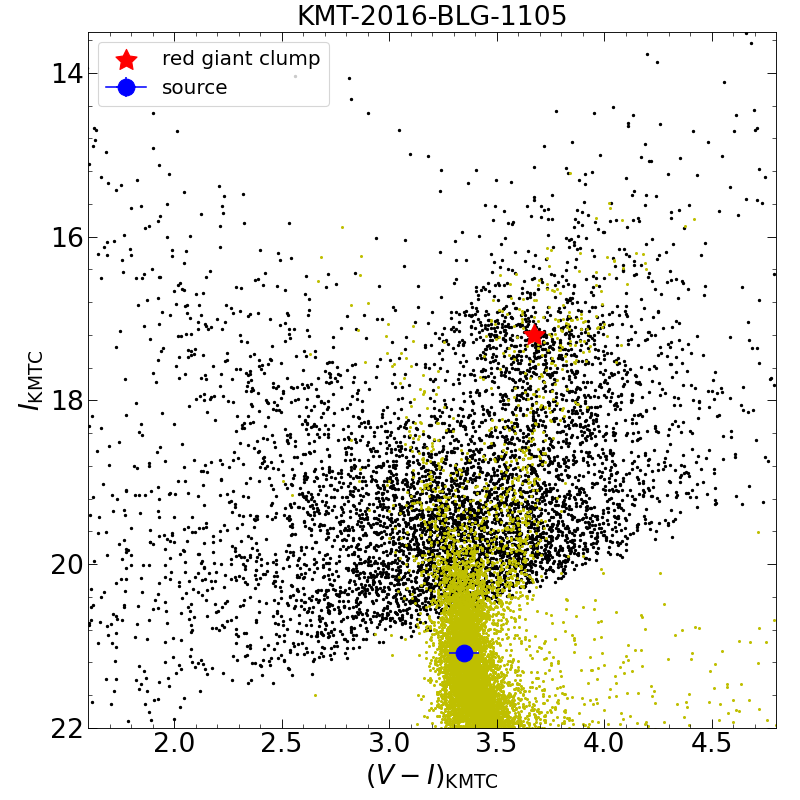}
    \caption{Color magnitude diagrams for the seven planetary events analyzed in this paper. The first five CMDs are constructed using the OGLE-III star catalog \citep{OGLEIII}, and the other two CMDs are constructed using the KMTC pyDIA photometry reduction. For each panel, the red asterisk and the blue dot are shown as the centroid of the red giant clump and the microlensed source star, respectively. The three green dots on the CMDs of \eventa, \eventc, and \eventd\ represent the blended light. For the bottom panel, the yellow dots represent the HST CMD of \citet{HSTCMD} whose red-clump centroid has been matched to that of KMTC using $(V - I, I)_{\rm cl, HST} = (1.62, 15.15)$ \citep{MB07192}.}
    \label{cmd}
\end{figure*}

Combining Equations (\ref{eqn:1}) and (\ref{equ:pie}), the mass $M_{\rm L}$ and distance $D_{\rm L}$ of a lens system are related to the angular Einstein radius $\thetae$ and the microlensing parallax $\pie$ by \citep{Gould1992, Gould2000}
\begin{equation}\label{eq:mass}
    M_{\rm L} = \frac{\thetae}{{\kappa}\pie};\qquad D_{\rm L} = \frac{\mathrm{au}}{\pie\thetae + \pi_{\rm S}},
\end{equation}
%where $\pi_{\rm S} = \mathrm{au}/D_{\rm S}$ is the source parallax, and $D_{\rm S}$ is the source distance. 

%which is astrometrically consistent with the sources (and thus the lenses)

\begin{table*}
    \renewcommand\arraystretch{1.5}
    \centering
    \caption{CMD parameters, $\theta_*$, $\thetae$ and $\mu_{\rm rel}$ for the five ``dip'' planetary events}
    \begin{tabular}{c|c|c|c|c c|c}
    \hline
    \hline
    Parameter & KB171194 & KB170428 & KB191806 & \multicolumn{2}{c|}{KB171003} & KB191367 \\
     &  & & & Inner & Outer &  \\
    \hline
    $(V - I)_{\rm cl}$ & $1.82 \pm 0.01$ & $1.95 \pm 0.01$ & $2.23 \pm 0.01$ & $2.39 \pm 0.01$ & $\gets$ & $1.70 \pm 0.01$ \\
    $I_{\rm cl}$ & $15.25 \pm 0.01$ & $15.39 \pm 0.01$ & $15.79 \pm 0.02$ & $16.04 \pm 0.01$ & $\gets$ & $15.13 \pm 0.01$ \\
    $I_{\rm cl,0}$ & $14.26 \pm 0.04$ & $14.36 \pm 0.04$ & $14.39 \pm 0.04$ & $14.34 \pm 0.04$ & $\gets$ & $14.37 \pm 0.04$ \\
    $(V - I)_{\rm S}$  & $1.47 \pm 0.07$ & $1.95 \pm 0.04$ & $1.93 \pm 0.03$ & $2.00 \pm 0.02$ & $\gets$ & $1.70 \pm 0.03$  \\
    $I_{\rm S}$ & $20.28 \pm 0.08$ & $20.43 \pm 0.05$ & $21.35 \pm 0.07$ & $19.30 \pm 0.04$ & $19.30 \pm 0.04$ & $21.47 \pm 0.13$ \\
    $(V - I)_{\rm S,0}$ & $0.71 \pm 0.08$ & $1.06 \pm 0.05$ & $0.76 \pm 0.05$ & $0.67 \pm 0.04$ & $\gets$ & $1.06 \pm 0.04$ \\
    $I_{\rm S,0}$ & $19.29 \pm 0.09$ & $19.40 \pm 0.07$ & $19.95 \pm 0.08$ & $17.60 \pm 0.06$ & $17.60 \pm 0.06$ & $20.71 \pm 0.14$ \\
    $\theta_*$ ($\mu$as) & $0.448 \pm 0.038$ & $0.578 \pm 0.034$ & $0.345 \pm 0.020$ & $0.942 \pm 0.046$ & $0.942 \pm 0.046$ & $0.316 \pm 0.023$ \\
    $\thetae$ (mas)  & $>0.17$ & $>0.09$ & $>0.19$ & $>0.14$ & $0.180 \pm 0.041$ & $>0.06$ \\ 
    $\mu_{\rm rel}$ (${\rm mas\,yr^{-1}}$) & $>1.3$ & $>0.74$ & $>0.51$ & $>2.0$ & $2.56 \pm 0.58$ & $>0.53$ \\
    \hline
    \hline
    \end{tabular}
    \tablecomments{$(V - I)_{\rm cl,0} = 1.06 \pm 0.03$. Event names are abbreviations, e.g., \eventa\ to KB171194.}
    \label{source1}
\end{table*}

\begin{table*}
    \renewcommand\arraystretch{1.5}
    \centering
    \caption{CMD parameters, $\theta_*$, $\thetae$ and $\mu_{\rm rel}$ for \eventf.}
    \begin{tabular}{c|c c| c c| c c}
    \hline
    \hline
    Parameter & \multicolumn{2}{c|}{Close A} &  \multicolumn{2}{c|}{Close B} & \multicolumn{2}{c}{Wide} \\
     & $u_0 > 0$ & $u_0 < 0$ & $u_0 > 0$ & $u_0 < 0$ & $u_0 > 0$ & $u_0 < 0$ \\
    \hline
    $(V - I)_{\rm cl}$ & $2.89 \pm 0.01$ & $\gets$ & $\gets$ & $\gets$ & $\gets$ & $\gets$\\
    $I_{\rm cl}$ & $16.42 \pm 0.02$ & $\gets$ & $\gets$ & $\gets$ & $\gets$ & $\gets$ \\
    $I_{\rm cl,0}$ & $14.33 \pm 0.04$ & $\gets$ & $\gets$ & $\gets$ & $\gets$ & $\gets$ \\
    $(V - I)_{\rm S}$ & $2.66 \pm 0.03$ & $\gets$ & $\gets$ & $\gets$ & $\gets$ & $\gets$ \\
    $I_{\rm S}$ & $21.12 \pm 0.07$ & $21.07 \pm 0.07$ & $21.10 \pm 0.08$ & $21.10 \pm 0.08$ & $21.03 \pm 0.07$ & $21.08 \pm 0.07$ \\
    $(V - I)_{\rm S,0}$ & $0.83 \pm 0.04$ & $\gets$ & $\gets$ & $\gets$ & $\gets$ & $\gets$ \\
    $I_{\rm S,0}$ & $19.03 \pm 0.08$ & $18.98 \pm 0.08$ & $19.01 \pm 0.09$ & $19.01 \pm 0.09$ & $18.94 \pm 0.08$ & $18.99 \pm 0.08$\\
    $\theta_*$ ($\mu$as) & $0.561 \pm 0.031$ & $0.574 \pm 0.031$ & $0.566 \pm 0.033$ & $0.566 \pm 0.033$ & $0.584 \pm 0.032$ & $0.571 \pm 0.032$ \\
    $\thetae$ (mas) & $0.322_{-0.145}^{+0.083}$ & $0.314_{-0.138}^{+0.087}$ & $0.377_{-0.157}^{+0.120}$ & $0.343_{-0.141}^{+0.105}$ & $>0.21$ & $>0.24$  \\ 
    $\mu_{\rm rel}$ (${\rm mas\,yr^{-1}}$) & $1.69^{+0.45}_{-0.77}$ & $1.72^{+0.49}_{-0.76}$ & $1.98^{+0.64}_{-0.84}$ & $2.08^{+0.65}_{-0.87}$ & $>1.2$ & $>1.3$\\
    \hline
    \hline
    \end{tabular}
    %\tablecomments{$(V - I, I)_{\rm cl,0} = 1.06 \pm 0.03$}
    \label{source2}
\end{table*}

\begin{table*}
    \renewcommand\arraystretch{1.5}
    \centering
    \caption{CMD parameters, $\theta_*$, $\thetae$ and $\mu_{\rm rel}$ for \eventg.}
    \begin{tabular}{c|c c c c c}
    \hline
    \hline
     & Wide A & Wide B & Wide C & Wide D & Close \\
    \hline
    $I_{\rm cl}$ & $17.20 \pm 0.01$ & $\gets$ & $\gets$ & $\gets$ & $\gets$ \\
    $I_{\rm cl,0}$ & $14.39 \pm 0.04$ & $\gets$ & $\gets$ & $\gets$ & $\gets$ \\
    $I_{\rm S}$ & $21.09 \pm 0.08$ & $21.20 \pm 0.05$ & $21.22 \pm 0.11$ & $21.22 \pm 0.11$ & $21.27 \pm 0.06$ \\
    $(V - I)_{\rm S,0}$ & $0.74 \pm 0.07$ & $0.75 \pm 0.07$ & $0.75 \pm 0.07$ & $0.75 \pm 0.07$ & $0.75 \pm 0.07$ \\
    $I_{\rm S,0}$ & $18.28 \pm 0.09$ & $18.39 \pm 0.07$ & $18.41 \pm 0.12$ & $18.41 \pm 0.12$ & $18.46 \pm 0.07$ \\
    $\theta_*$ ($\mu$as) & $0.732 \pm 0.057$ & $0.702 \pm 0.051$ & $0.696 \pm 0.061$ & $0.696 \pm 0.061$ & $0.680 \pm 0.050$ \\
    $\thetae$ (mas) & $>0.31$ & $0.240 \pm 0.070$ & $>0.15$ & $>0.13$ & $0.907 \pm 0.182$ \\ 
    $\mu_{\rm rel}$ (${\rm mas\,yr^{-1}}$) & $>2.9$ & $2.07 \pm 0.62$ & $>1.3$ & $>1.1$ & $7.65 \pm 1.54$ \\
    \hline
    \hline
    \end{tabular}
    %\tablecomments{$(V - I, I)_{\rm cl,0} = 1.06 \pm 0.03$}
    \label{source3}
\end{table*}

To obtain the angular Einstein radius through $\thetae = \theta_*/\rho$, we first estimate the angular source radius $\theta_*$ by locating the source on a color-magnitude diagram (CMD, \citealt{Yoo2004}). For each event, we construct a $V - I$ versus $I$ CMD using the ambient stars of the OGLE-III catalog \citep{OGLEIII} or the KMTC images with the pyDIA reductions. See Figure \ref{cmd} for the CMDs of the seven planetary events. We estimate the centroid of the red giant clump as $(V - I, I)_{\rm cl}$ from CMDs and adopt the de-reddened color and magnitude of the red giant clump, $(V - I, I)_{\rm cl,0}$, from \cite{Bensby2013} and Table 1 of \cite{Nataf2013}. We obtain the source apparent magnitude from the light-curve analysis of Section \ref{model}, and the source color by a regression of the KMTC $V$ versus $I$ flux with the change of the lensing magnification.

We find that the $V$-band observations of \eventg\ have insufficient signal-to-noise ratio to determine the source color, so we estimate the source color by the {\it Hubble} Space Telescope ({\it HST}) CMD of \cite{HSTCMD} (see Section \ref{lensf} for details). Finally, using the color/surface-brightness relation of \cite{Adams2018}, we obtain the angular source radius $\theta_*$. Tables \ref{source1}, \ref{source2} and \ref{source3} present the CMD values and ($\theta_*$, $\thetae$, $\mu_{\rm rel}$) from the procedures above. 

Because the blended light could provide additional constraints on the lens properties (e.g., the upper limits for the lens brightness), we also check the brightness and the astrometric alignment of the baseline object. For \eventb\ and \eventc, we adopt the $i'$-band baseline images taken by the 3.6m Canada-France-Hawaii Telescope (CFHT) from 2020 to 2022, whose seeing FWHM is $0.''55$--$0.''70$. For the other five events which do not have any CFHT image, we check the baseline objects from the KMTC pyDIA reduction, whose seeing FWHM is about $1.''0$. 

Because none of the seven planetary events have simultaneous measurements of $\pie$ and $\thetae$ at the $>3\sigma$ level, the lens masses and distances cannot be directly determined by Equation (\ref{eq:mass}). We conduct a Bayesian analysis using a Galactic model to estimate the lens properties. The Galactic model and the procedures we adopt are the same as described in \cite{OB191053}. We refer the reader to that work for details. The only exception is that we include upper limits of the lens light, $I_{\rm L, limit}$, from the analysis of the blended light. We adopt the mass-luminosity relation of \cite{OB171130},
\begin{equation}\label{equ:ML1}
    M_I = 4.4 - 8.5\log{\frac{M_{\rm L}}{M_{\odot}}},
\end{equation}
where $M_I$ is the absolute magnitude in the $I$ band, and we reject trial events for which the lens properties obey
\begin{equation}\label{equ:ML2}
    M_I + 5 \log{\frac{D_{\rm L}}{10 {\rm pc}}} + A_{I, D_{\rm L}} < I_{\rm L,limit},
\end{equation}
where $A_{I, D_{\rm L}}$ is the extinction at $D_{\rm L}$. We adopt an extinction curve with a scale height of 120 pc. For the five events with OGLE CMDs, the total extinction is derived from the CMD analysis, $A_I = I_{\rm cl} - I_{\rm cl, 0}$. For the other two events with KMTC CMDs, we adopt the extinction in the $K$ band from \cite{Gonzalez2012} and $A_I = 7.26~A_K$ from \cite{Nataf2016}.

\begin{table*}
    \renewcommand\arraystretch{1.5}
    \centering
    \caption{Physical parameters of the six planetary events from a Bayesian analysis.}
    \begin{tabular}{c c|c c c c c|c c}
    \hline
    \hline
    Event & Solution & \multicolumn{5}{c|}{Physical Properties} & \multicolumn{2}{c}{Relative Weights} \\
    \hline
     &  & $M_{\rm host}[M_{\odot}]$ & $M_{\rm planet}[M_{\oplus}]$ & $D_{\rm L}$[kpc] & $a_{\bot}$[au] & $\mu_{\rm rel}$[${\rm mas\,yr^{-1}}$] & Gal.Mod. & $\chi^2$ \\
    \hline
    KB171194 &  & $0.41_{-0.19}^{+0.23}$ & $3.54_{-1.63}^{+1.95}$ & $4.24_{-1.71}^{+2.16}$ & $1.78_{-0.46}^{+0.45}$ & $4.29_{-1.66}^{+2.50}$ & ... & ... \\
    \hline
    KB170428 & Inner & $0.34_{-0.17}^{+0.22}$ & $5.63_{-2.85}^{+3.59}$ & $5.40_{-2.60}^{+1.82}$ & $1.78_{-0.58}^{+0.54}$ & $3.27_{-1.32}^{+2.26}$ & 0.99 & 1.00 \\
    & Outer & $0.34_{-0.17}^{+0.22}$ & $5.55_{-2.81}^{+3.53}$ & $5.40_{-2.60}^{+1.82}$ & $1.85_{-0.60}^{+0.55}$ & $3.28_{-1.34}^{+2.24}$ & 1.00 & 0.95 \\
    & Combined & $0.34_{-0.17}^{+0.22}$ & $5.59_{-2.83}^{+3.57}$ & $5.40_{-2.60}^{+1.82}$ & $1.81_{-0.59}^{+0.55}$ & $3.27_{-1.32}^{+2.26}$ & ... & ... \\ 
    \hline
    KB191806 & Inner ($u_0 > 0$) & $0.75_{-0.25}^{+0.24}$ & $4.67_{-1.52}^{+1.52}$ & $6.62_{-1.93}^{+0.72}$ & $2.87_{-0.66}^{+0.64}$ & $1.17_{-0.34}^{+0.70}$ & 1.00 & 0.70 \\
    & Inner ($u_0 < 0$) &  $0.74_{-0.26}^{+0.25}$ & $4.47_{-1.52}^{+1.56}$ & $6.63_{-2.01}^{+0.73}$ & $2.85_{-0.70}^{+0.66}$ & $1.11_{-0.34}^{+0.74}$ & 0.84 & 0.58 \\
    & Outer ($u_0 > 0$) & $0.73_{-0.26}^{+0.25}$ & $4.63_{-1.64}^{+1.60}$ & $6.68_{-1.97}^{+0.72}$ & $3.11_{-0.79}^{+0.73}$ & $1.13_{-0.36}^{+0.70}$ & 0.98 & 0.82 \\
    & Outer ($u_0 < 0$) & $0.75_{-0.25}^{+0.24}$ & $4.79_{-1.60}^{+1.56}$ & $6.62_{-2.09}^{+0.74}$ & $3.17_{-0.75}^{+0.70}$ & $1.15_{-0.36}^{+0.74}$ & 0.98 & 1.00 \\
    & Combined & $0.74_{-0.25}^{+0.25}$ & $4.67_{-1.60}^{+1.56}$ & $6.64_{-2.01}^{+0.72}$ & $3.02_{-0.73}^{+0.70}$ & $1.13_{-0.34}^{+0.74}$ & ... & ... \\
    \hline
    KB171003 & Inner & $0.37_{-0.19}^{+0.32}$ & $6.75_{-3.44}^{+5.79}$ & $7.03_{-0.74}^{+0.61}$ & $1.54_{-0.37}^{+0.38}$ & $3.55_{-0.84}^{+0.88}$ & 1.00 & 0.90 \\
    & Outer & $0.27_{-0.13}^{+0.26}$ & $3.72_{-1.80}^{+3.71}$ & $7.16_{-0.65}^{+0.61}$ & $1.25_{-0.25}^{+0.27}$ & $2.75_{-0.54}^{+0.56}$ & 0.74 & 1.00 \\
    & Combined & $0.32_{-0.17}^{+0.31}$ & $5.19_{-2.80}^{+5.39}$ & $7.09_{-0.70}^{+0.61}$ & $1.38_{-0.32}^{+0.39}$ & $3.11_{-0.72}^{+0.94}$ & ... & ... \\
    \hline
    KB191367 & Inner & $0.25_{-0.13}^{+0.16}$ & $4.06_{-2.08}^{+2.56}$ & $4.68_{-2.10}^{+2.44}$ & $1.67_{-0.55}^{+0.49}$ & $3.92_{-1.71}^{+2.57}$ & 1.00 & 1.00 \\ 
    & Outer & $0.25_{-0.13}^{+0.16}$ & $4.12_{-2.10}^{+2.58}$ & $4.67_{-2.10}^{+2.45}$ & $1.73_{-0.57}^{+0.51}$ & $3.89_{-1.71}^{+2.55}$ & 0.96 & 0.90 \\
    & Combined & $0.25_{-0.13}^{+0.16}$ & $4.08_{-2.08}^{+2.58}$ & $4.67_{-2.10}^{+2.45}$ & $1.70_{-0.56}^{+0.50}$ & $3.91_{-1.71}^{+2.56}$ & ... & ... \\
    \hline
    \hline
    OB171806 & Close A ($u_0 > 0$) & $0.44_{-0.23}^{+0.33}$ & $5.87_{-3.04}^{+4.43}$ & $6.60_{-1.06}^{+0.65}$ & $1.84_{-0.51}^{+0.44}$ & $1.95_{-0.54}^{+0.46}$ & 0.85 & 0.90 \\
    & Close A ($u_0 < 0$) & $0.33_{-0.16}^{+0.33}$ & $4.83_{-2.44}^{+4.83}$ & $6.17_{-1.57}^{+0.92}$ & $1.69_{-0.43}^{+0.46}$ & $2.13_{-0.58}^{+0.56}$ & 1.00 & 1.00 \\
    & Close B ($u_0 > 0$) & $0.44_{-0.26}^{+0.39}$ & $2.40_{-1.48}^{+2.24}$ & $5.95_{-2.13}^{+1.11}$ & $1.89_{-0.62}^{+0.58}$ & $2.39_{-0.70}^{+0.78}$ & 0.21 & $10^{-3.1}$ \\
    & Close B ($u_0 < 0$) & $0.48_{-0.26}^{+0.35}$ & $2.68_{-1.48}^{+1.96}$ & $6.53_{-1.41}^{+0.69}$ & $1.91_{-0.58}^{+0.50}$ & $2.09_{-0.60}^{+0.58}$ & 0.16 & $10^{-3.2}$ \\
    & Wide ($u_0 > 0$) & $0.34_{-0.16}^{+0.31}$ & $5.47_{-2.60}^{+4.87}$ & $3.01_{-1.09}^{+2.22}$ & $2.53_{-0.78}^{+1.06}$ & $4.92_{-1.94}^{+1.74}$ & $10^{-1.5}$ & $10^{-1.8}$ \\
    & Wide ($u_0 < 0$) & $0.41_{-0.17}^{+0.23}$ & $4.87_{-2.00}^{+2.72}$ & $2.87_{-0.86}^{+1.13}$ & $2.82_{-0.78}^{+0.85}$ & $5.48_{-1.42}^{+1.30}$ & 0.24 & $10^{-1.8}$ \\
    & Combined & $0.38_{-0.20}^{+0.34}$ & $5.27_{-2.72}^{+4.71}$ & $6.40_{-1.51}^{+0.77}$ & $1.75_{-0.47}^{+0.46}$ & $2.05_{-0.56}^{+0.52}$ & ... & ... \\
    \hline
    KB161105 & Wide A & $0.43_{-0.20}^{+0.22}$ & $0.92_{-0.44}^{+0.44}$ & $3.79_{-1.44}^{+1.38}$ & $2.93_{-0.91}^{+0.69}$ & $6.48_{-1.30}^{+2.08}$ & 0.37 & 1.00 \\
    & Wide B & $0.37_{-0.21}^{+0.31}$ & $4.67_{-2.64}^{+3.75}$ & $7.12_{-1.10}^{+0.65}$ & $2.03_{-0.53}^{+0.52}$ & $2.29_{-0.56}^{+0.62}$ & 1.00 & 0.32 \\
    & Wide C & $0.43_{-0.23}^{+0.27}$ & $12.14_{-6.31}^{+7.63}$ & $5.42_{-2.33}^{+1.85}$ & $2.63_{-0.83}^{+0.79}$ & $3.97_{-1.58}^{+2.68}$ & 0.66 & 0.11 \\
    & Wide D & $0.44_{-0.23}^{+0.27}$ & $9.51_{-4.87}^{+5.75}$ & $5.28_{-2.23}^{+1.92}$ & $2.56_{-0.74}^{+0.73}$ & $4.11_{-1.56}^{+2.64}$ & 0.64 & 0.26 \\
    & Close & $0.43_{-0.18}^{+0.18}$ & $1.32_{-0.56}^{+0.56}$ & $3.27_{-1.15}^{+1.28}$ & $2.26_{-0.69}^{+0.51}$ & $6.74_{-1.58}^{+1.74}$ & 0.29 & 0.17 \\
    & Combined & $0.41_{-0.21}^{+0.25}$ & $2.32_{-1.56}^{+7.43}$ &  $5.08_{-2.18}^{+2.24}$ & $2.44_{-0.75}^{+0.88}$ & $4.68_{-2.50}^{+2.76}$ & ... & ... \\
    \hline
    \hline
    \end{tabular}\\
    \tablecomments{The combined solution is obtained by a combination of all solutions weighted by the probability for the Galactic model (Gal.Mod.) and ${\rm exp}(-\Delta\chi^2/2)$.}
    \label{phy}
\end{table*}

Table \ref{phy} presents the resulting Bayesian estimates of the host mass $M_{\rm host}$, the planet mass $M_{\rm planet}$, the lens distance $D_{\rm L}$, the projected planet-host separation $a_{\bot}$ and the lens-source relative proper motion $\mu_{\rm rel}$. For events with multiple solutions, we show the results for each solution and the ``combined results'' of combining all solutions weighted by their Galactic-model likelihood and ${\rm exp}(-\Delta\chi^2/2)$, where $\Delta\chi^2$ is the $\chi^2$ difference compared to the best-fit solution. Here the Galactic-model likelihood represents the total weight for the simulated events given the error distributions of $\te, \thetae$ and $\pie$. See Equation (16) of \cite{OB191053} for the weight procedures. 

We do not adopt the ``combined results'' as the final physical parameters but just show them for consideration, because there is no conclusion about how to combine degenerate solutions. We note that the ${\rm exp}(-\Delta\chi^2/2)$ probability might be suffered from systematic errors of the observed data. However, the weight from $\Delta\chi^2$ only has minor effects on the ``combined results''. Except for \eventg\ the degenerate solutions have similar physical interpretations and except for \eventf\ the $\Delta\chi^2$ is small, but for \eventf\ the ``combined results'' are already dominated by the ``Close A'' solutions due to their Galactic-model likelihoods. Due to similar reasons, whether to include the phase-space factors also has a minor impact on the ``combined results''.

\subsection{\eventa}\label{lensa}

The corresponding CMD shown in Figure \ref{cmd} is constructed from the OGLE-III field stars within $240''$ centered on the event. The baseline object has $(V, I)_{\rm base} = (21.343 \pm 0.085, 19.608 \pm 0.051)$, yielding a blend of $(V - I, I)_{\rm B} = (2.15 \pm 0.39, 20.45 \pm 0.14)$. We display the blend on the CMD. The source position measured by the difference imaging analysis is displaced from the baseline object by $\Delta\theta(N, E) = (-26, 41)$ mas. We estimate the error of the baseline position by the fractional astrometric error being equal to the fractional photometric error \citep{KB190842}, which yields $\sigma_{\rm ast} = 0.39\sigma_{I}{\rm FWHM} = 20$ mas. We note that the astrometric error should be underestimated due to the mottled background from unresolved stars and other systematic errors, but the whole astrometric error should be not more than twice our estimate. Thus, the baseline object is astrometrically consistent with the source and the lens within $2\sigma$. The blend does not have a useful color constraint. We adopt the $3\sigma$ upper limit of the blended light, $I_{\rm L, limit} = 20.03$, as the upper limit of the lens brightness. 

As given in Table \ref{phy}, the preferred host star is an M dwarf located in the Galactic disk, and the planet is probably a super-Earth beyond the snow line of the lens system (assuming a snow line radius $a_{\rm SL} = 2.7(M/M_{\odot})$~{\rm au}, \citealt{snowline}).

\subsection{\eventb}\label{lensb}

The corresponding CMD shown in Figure \ref{cmd} consists of the OGLE-III field stars within $150''$ centered on the event. The baseline object on the CFHT images has $I_{\rm base} = 20.056 \pm 0.063$, with an astrometric offset of $\Delta\theta(N, E) = (6, -2)$ mas and an astrometric error of $\sigma_{\rm ast} \sim 5$ mas. Thus, the baseline object is astrometrically consistent with the source at about $1\sigma$. Because the CFHT images do not contain color information, we do not display the blend on the CMD. We also adopt the $3\sigma$ upper limit of the blended light, $I_{\rm L, limit} = 20.81$, as the upper limit of the lens brightness. 

As shown in Table \ref{phy}, the Bayesian analysis indicates another cold super-Earth orbiting an M dwarf. 

\subsection{\eventc}\label{lensc}

The CMD of this event is constructed from the OGLE-III field stars within $150''$ centered on the event, shown in Figure \ref{cmd}. The baseline object on the KMTC images has $(V, I)_{\rm base} = (20.155 \pm 0.125, 18.685 \pm 0.076)$. We plot the blend on the CMD and find that the blend probably belongs to the foreground main-sequence branch and thus could be the lens. However, the astrometric offset is $\Delta\theta(N, E) = (433, -76)$ mas and $\Delta\theta(N, E) = (416, -96)$ mas on the CFHT and KMTC images, respectively, so the majority of the blended light is unrelated to the lens. We adopt the median value of the blended light, $I_{\rm L, limit} = 18.8$, as the upper limit of the lens brightness. 

The results of the Bayesian analysis are given in Table \ref{phy}. The planet is another cold super-Earth, and the preferred host is a K dwarf.

\subsection{\eventd}\label{lensd}

We use the OGLE-III field stars within $180''$ centered on the event to build the CMD. Combining the measured $\rho$ from the light-curve analysis, we obtain $\thetae = 0.180 \pm 0.041$ mas for the ``Outer'' solution and $\thetae > 0.14$ mas ($3\sigma$) for the ``Inner'' solution. The KMTNet baseline object has $(V, I)_{\rm base} = (20.968 \pm 0.046, 18.780 \pm 0.028)$, corresponding to a blend of $(V - I, I)_{\rm B} = (2.54 \pm 0.20, 19.83 \pm 0.10)$, and we display the blend on the CMD. The source-baseline astrometric offset is $\Delta\theta(N, E) = (-64, -77)$ mas, with an astrometric error of $\sigma_{\rm ast} \sim 12$ mas, implying that most of the blend light should be unrelated to the event. We adopt the median value of the blended light, $I_{\rm L, limit} = 19.83$, as the upper limit of the lens brightness. 

The Bayesian analysis shows that the host star is probably an M dwarf located in the Galactic bulge. Again, the preferred planet is a cold super-Earth. 

%The lens-source relative proper motion is $<3.3$ mas/yr ($1\sigma$), indicating that the lens system is probably located in the Galactic bulge. 

\subsection{\evente}\label{lense}

In Figure \ref{cmd}, we display the position of the source on the CMD of stars within $180''$ around the source. On the KMTC images, there is no star within $1.''4$ around the source position. We thus adopt the detection limit of the KMTC images, $I = 21.0$, as the upper limit of the baseline brightness, yielding the $3\sigma$ upper limit of the blended light, $I_{\rm L, limit} = 21.6$. Applying Equations (\ref{equ:ML1}) and (\ref{equ:ML2}) and assuming $D_{\rm L} < 8$ kpc, this flux constraint corresponds to an upper limit of the lens mass of $0.6M_{\odot}$.

As shown in Table \ref{phy}, the Bayesian estimate shows another cold super-Earth orbiting an M dwarf.

\subsection{\eventf}\label{lensf}

The CMD of this event is constructed from KMTC field stars within a $300''$ square centered on the event position. The baseline object, $(V, I)_{\rm base} = (22.300 \pm 0.308, 20.042 \pm 0.128)$, is displaced from the source by 835 mas. Thus, most of the blend light should be unrelated to the event. We do not show the blend on the CMD and adopt the median value of the blended light, $I_{\rm L, limit} = 20.5$, as the upper limit of the lens brightness.

The results of the Bayesian analysis are presented in Table \ref{phy}, and all solutions indicate a cold super-Earth orbiting a low-mass star. The constraints on $\pi_{\rm E, \bot}$ from the light-curve analysis are useful. The ``Wide'' solution has a relatively large $\thetae$, with a $2\sigma$ lower limit of $0.60$ mas and the best-fit value of $\sim 1.1$ mas, so the corresponding lens system is located in the Galactic disk. Then, the ``Wide ($u_0 > 0$)'' solution has $\pi_{\rm E, \bot} < 0$ and thus a lens velocity in Galactic coordinates of $v_{\ell} \sim 100$ km s$^{-1}$, so this solution is strongly disfavored. For the two ``Close'' solutions, both the $\pi_{\rm E, \bot} < 0$ solutions are slightly disfavored and have relatively higher probabilities of a bulge lens system.

For the ``Wide'' solution, the predicted apparent magnitude of the lens system is fainter than the source by $\sim 0.8$ mag in the $H$ band. In the case of OGLE-2012-BLG-0950, the source and the lens have roughly equal brightness and were resolved by the Keck AO imaging and the {\it HST} imaging when they were separated by about 34 mas \citep{OB120950}. For \eventf, we estimate that resolving the lens and source probably requires a separation of 45 mas for the ``Wide'' solution. We note that the proper motions of the two ``Close'' solutions are $\sim 2~{\rm mas\,yr^{-1}}$. If high-resolution observations resolve the lens and the source and find that $\mu_{\rm rel}$ (e.g., $\sim 5~{\rm mas\,yr^{-1}}$) is much higher than that of the ``Close'' solutions, the three solutions can be distinguished. Such observations can be taken in 2026 or earlier.

\subsection{\eventg}\label{lensg}

To collect enough red-giant stars to determine the centroid of the red giant clump, the CMD of this event shown in Figure \ref{cmd} contains KMTC field stars within a $280'' \times 300''$ rectangle region. Because the event lies about $80''$ from the edge of the CCD chip, it is displaced from the center of the rectangle region by about $70''$. The $V$-band data have insufficient signal-to-noise ratio to determine the source color, so we adopt the method of \cite{MB07192} to estimate the source color. We first calibrate the CMD of \cite{HSTCMD} HST observations to the KMTC CMD using the centroids of red giant clumps. We then estimate the source color by taking the color of the HST field stars whose brightness are within the $5\sigma$ of the source star. 

The baseline object has $I_{\rm base} = 20.729 \pm 0.125$ without color information, so we do not plot the blend on the CMD. The source-baseline astrometric offset is $\Delta\theta(N, E) = (73, 166)$ mas, at about $3\sigma$. Because the baseline object is marginally detected on the KMTC images, we adopt the median value of the blended light, $I_{\rm L, limit} = 21.7$, as the upper limit of the lens brightness. 

%with an astrometric error of $\sigma_{\rm ast} \sim 50$ mas. Although the astrometric offset is at about $3\sigma$, 

The Bayesian analysis indicates that the host star is probably an M dwarf. Due to a factor of $\sim 13$ differences within the mass ratios of the five degenerate solutions, there is a wide range for the planetary mass, from sub-Earth-mass to sub-Neptune-mass. Because no solution has a very different proper motion from other solutions, future high-resolution observations cannot break the degeneracy. However, such observations are still important because the measurements of the host brightness can yield the host mass and distance, which could be used for studying the relation between the planetary occurrence rate and the host properties. For the ``Wide A'' and ``Close'' solutions, the predicted apparent magnitude of the lens system is fainter than the source by $\sim 2$ mag and $\sim 3$ mag in the $H$ and $I$ bands, respectively. In 2025, the lens and the source will be separated by $\gtrsim 50$ mas and may be resolved.

\section{Discussion}\label{dis}

In this paper, we have presented the analysis of seven $q < 10^{-4}$ planets. Together with 17 already published and three that will be published elsewhere, the KMTNet AnomalyFinder algorithm has found 27 events that can be fit by $q < 10^{-4}$ models from 2016--2019 KMTNet data. For the analysis above and in other published papers, all of the local minima are investigated, but here for each planet, we only consider the models with $\Delta\chi^2 < 10$ compared to the best-fit model.

%Here these $q < 10^{-4}$ models have $\Delta\chi^2 < 10$ compared to their respective best-fit models. 

Table \ref{tab:-4planet} presents the event name, $\log q$, $s$, $u_0$, discovery method, $\Delta\chi^2$ compared to the best-fit models, whether it has a caustic crossing, anomaly type (bump or dip), and the KMTNet fields (prime or sub-prime) of each planet, ranked-ordered by $\log q$ of the best-fit models. Of them, 15 were solely detected using AnomalyFinder, and 12 were first discovered from by-eye searches and then recovered by AnomalyFinder, which illustrates the importance of systematic planetary anomaly searches in finding low mass-ratio microlensing planets. The seasonal distribution, (5, 8, 8, 6) for 2016--2019, is consistent with normal Poisson variations.

Among the 27 planets, four have alternative possible models with $q > 10^{-4}$, and 23 are secure $q < 10^{-4}$ planets. Because the detection of $q < 10^{-4}$ planets is one of the major scientific goals of the ongoing KMTNet survey and future space-based microlensing projects \citep{MatthewWFIRSTI,ET,CSST_Wei}, it is worthwhile to review the properties of the 27 planetary events and study how to detect more such planets.

\subsection{The Missing Planetary Caustics Problem}

\begin{figure}[htb] 
    \centering
    \includegraphics[width=1.00\columnwidth]{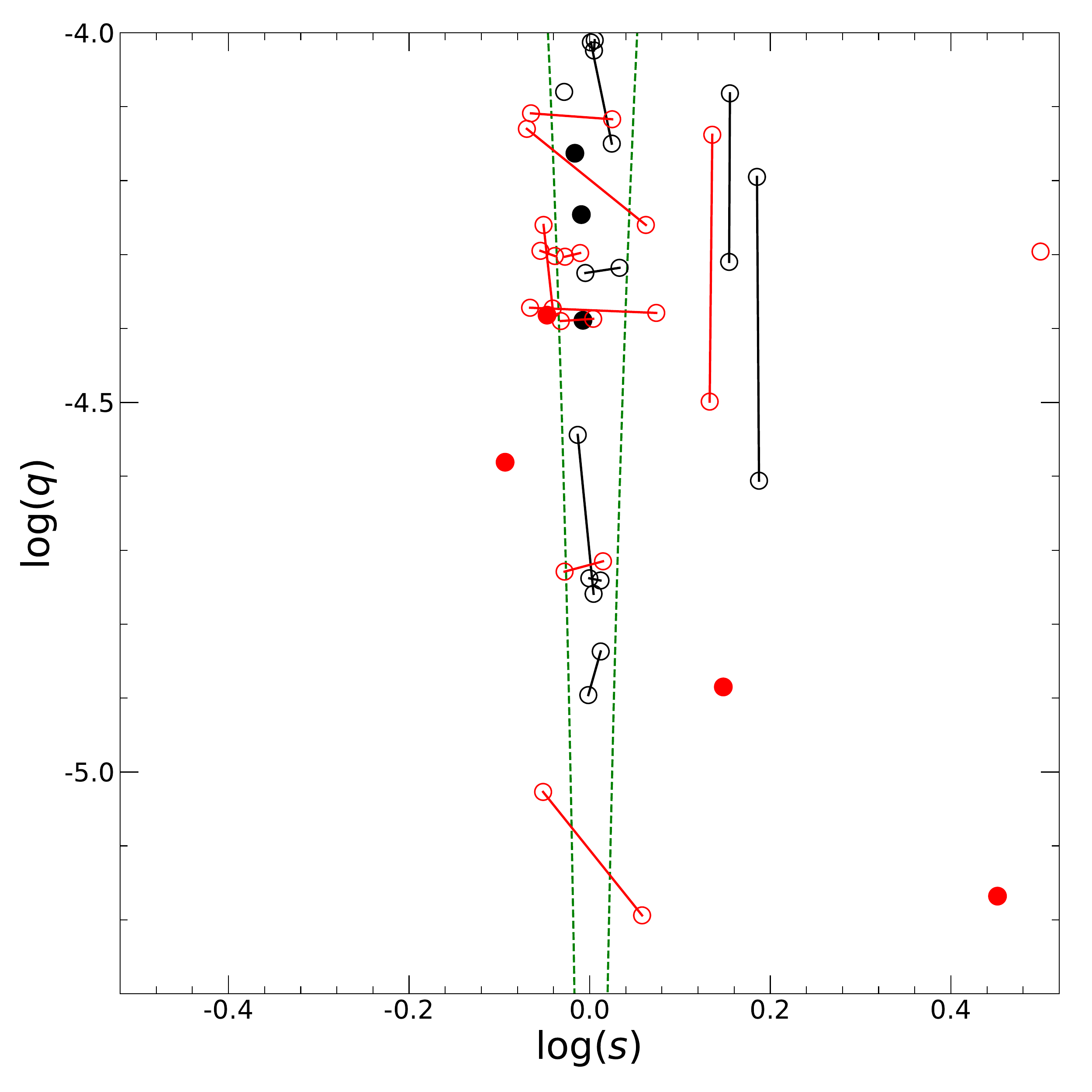}
    \caption{$\log q$ vs. $\log s$ distribution for the 27 planetary events with $q < 10^{-4}$ shown in Table \ref{tab:-4planet}, adapted from Figure 11 of \cite{OB190960}. The red points represent planets that were solely detected by AnomalyFinder, and the black points represent planets that were first discovered from by-eye searches and then recovered by AnomalyFinder. Solutions are considered to be ``unique'' (filled points) if there are no competing solutions within $\Delta\chi^2 < 10$. Otherwise, they are shown by open circles. The event KMT-2016-BLG-1105 has five degenerate solutions, but we only plot the best-fit $s > 1$ and $s < 1$ solutions for simplicity. For two solutions that are subject to the $u_0 > 0$ and $u_0 < 0$ degeneracy, we show them as one solution and take the average values. The two green dashed lines indicate the boundaries for ``near-resonant'' caustics \citep{Dominik1999}.}
    \label{qs}
\end{figure}

As illustrated by \cite{OB191053}, the motivation for building the KMTNet AnomalyFinder algorithm is to exhume the buried signatures of ``missing planetary caustics'' in the KMTNet data. \cite{Zhu2014ApJ} predicted that $\sim 50\%$ of the KMTNet $q < 10^{-4}$ planets should be detected by caustics outside of the near-resonant \citep{Dominik1999,OB190960} range. Below we follow the definitions of \cite{OB191053} and refer to caustics inside and outside of the near-resonant range as near-resonant caustics and pure-planetary caustics. Contrary to the prediction of \cite{Zhu2014ApJ}, before the application of AnomalyFinder only two of ten $q < 10^{-4}$ KMTNet planets were detected by pure-planetary caustics. The two cases are OGLE-2017-BLG-0173Lb \citep{OB170173} and KMT-2016-BLG-0212Lb \citep{KB160212}. Hence, it is necessary to check the caustic types for the planetary sample of AnomalyFinder. 

Figure \ref{qs} shows the $\log q$ versus $\log s$ plot for the 27 planets. The red and black points represent planets that were first discovered using AnomalyFinder and by-eye searches, respectively. The two green dashed lines indicate the boundaries for the near-resonant range. A striking feature is that in constrast to the locations of the by-eye planets, of the 15 AnomalyFinder planets 11 have pure-planetary caustics, two have both pure-planetary and near-resonant caustics, and only two are fully located inside the near-resonant range. In total, at least 13 planets were detected by pure-planetary caustics. Thus, the caustic types of the AnomalyFinder planetary sample agree with the expectation of \cite{Zhu2014ApJ}, and the missing planetary caustics problem has been solved by the systematic planetary anomaly search. 

\subsection{Caustic Crossing and Anomaly Type}

\cite{Zhu2014ApJ} predicted that about half of the KMTNet planets will be detected by caustic-crossing anomalies. \cite{2018_subprime} found that 16/33 of 2018 KMTNet AnomalyFinder planets have caustic-crossing anomalies. As shown in Table \ref{tab:-4planet}, 14/27 of the $q < 10^{-4}$ planets have caustic-crossing anomalies, in good agreement with the expectation of \cite{Zhu2014ApJ}. Thus, the $\sim 50\%$ probability of caustic-crossing anomalies is likely applicable down to $q \sim 10^{-5}$. 

\cite{OB191053} and \cite{KB190253} applied the AnomalyFinder algorithm to 2018--2019 KMTNet prime-field events and found seven newly discovered $q < 2 \times 10^{-4}$ planets. Among them, only OGLE-2019-BLG-1053Lb has a bump-type anomaly and the other six planets were detected by dip-type anomalies. Thus, it is necessary to check whether dip-type anomalies dominate the detection of low-$q$ planets. As presented in Table \ref{tab:-4planet}, the ratio of bump-type to dip-type anomalies for the $q < 10^{-4}$ planets is 15 to 12, so the two types of anomalies play roughly equal roles in the low-$q$ detection. However, of the 12 dip-type anomalies, nine were solely detected by AnomalyFinder, including eight non-caustic-crossing anomalies. KMT-2018-BLG-1988 \citep{KB181988} is the only case that the anomaly is a non-caustic-crossing dip and was first discovered from by-eye searches. Unlike the dip-type anomalies, the four non-caustic-crossing bumps were all first noticed from by-eye searches. Hence, by-eye searches have proved to be quite insensitive to non-caustic-crossing dip-type anomalies for low-$q$ planets.

\subsection{A Desert of High-magnification Planetary Signals}

\begin{figure}[htb] 
    \centering
    \includegraphics[width=1.00\columnwidth]{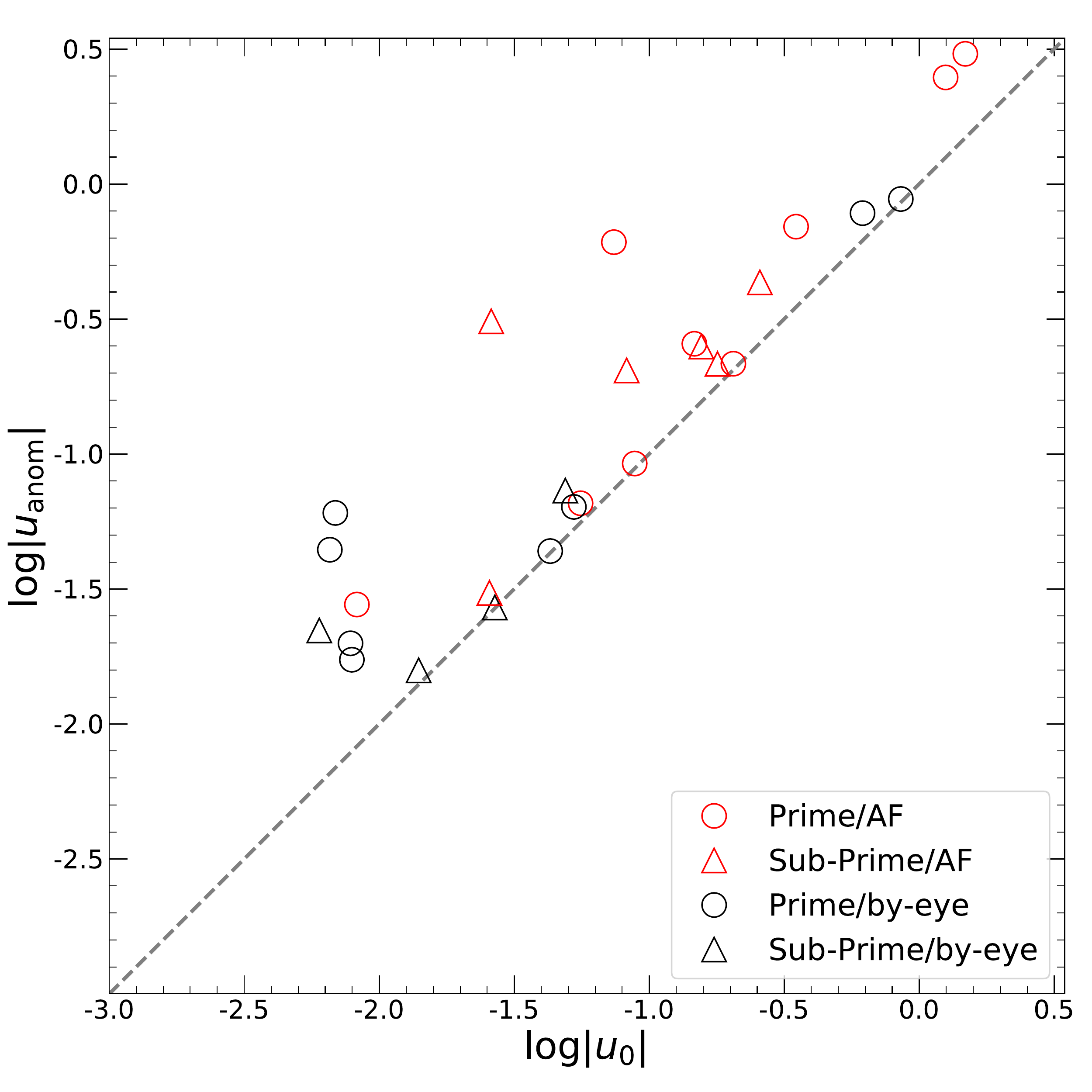}
    \caption{$\log|u_{\rm anom}|$ vs. $\log|u_0|$ distribution for the 27 planetary events with $q < 10^{-4}$ shown in Table \ref{tab:-4planet}. Colors are the same as the colors of Figure \ref{qs}. Circles and triangles represent prime-field and sub-prime-field planets, respectively. The grey dashed line indicates $|u_0| = |u_{\rm anom}|$.}
    \label{uanom}
\end{figure}

\cite{OB191053} suggested that the missing planetary caustics problem was caused by the way that modelers searched for planetary signatures. Because high-magnification events are intrinsically more sensitive to planets \citep{Griest1998}, by-eye searches paid more attention to them, while pure-planetary caustics are mainly detected in low-magnification events. If this hypothesis is correct, we expect that by-eye planets and AnomalyFinder planets will have different $|u_0|$ and $|u_{\rm anom}|$ distributions. The $\log|u_{\rm anom}|$ versus $\log|u_0|$ distribution of Figure \ref{uanom} confirms our expectation. Except for the two planets that were detected by pure-planetary caustics, all the other by-eye planets, which are located inside the near-resonant range, were detected with $|u_0| \lesssim 0.05$ and $|u_{\rm anom}| \lesssim 0.07$. The roughly one-dex gap of the by-eye planets, at $0.05 \lesssim |u_0| \lesssim 0.62$ and $0.07 \lesssim |u_{\rm anom}| \lesssim 0.78$, is filled by the AnomalyFinder planets\footnote{Although it might seem that the correlation could be with anomaly brightness rather than $|u_0|$ (because smaller $|u_0|$ implies a more highly magnified event), \cite{2018_subprime} showed that there is no correlation with event brightness at the time of the anomaly between by-eye vs. AnomalyFinder detections. On the other hand, \cite{KB190253} and \citet{2019_prime} have shown that AnomalyFinder is much better at finding anomalies with smaller $\Delta\chi^2$.}.

However, there is no planet located at the left lower corner of Figure \ref{uanom}, with $|u_{0,{\rm limit}}| = 0.0060$ and $|u_{\rm anom, limit}| = 0.0158$. Although six of the planets were detected in high-magnification events ($|u_0| < 0.01$), all the planetary signals occurred on the low- and median-magnification regions. This desert of high-magnification planetary signals could be caused by the insufficient observing cadences of the current KMTNet survey. High-magnification planetary signals for $q < 10^{-4}$ events are weak and thus require dense observations over the peak. There are three known $q < 10^{-4}$ events whose planetary signals occurred on the high-magnification regions ($|u_{\rm anom}| < 0.01$). They are OGLE-2005-BLG-169 with $u_{\rm anom} = 0.0012$ \citep{OB05169}, KMT-2021-BLG-0171 with $|u_{\rm anom}| = 0.0066$ \citep{KB210171}, and KMT-2022-BLG-0440 with $|u_{\rm anom}| = 0.0041$ (Zhang et al. in prep). The follow-up data played decisive roles in these detections and the combined cadences of survey and follow-up data are higher than $30~{\rm hr}^{-1}$, while the highest cadence of the current KMTNet survey is $8~{\rm hr}^{-1}$ for about 0.4 deg$^2$ from the overlap of two $\Gamma = 4~{\rm hr}^{-1}$ fields.

However, we note that AnomalyFinder used the KMTNet end-of-year pipeline light curves, for which the photometric quality is not as good as that of TLC re-reductions. For the three follow-up planets, the planetary signals only have $\Delta I < 0.05$ mag. Thus, TLC re-reductions may be needed to recover such weak signals in the KMTNet data, and we cannot rule out the possibility that the desert may also be due to the imperfect KMTNet photometric quality. Each year there are about 20 events with $|u_0| < 0.01$ observed by KMTNet with $\Gamma \geq 4~{\rm hr}^{-1}$. The current KMTNet quasi-automated TLC re-reductions pipeline takes $<$1 hr of human effort for each event (H. Yang et al. in prep), so an optimized systematic search for $q < 10^{-4}$ planets in the KMTNet high-magnification events can be done very quickly. This search could have important implications for future space-based microlensing projects, because their tentative cadences are similar to or lower than $\Gamma = 4~{\rm hr}^{-1}$ \citep{MatthewWFIRSTI,ET,CSST_Wei}. If this search demonstrates that high-magnification events need denser observations to capture the weak planetary signals for low-$q$ planets, one could consider conducting (if feasible) ground-based follow-up projects for high-magnification events that are discovered by space-based telescopes. We also note that for the 2018 AnomalyFinder planets \citep{2018_prime,2018_subprime} and 2019 prime-field AnomalyFinder planets \citep{2019_prime}, which are complete now, only one $q > 10^{-4}$ planet, KMT-2019-BLG-1953Lb, has $|u_{\rm anom}| < |u_{\rm anom, limit}| $. Future analysis of all the 2016--2019 KMTNet should check whether the desert is obvious for more massive planets.

\subsection{Prime and sub-Prime Fields}

In its 2015 commissioning season, KMTNet observed four fields at a cadence of $\Gamma = 6~{\rm hr}^{-1}$. To support the 2016--2019 \Sp\ microlensing campaign \citep{GouldSp1,GouldSp2,GouldSp4,GouldSp3,GouldSp5,GouldSp6} and find more planets, KMTNet monitored a wider area, with a total of (3, 7, 11, 3) fields at cadences of $\Gamma \sim (4, 1, 0.4, 0.2)~ {\rm hr}^{-1}$. The three fields with the highest cadence are the KMTNet prime fields and the other 21 are the KMTNet sub-prime fields. See Figure 12 of \cite{KMTeventfinder} for the field placement. As shown in Table \ref{tab:-4planet} and Figure \ref{uanom}, the prime fields played the main role in the detection of $q < 10^{-4}$ planets, as predicted by \cite{Henderson2014}, and 17 of 27 planets were detected therein. However, the sub-prime fields are also important and six of the ten lowest-$q$ planets were discovered therein. 

For the six planets with $|u_0| < 0.01$, there is a clear bias in cadences, and only one of them was detected from the sub-prime fields. For the prime and sub-prime fields, the current detection rates are 1.25 and 0.25 per year, respectively. Because $\sim 60\%$ of the KMTNet microlensing events are located in the sub-prime fields, if the sub-prime-field events with $|u_0| < 0.01$ can had the same cadence as the prime-field events from follow-up observations, each year there would be $(1.25 \times (60\% / 40\%) - 0.25) = 1.6$ more $q < 10^{-4}$ planets. Because follow-up observations can have higher cadences and capture the high-magnification planetary signals (e.g., \citealt{KB210171}), the yield of a follow-up project can be at least two $q < 10^{-4}$ planets per year. The reward is not only enlarging the low-$q$ planetary sample, but also an independent check of the statistical results from AnomalyFinder if the follow-up planets can form a homogeneous statistical sample \citep{mufun}. However, this reward requires that the KMTNet alert-finder system should alert new events before they reach the high-magnification regions (e.g., $A > 20$).

\acknowledgments
We appreciate the anonymous referee for helping to improve the paper. W.Zang, H.Y., S.M., J.Z., and W.Zhu acknowledge support by the National Science Foundation of China (Grant No. 12133005). W.Zang acknowledges the support from the Harvard-Smithsonian Center for Astrophysics through the CfA Fellowship. This research was supported by the Korea Astronomy and Space Science Institute under the R\&D program (Project No. 2022-1-830-04) supervised by the Ministry of Science and ICT. This research has made use of the KMTNet system operated by the Korea Astronomy and Space Science Institute (KASI) and the data were obtained at three host sites of CTIO in Chile, SAAO in South Africa, and SSO in Australia. The authors acknowledge the Tsinghua Astrophysics High-Performance Computing platform at Tsinghua University for providing computational and data storage resources that have contributed to the research results reported within this paper. Work by J.C.Y. acknowledges support from N.S.F Grant No. AST-2108414. Work by C.H. was supported by the grants of National Research Foundation of Korea (2019R1A2C2085965 and 2020R1A4A2002885). Y.S. acknowledges support from BSF Grant No. 2020740. W.Zhu acknowledges the science research grants from the China Manned Space Project with No.\ CMS-CSST-2021-A11. This research uses data obtained through the Telescope Access Program (TAP), which has been funded by the TAP member institutes. This research is supported by Tsinghua University Initiative Scientific Research Program (Program ID 2019Z07L02017). This research has made use of the NASA Exoplanet Archive, which is operated by the California Institute of Technology, under contract with the National Aeronautics and Space Administration under the Exoplanet Exploration Program.

\bibliography{Zang.bib}

\begin{thebibliography}{}
\expandafter\ifx\csname natexlab\endcsname\relax\def\natexlab#1{#1}\fi
\providecommand{\url}[1]{\href{#1}{#1}}
\providecommand{\dodoi}[1]{doi:~\href{http://doi.org/#1}{\nolinkurl{#1}}}
\providecommand{\doeprint}[1]{\href{http://ascl.net/#1}{\nolinkurl{http://ascl.net/#1}}}
\providecommand{\doarXiv}[1]{\href{https://arxiv.org/abs/#1}{\nolinkurl{https://arxiv.org/abs/#1}}}

\bibitem[{{Adams} {et~al.}(2018){Adams}, {Boyajian}, \& {von
  Braun}}]{Adams2018}
{Adams}, A.~D., {Boyajian}, T.~S., \& {von Braun}, K. 2018, \mnras, 473, 3608,
  \dodoi{10.1093/mnras/stx2367}

\bibitem[{{Alard} \& {Lupton}(1998)}]{Alard1998}
{Alard}, C., \& {Lupton}, R.~H. 1998, \apj, 503, 325, \dodoi{10.1086/305984}

\bibitem[{{Albrow}(2017)}]{pyDIA}
{Albrow}, M.~D. 2017, {Michaeldalbrow/Pydia: Initial Release On Github.},
  v1.0.0,  Zenodo, \dodoi{10.5281/zenodo.268049}

\bibitem[{{Albrow} {et~al.}(2009){Albrow}, {Horne}, {Bramich}, {Fouqu{\'e}},
  {Miller}, {Beaulieu}, {Coutures}, {Menzies}, {Williams}, {Batista},
  {Bennett}, {Brillant}, {Cassan}, {Dieters}, {Dominis Prester}, {Donatowicz},
  {Greenhill}, {Kains}, {Kane}, {Kubas}, {Marquette}, {Pollard}, {Sahu},
  {Tsapras}, {Wambsganss}, \& {Zub}}]{pysis}
{Albrow}, M.~D., {Horne}, K., {Bramich}, D.~M., {et~al.} 2009, \mnras, 397,
  2099, \dodoi{10.1111/j.1365-2966.2009.15098.x}

\bibitem[{{An} {et~al.}(2002){An}, {Albrow}, {Beaulieu}, {Caldwell}, {DePoy},
  {Dominik}, {Gaudi}, {Gould}, {Greenhill}, {Hill}, {Kane}, {Martin},
  {Menzies}, {Pogge}, {Pollard}, {Sackett}, {Sahu}, {Vermaak}, {Watson}, \&
  {Williams}}]{An2002}
{An}, J.~H., {Albrow}, M.~D., {Beaulieu}, J.-P., {et~al.} 2002, \apj, 572, 521,
  \dodoi{10.1086/340191}

\bibitem[{{Batista} {et~al.}(2011){Batista}, {Gould}, {Dieters}, {Dong},
  {Bond}, {Beaulieu}, {Maoz}, {Monard}, {Christie}, {McCormick}, {Albrow},
  {Horne}, {Tsapras}, {Burgdorf}, {Calchi Novati}, {Skottfelt}, {Caldwell},
  {Koz{\l}owski}, {Kubas}, {Gaudi}, {Han}, {Bennett}, {An}, {MOA
  Collaboration}, {Abe}, {Botzler}, {Douchin}, {Freeman}, {Fukui}, {Furusawa},
  {Hearnshaw}, {Hosaka}, {Itow}, {Kamiya}, {Kilmartin}, {Korpela}, {Lin},
  {Ling}, {Makita}, {Masuda}, {Matsubara}, {Miyake}, {Muraki}, {Nagaya},
  {Nishimoto}, {Ohnishi}, {Okumura}, {Perrott}, {Rattenbury}, {Saito},
  {Sullivan}, {Sumi}, {Sweatman}, {Tristram}, {von Seggern}, {Yock}, {PLANET
  Collaboration}, {Brillant}, {Calitz}, {Cassan}, {Cole}, {Cook}, {Coutures},
  {Dominis Prester}, {Donatowicz}, {Greenhill}, {Hoffman}, {Jablonski}, {Kane},
  {Kains}, {Marquette}, {Martin}, {Martioli}, {Meintjes}, {Menzies},
  {Pedretti}, {Pollard}, {Sahu}, {Vinter}, {Wambsganss}, {Watson}, {Williams},
  {Zub}, {FUN Collaboration}, {Allen}, {Bolt}, {Bos}, {DePoy}, {Drummond},
  {Eastman}, {Gal-Yam}, {Gorbikov}, {Higgins}, {Janczak}, {Kaspi}, {Lee},
  {Mallia}, {Maury}, {Monard}, {Moorhouse}, {Morgan}, {Natusch}, {Ofek},
  {Park}, {Pogge}, {Polishook}, {Santallo}, {Shporer}, {Spector}, {Thornley},
  {Yee}, {MiNDSTEp Consortium}, {Bozza}, {Browne}, {Dominik}, {Dreizler},
  {Finet}, {Glitrup}, {Grundahl}, {Harps{\o}e}, {Hessman}, {Hinse},
  {Hundertmark}, {J{\o}rgensen}, {Liebig}, {Maier}, {Mancini}, {Mathiasen},
  {Rahvar}, {Ricci}, {Scarpetta}, {Southworth}, {Surdej}, {Zimmer}, {RoboNet
  Collaboration}, {Allan}, {Bramich}, {Snodgrass}, {Steele}, \&
  {Street}}]{MB09387}
{Batista}, V., {Gould}, A., {Dieters}, S., {et~al.} 2011, \aap, 529, A102,
  \dodoi{10.1051/0004-6361/201016111}

\bibitem[{{Beaulieu} {et~al.}(2006){Beaulieu}, {Bennett}, {Fouqu{\'e}},
  {Williams}, {Dominik}, {J{\o}rgensen}, {Kubas}, {Cassan}, {Coutures},
  {Greenhill}, {Hill}, {Menzies}, {Sackett}, {Albrow}, {Brillant}, {Caldwell},
  {Calitz}, {Cook}, {Corrales}, {Desort}, {Dieters}, {Dominis}, {Donatowicz},
  {Hoffman}, {Kane}, {Marquette}, {Martin}, {Meintjes}, {Pollard}, {Sahu},
  {Vinter}, {Wambsganss}, {Woller}, {Horne}, {Steele}, {Bramich}, {Burgdorf},
  {Snodgrass}, {Bode}, {Udalski}, {Szyma{\'n}ski}, {Kubiak}, {Wi{\c e}ckowski},
  {Pietrzy{\'n}ski}, {Soszy{\'n}ski}, {Szewczyk}, {Wyrzykowski},
  {Paczy{\'n}ski}, {Abe}, {Bond}, {Britton}, {Gilmore}, {Hearnshaw}, {Itow},
  {Kamiya}, {Kilmartin}, {Korpela}, {Masuda}, {Matsubara}, {Motomura},
  {Muraki}, {Nakamura}, {Okada}, {Ohnishi}, {Rattenbury}, {Sako}, {Sato},
  {Sasaki}, {Sekiguchi}, {Sullivan}, {Tristram}, {Yock}, \&
  {Yoshioka}}]{OB05390}
{Beaulieu}, J.-P., {Bennett}, D.~P., {Fouqu{\'e}}, P., {et~al.} 2006, \nat,
  439, 437, \dodoi{10.1038/nature04441}

\bibitem[{{Bennett} {et~al.}(2008){Bennett}, {Bond}, {Udalski}, {Sumi}, {Abe},
  {Fukui}, {Furusawa}, {Hearnshaw}, {Holderness}, {Itow}, {Kamiya}, {Korpela},
  {Kilmartin}, {Lin}, {Ling}, {Masuda}, {Matsubara}, {Miyake}, {Muraki},
  {Nagaya}, {Okumura}, {Ohnishi}, {Perrott}, {Rattenbury}, {Sako}, {Saito},
  {Sato}, {Skuljan}, {Sullivan}, {Sweatman}, {Tristram}, {Yock}, {Kubiak},
  {Szyma{\'n}ski}, {Pietrzy{\'n}ski}, {Soszy{\'n}ski}, {Szewczyk},
  {Wyrzykowski}, {Ulaczyk}, {Batista}, {Beaulieu}, {Brillant}, {Cassan},
  {Fouqu{\'e}}, {Kervella}, {Kubas}, \& {Marquette}}]{MB07192}
{Bennett}, D.~P., {Bond}, I.~A., {Udalski}, A., {et~al.} 2008, \apj, 684, 663,
  \dodoi{10.1086/589940}

\bibitem[{{Bensby} {et~al.}(2013){Bensby}, {Yee}, {Feltzing}, {Johnson},
  {Gould}, {Cohen}, {Asplund}, {Mel{\'e}ndez}, {Lucatello}, {Han}, {Thompson},
  {Gal-Yam}, {Udalski}, {Bennett}, {Bond}, {Kohei}, {Sumi}, {Suzuki}, {Suzuki},
  {Takino}, {Tristram}, {Yamai}, \& {Yonehara}}]{Bensby2013}
{Bensby}, T., {Yee}, J.~C., {Feltzing}, S., {et~al.} 2013, \aap, 549, A147,
  \dodoi{10.1051/0004-6361/201220678}

\bibitem[{{Bhattacharya} {et~al.}(2018){Bhattacharya}, {Beaulieu}, {Bennett},
  {Anderson}, {Koshimoto}, {Lu}, {Batista}, {Blackman}, {Bond}, {Fukui},
  {Henderson}, {Hirao}, {Marquette}, {Mroz}, {Ranc}, \& {Udalski}}]{OB120950}
{Bhattacharya}, A., {Beaulieu}, J.~P., {Bennett}, D.~P., {et~al.} 2018, \aj,
  156, 289, \dodoi{10.3847/1538-3881/aaed46}

\bibitem[{{Bond} {et~al.}(2004){Bond}, {Udalski}, {Jaroszy{\'n}ski},
  {Rattenbury}, {Paczy{\'n}ski}, {Soszy{\'n}ski}, {Wyrzykowski},
  {Szyma{\'n}ski}, {Kubiak}, {Szewczyk}, {{\.Z}ebru{\'n}}, {Pietrzy{\'n}ski},
  {Abe}, {Bennett}, {Eguchi}, {Furuta}, {Hearnshaw}, {Kamiya}, {Kilmartin},
  {Kurata}, {Masuda}, {Matsubara}, {Muraki}, {Noda}, {Okajima}, {Sako},
  {Sekiguchi}, {Sullivan}, {Sumi}, {Tristram}, {Yanagisawa}, {Yock}, \& {OGLE
  Collaboration}}]{OB03235}
{Bond}, I.~A., {Udalski}, A., {Jaroszy{\'n}ski}, M., {et~al.} 2004, \apjl, 606,
  L155, \dodoi{10.1086/420928}

\bibitem[{{Bond} {et~al.}(2017){Bond}, {Bennett}, {Sumi}, {Udalski}, {Suzuki},
  {Rattenbury}, {Bozza}, {Koshimoto}, {Abe}, {Asakura}, {Barry},
  {Bhattacharya}, {Donachie}, {Evans}, {Fukui}, {Hirao}, {Itow}, {Li}, {Ling},
  {Masuda}, {Matsubara}, {Muraki}, {Nagakane}, {Ohnishi}, {Ranc}, {Saito},
  {Sharan}, {Sullivan}, {Tristram}, {Yamada}, {Yamada}, {Yonehara}, {Skowron},
  {Szyma{\'n}ski}, {Poleski}, {Mr{\'o}z}, {Soszy{\'n}ski}, {Pietrukowicz},
  {Koz{\l}owski}, {Ulaczyk}, \& {Pawlak}}]{OB161195_MOA}
{Bond}, I.~A., {Bennett}, D.~P., {Sumi}, T., {et~al.} 2017, \mnras, 469, 2434,
  \dodoi{10.1093/mnras/stx1049}

\bibitem[{{Bozza}(2010)}]{Bozza2010}
{Bozza}, V. 2010, \mnras, 408, 2188, \dodoi{10.1111/j.1365-2966.2010.17265.x}

\bibitem[{{Bozza} {et~al.}(2018){Bozza}, {Bachelet}, {Bartoli{\'c}}, {Heintz},
  {Hoag}, \& {Hundertmark}}]{Bozza2018}
{Bozza}, V., {Bachelet}, E., {Bartoli{\'c}}, F., {et~al.} 2018, \mnras, 479,
  5157, \dodoi{10.1093/mnras/sty1791}

\bibitem[{{Dominik}(1999)}]{Dominik1999}
{Dominik}, M. 1999, \aap, 349, 108

\bibitem[{{Dong} {et~al.}(2009){Dong}, {Gould}, {Udalski}, {Anderson},
  {Christie}, {Gaudi}, {OGLE Collaboration}, {Jaroszy{\'n}ski}, {Kubiak},
  {Szyma{\'n}ski}, {Pietrzy{\'n}ski}, {Soszy{\'n}ski}, {Szewczyk}, {Ulaczyk},
  {Wyrzykowski}, {{$\mu$}FUN Collaboration}, {DePoy}, {Fox}, {Gal-Yam}, {Han},
  {L{\'e}pine}, {McCormick}, {Ofek}, {Park}, {Pogge}, {MOA Collaboration},
  {Abe}, {Bennett}, {Bond}, {Britton}, {Gilmore}, {Hearnshaw}, {Itow},
  {Kamiya}, {Kilmartin}, {Korpela}, {Masuda}, {Matsubara}, {Motomura},
  {Muraki}, {Nakamura}, {Ohnishi}, {Okada}, {Rattenbury}, {Saito}, {Sako},
  {Sasaki}, {Sullivan}, {Sumi}, {Tristram}, {Yanagisawa}, {Yock}, {Yoshoika},
  {PLANET/RoboNet Collaborations}, {Albrow}, {Beaulieu}, {Brillant}, {Calitz},
  {Cassan}, {Cook}, {Coutures}, {Dieters}, {Dominis Prester}, {Donatowicz},
  {Fouqu{\'e}}, {Greenhill}, {Hill}, {Hoffman}, {Horne}, {J{\o}rgensen},
  {Kane}, {Kubas}, {Marquette}, {Martin}, {Meintjes}, {Menzies}, {Pollard},
  {Sahu}, {Vinter}, {Wambsganss}, {Williams}, {Bode}, {Bramich}, {Burgdorf},
  {Snodgrass}, {Steele}, {Doublier}, \& {Foellmi}}]{OB050071D}
{Dong}, S., {Gould}, A., {Udalski}, A., {et~al.} 2009, \apj, 695, 970,
  \dodoi{10.1088/0004-637X/695/2/970}

\bibitem[{{Foreman-Mackey} {et~al.}(2013){Foreman-Mackey}, {Hogg}, {Lang}, \&
  {Goodman}}]{emcee}
{Foreman-Mackey}, D., {Hogg}, D.~W., {Lang}, D., \& {Goodman}, J. 2013, \pasp,
  125, 306, \dodoi{10.1086/670067}

\bibitem[{{Gaudi}(1998)}]{Gaudi1998}
{Gaudi}, B.~S. 1998, \apj, 506, 533, \dodoi{10.1086/306256}

\bibitem[{{Gaudi} \& {Gould}(1997)}]{GG1997}
{Gaudi}, B.~S., \& {Gould}, A. 1997, \apj, 486, 85, \dodoi{10.1086/304491}

\bibitem[{{Ge} {et~al.}(2022){Ge}, {Zhang}, {Zang}, {Deng}, {Mao}, {Xie},
  {Liu}, {Zhou}, {Willis}, {Huang}, {Howell}, {Feng}, {Zhu}, {Yao}, {Liu},
  {Aizawa}, {Zhu}, {Li}, {Ma}, {Ye}, {Yu}, {Xiang}, {Yu}, {Liu}, {Yang},
  {Wang}, {Shi}, {Fang}, {Zong}, {Liu}, {Zhang}, {Zhang}, {El-Badry}, {Shen},
  {Tam}, {Hu}, {Yang}, {Zou}, {Wu}, {Lei}, {Wei}, {Wu}, {Sun}, {Wang}, {Zhang},
  {Xu}, {Yang}, {Li}, {Xiang}, {Wang}, {Wang}, {Zhang}, {Jia}, {Yuan}, {Zhang},
  {Xuesong Wang}, {Gan}, {Wang}, {Zhao}, {Liu}, {Wei}, {Kang}, {Yang}, {Qi},
  {Liu}, {Zhang}, {Zhu}, {Zhou}, {Zhang}, {Yu}, {Zhang}, {Li}, {Tang}, {Wang},
  {Wang}, {Li}, {Cheng}, {Shen}, {Li}, {Pan}, {Yang}, {Gao}, {Song}, {Wang},
  {Zhang}, {Chen}, {Wang}, {Zhang}, {Wang}, {Zeng}, {Zheng}, {Zhu}, {Guo},
  {Zhang}, {Li}, {Wen}, {Feng}, {Chen}, {Chen}, {Han}, {Yang}, {Wang}, {Duan},
  {Huang}, {Liang}, {Bi}, {Gai}, {Ge}, {Guo}, {Huang}, {Li}, {Li}, {Li},
  {Yuxi}, {Lu}, {Rix}, {Shi}, {Song}, {Tang}, {Ting}, {Wu}, {Wu}, {Yang},
  {Yin}, {Gould}, {Lee}, {Dong}, {Yee}, {Shvartzvald}, {Yang}, {Kuang},
  {Zhang}, {Liao}, {Qi}, {Yang}, {Zhang}, {Jiang}, {Ou}, {Li}, {Beck},
  {Bedding}, {Campante}, {Chaplin}, {Christensen-Dalsgaard}, {Garc{\'\i}a},
  {Gaulme}, {Gizon}, {Hekker}, {Huber}, {Khanna}, {Li}, {Mathur}, {Miglio},
  {Mosser}, {Ong}, {Santos}, {Stello}, {Bowman}, {Lares-Martiz}, {Murphy},
  {Niu}, {Ma}, {Moln{\'a}r}, {Fu}, {De Cat}, {Su}, \& {consortium}}]{ET}
{Ge}, J., {Zhang}, H., {Zang}, W., {et~al.} 2022, arXiv e-prints,
  arXiv:2206.06693.
\newblock \doarXiv{2206.06693}

\bibitem[{{Gonzalez} {et~al.}(2012){Gonzalez}, {Rejkuba}, {Zoccali}, {Valenti},
  {Minniti}, {Schultheis}, {Tobar}, \& {Chen}}]{Gonzalez2012}
{Gonzalez}, O.~A., {Rejkuba}, M., {Zoccali}, M., {et~al.} 2012, \aap, 543, A13,
  \dodoi{10.1051/0004-6361/201219222}

\bibitem[{{Gould}(1992)}]{Gould1992}
{Gould}, A. 1992, \apj, 392, 442, \dodoi{10.1086/171443}

\bibitem[{{Gould}(1996)}]{Gould2D}
---. 1996, \apj, 470, 201, \dodoi{10.1086/177861}

\bibitem[{{Gould}(2000)}]{Gould2000}
---. 2000, \apj, 542, 785, \dodoi{10.1086/317037}

\bibitem[{{Gould}(2004)}]{Gouldpies2004}
---. 2004, \apj, 606, 319, \dodoi{10.1086/382782}

\bibitem[{{Gould} {et~al.}(2013){Gould}, {Carey}, \& {Yee}}]{GouldSp1}
{Gould}, A., {Carey}, S., \& {Yee}, J. 2013, {Spitzer Microlens Planets and
  Parallaxes}, Spitzer Proposal

\bibitem[{{Gould} {et~al.}(2014{\natexlab{a}}){Gould}, {Carey}, \&
  {Yee}}]{GouldSp2}
---. 2014{\natexlab{a}}, {Galactic Distribution of Planets from Spitzer
  Microlens Parallaxes}, Spitzer Proposal

\bibitem[{{Gould} {et~al.}(2016){Gould}, {Carey}, \& {Yee}}]{GouldSp5}
---. 2016, {Galactic Distribution of Planets Spitzer Microlens Parallaxes},
  Spitzer Proposal

\bibitem[{{Gould} \& {Loeb}(1992)}]{Andy1992}
{Gould}, A., \& {Loeb}, A. 1992, \apj, 396, 104, \dodoi{10.1086/171700}

\bibitem[{{Gould} {et~al.}(2015{\natexlab{a}}){Gould}, {Yee}, \&
  {Carey}}]{GouldSp4}
{Gould}, A., {Yee}, J., \& {Carey}, S. 2015{\natexlab{a}}, {Degeneracy Breaking
  for K2 Microlens Parallaxes}, Spitzer Proposal

\bibitem[{{Gould} {et~al.}(2015{\natexlab{b}}){Gould}, {Yee}, \&
  {Carey}}]{GouldSp3}
---. 2015{\natexlab{b}}, {Galactic Distribution of Planets From
  High-Magnification Microlensing Events}, Spitzer Proposal

\bibitem[{{Gould} {et~al.}(2018){Gould}, {Yee}, {Carey}, \&
  {Shvartzvald}}]{GouldSp6}
{Gould}, A., {Yee}, J., {Carey}, S., \& {Shvartzvald}, Y. 2018, {The Galactic
  Distribution of Planets via Spitzer Microlensing Parallax}, Spitzer Proposal

\bibitem[{{Gould} {et~al.}(2006){Gould}, {Udalski}, {An}, {Bennett}, {Zhou},
  {Dong}, {Rattenbury}, {Gaudi}, {Yock}, {Bond}, {Christie}, {Horne},
  {Anderson}, {Stanek}, {DePoy}, {Han}, {McCormick}, {Park}, {Pogge},
  {Poindexter}, {Soszy{\'n}ski}, {Szyma{\'n}ski}, {Kubiak}, {Pietrzy{\'n}ski},
  {Szewczyk}, {Wyrzykowski}, {Ulaczyk}, {Paczy{\'n}ski}, {Bramich},
  {Snodgrass}, {Steele}, {Burgdorf}, {Bode}, {Botzler}, {Mao}, \&
  {Swaving}}]{OB05169}
{Gould}, A., {Udalski}, A., {An}, D., {et~al.} 2006, \apjl, 644, L37,
  \dodoi{10.1086/505421}

\bibitem[{{Gould} {et~al.}(2010){Gould}, {Dong}, {Gaudi}, {Udalski}, {Bond},
  {Greenhill}, {Street}, {Dominik}, {Sumi}, {Szyma{\'n}ski}, {Han}, {Allen},
  {Bolt}, {Bos}, {Christie}, {DePoy}, {Drummond}, {Eastman}, {Gal-Yam},
  {Higgins}, {Janczak}, {Kaspi}, {Koz{\l}owski}, {Lee}, {Mallia}, {Maury},
  {Maoz}, {McCormick}, {Monard}, {Moorhouse}, {Morgan}, {Natusch}, {Ofek},
  {Park}, {Pogge}, {Polishook}, {Santallo}, {Shporer}, {Spector}, {Thornley},
  {Yee}, {{$\mu$}FUN Collaboration}, {Kubiak}, {Pietrzy{\'n}ski},
  {Soszy{\'n}ski}, {Szewczyk}, {Wyrzykowski}, {Ulaczyk}, {Poleski}, {OGLE
  Collaboration}, {Abe}, {Bennett}, {Botzler}, {Douchin}, {Freeman}, {Fukui},
  {Furusawa}, {Hearnshaw}, {Hosaka}, {Itow}, {Kamiya}, {Kilmartin}, {Korpela},
  {Lin}, {Ling}, {Makita}, {Masuda}, {Matsubara}, {Miyake}, {Muraki}, {Nagaya},
  {Nishimoto}, {Ohnishi}, {Okumura}, {Perrott}, {Philpott}, {Rattenbury},
  {Saito}, {Sako}, {Sullivan}, {Sweatman}, {Tristram}, {von Seggern}, {Yock},
  {MOA Collaboration}, {Albrow}, {Batista}, {Beaulieu}, {Brillant}, {Caldwell},
  {Calitz}, {Cassan}, {Cole}, {Cook}, {Coutures}, {Dieters}, {Dominis Prester},
  {Donatowicz}, {Fouqu{\'e}}, {Hill}, {Hoffman}, {Jablonski}, {Kane}, {Kains},
  {Kubas}, {Marquette}, {Martin}, {Martioli}, {Meintjes}, {Menzies},
  {Pedretti}, {Pollard}, {Sahu}, {Vinter}, {Wambsganss}, {Watson}, {Williams},
  {Zub}, {PLANET Collaboration}, {Allan}, {Bode}, {Bramich}, {Burgdorf},
  {Clay}, {Fraser}, {Hawkins}, {Horne}, {Kerins}, {Lister}, {Mottram},
  {Saunders}, {Snodgrass}, {Steele}, {Tsapras}, {RoboNet Collaboration},
  {J{\o}rgensen}, {Anguita}, {Bozza}, {Calchi Novati}, {Harps{\o}e}, {Hinse},
  {Hundertmark}, {Kj{\ae}rgaard}, {Liebig}, {Mancini}, {Masi}, {Mathiasen},
  {Rahvar}, {Ricci}, {Scarpetta}, {Southworth}, {Surdej}, {Th{\"o}ne}, \&
  {MiNDSTEp Consortium}}]{mufun}
{Gould}, A., {Dong}, S., {Gaudi}, B.~S., {et~al.} 2010, \apj, 720, 1073,
  \dodoi{10.1088/0004-637X/720/2/1073}

\bibitem[{{Gould} {et~al.}(2014{\natexlab{b}}){Gould}, {Udalski}, {Shin},
  {Porritt}, {Skowron}, {Han}, {Yee}, {Koz{\l}owski}, {Choi}, {Poleski},
  {Wyrzykowski}, {Ulaczyk}, {Pietrukowicz}, {Mr{\'o}z}, {Szyma{\'n}ski},
  {Kubiak}, {Soszy{\'n}ski}, {Pietrzy{\'n}ski}, {Gaudi}, {Christie},
  {Drummond}, {McCormick}, {Natusch}, {Ngan}, {Tan}, {Albrow}, {DePoy},
  {Hwang}, {Jung}, {Lee}, {Park}, {Pogge}, {Abe}, {Bennett}, {Bond}, {Botzler},
  {Freeman}, {Fukui}, {Fukunaga}, {Itow}, {Koshimoto}, {Larsen}, {Ling},
  {Masuda}, {Matsubara}, {Muraki}, {Namba}, {Ohnishi}, {Philpott},
  {Rattenbury}, {Saito}, {Sullivan}, {Sumi}, {Suzuki}, {Tristram}, {Tsurumi},
  {Wada}, {Yamai}, {Yock}, {Yonehara}, {Shvartzvald}, {Maoz}, {Kaspi}, \&
  {Friedmann}}]{OB130341}
{Gould}, A., {Udalski}, A., {Shin}, I.~G., {et~al.} 2014{\natexlab{b}},
  Science, 345, 46, \dodoi{10.1126/science.1251527}

\bibitem[{{Gould} {et~al.}(2020){Gould}, {Ryu}, {Calchi Novati}, {Zang},
  {Albrow}, {Chung}, {Han}, {Hwang}, {Jung}, {Shin}, {Shvartzvald}, {Yee},
  {Cha}, {Kim}, {Kim}, {Kim}, {Lee}, {Lee}, {Lee}, {Park}, {Pogge}, {Beichman},
  {Bryden}, {Carey}, {Gaudi}, {Henderson}, {Zhu}, {Fouque}, {Penny}, {Petric},
  {Burdullis}, \& {Mao}}]{KB180029}
{Gould}, A., {Ryu}, Y.-H., {Calchi Novati}, S., {et~al.} 2020, Journal of
  Korean Astronomical Society, 53, 9.
\newblock \doarXiv{1906.11183}

\bibitem[{{Gould} {et~al.}(2022){Gould}, {Han}, {Zang}, {Yang}, {Hwang},
  {Udalski}, {Bond}, {Albrow}, {Chung}, {Jung}, {Ryu}, {Shin}, {Shvartzvald},
  {Yee}, {Cha}, {Kim}, {Kim}, {Kim}, {Lee}, {Lee}, {Lee}, {Park}, {Pogge},
  {KMTNet Collaboration}, {Mr{\'o}z}, {Szyma{\'n}ski}, {Skowron}, {Poleski},
  {Soszy{\'n}ski}, {Pietrukowicz}, {Koz{\l}owski}, {Ulaczyk}, {Rybicki},
  {Iwanek}, {Wrona}, {OGLE Collaboration}, {Abe}, {Barry}, {Bennett},
  {Bhattacharya}, {Fujii}, {Fukui}, {Hirao}, {Silva}, {Kirikawa}, {Kondo},
  {Koshimoto}, {Matsubara}, {Matsumoto}, {Miyazaki}, {Muraki}, {Okamura},
  {Olmschenk}, {Ranc}, {Rattenbury}, {Satoh}, {Sumi}, {Suzuki}, {Toda},
  {Tristram}, {Vandorou}, {Yama}, {Moa Collaboration}, {Beichman}, {Bryden},
  {Novati}, {Gaudi}, {Henderson}, {Penny}, {Jacklin}, {Stassun}, \& {Ukirt
  Microlensing Team}}]{2018_prime}
{Gould}, A., {Han}, C., {Zang}, W., {et~al.} 2022, \aap, 664, A13,
  \dodoi{10.1051/0004-6361/202243744}

\bibitem[{{Griest} \& {Safizadeh}(1998)}]{Griest1998}
{Griest}, K., \& {Safizadeh}, N. 1998, \apj, 500, 37, \dodoi{10.1086/305729}

\bibitem[{{Han} {et~al.}(2021){Han}, {Udalski}, {Lee}, {Albrow}, {Chung},
  {Gould}, {Hwang}, {Jung}, {Kim}, {Kim}, {Ryu}, {Shin}, {Shvartzvald}, {Yee},
  {Zang}, {Cha}, {Kim}, {Kim}, {Lee}, {Lee}, {Park}, {Pogge}, {Kim}, {Kim},
  {Mr{\'o}z}, {Szyma{\'n}ski}, {Skowron}, {Poleski}, {Soszy{\'n}ski},
  {Pietrukowicz}, {Koz{\l}owski}, {Ulaczyk}, {Rybicki}, {Iwanek}, \&
  {Wrona}}]{KB181025}
{Han}, C., {Udalski}, A., {Lee}, C.-U., {et~al.} 2021, \aap, 649, A90,
  \dodoi{10.1051/0004-6361/202039817}

\bibitem[{{Han} {et~al.}(2022{\natexlab{a}}){Han}, {Gould}, {Albrow}, {Chung},
  {Hwang}, {Kil Jung}, {Kim}, {Lee}, {Ryu}, {Shin}, {Shvartzvald}, {Yee},
  {Zang}, {Cha}, {Kim}, {Kim}, {Kim}, {Lee}, {Lee}, {Park}, {Pogge}, \&
  {Kim}}]{KB181988}
{Han}, C., {Gould}, A., {Albrow}, M.~D., {et~al.} 2022{\natexlab{a}}, \aap,
  658, A62, \dodoi{10.1051/0004-6361/202142077}

\bibitem[{{Han} {et~al.}(2022{\natexlab{b}}){Han}, {Kim}, {Gould}, {Udalski},
  {Bond}, {Bozza}, {Jung}, {Albrow}, {Chung}, {Hwang}, {Ryu}, {Shin},
  {Shvartzvald}, {Yee}, {Zang}, {Cha}, {Kim}, {Kim}, {Lee}, {Lee}, {Lee},
  {Park}, {Pogge}, {KMTNet Collaboration}, {Mr{\'o}z}, {Szyma{\'n}ski},
  {Skowron}, {Poleski}, {Soszy{\'n}ski}, {Pietrukowicz}, {Koz{\l}owski},
  {Ulaczyk}, {Rybicki}, {Iwanek}, {OGLE Collaboration}, {Abe}, {Barry},
  {Bennett}, {Bhattacharya}, {Fujii}, {Fukui}, {Hirao}, {Itow}, {Kirikawa},
  {Koshimoto}, {Kondo}, {Matsubara}, {Matsumoto}, {Miyazaki}, {Muraki},
  {Olmschenk}, {Okamura}, {Ranc}, {Rattenbury}, {Satoh}, {Silva}, {Sumi},
  {Suzuki}, {Toda}, {Tristram}, {Vandorou}, {Yama}, \& {MOA
  Collaboration}}]{OB171691}
{Han}, C., {Kim}, D., {Gould}, A., {et~al.} 2022{\natexlab{b}}, \aap, 664, A33,
  \dodoi{10.1051/0004-6361/202243484}

\bibitem[{{Hayashi}(1981)}]{Hayashi1981}
{Hayashi}, C. 1981, in Fundamental Problems in the Theory of Stellar Evolution,
  ed. D.~{Sugimoto}, D.~Q. {Lamb}, \& D.~N. {Schramm}, Vol.~93, 113--126

\bibitem[{{Henderson} {et~al.}(2014){Henderson}, {Gaudi}, {Han}, {Skowron},
  {Penny}, {Nataf}, \& {Gould}}]{Henderson2014}
{Henderson}, C.~B., {Gaudi}, B.~S., {Han}, C., {et~al.} 2014, \apj, 794, 52,
  \dodoi{10.1088/0004-637X/794/1/52}

\bibitem[{{Holtzman} {et~al.}(1998){Holtzman}, {Watson}, {Baum}, {Grillmair},
  {Groth}, {Light}, {Lynds}, \& {O'Neil}}]{HSTCMD}
{Holtzman}, J.~A., {Watson}, A.~M., {Baum}, W.~A., {et~al.} 1998, \aj, 115,
  1946, \dodoi{10.1086/300336}

\bibitem[{{Hwang} {et~al.}(2013){Hwang}, {Choi}, {Bond}, {Sumi}, {Han},
  {Gaudi}, {Gould}, {Bozza}, {Beaulieu}, {Tsapras}, {Abe}, {Bennett},
  {Botzler}, {Chote}, {Freeman}, {Fukui}, {Fukunaga}, {Harris}, {Itow},
  {Koshimoto}, {Ling}, {Masuda}, {Matsubara}, {Muraki}, {Namba}, {Ohnishi},
  {Rattenbury}, {Saito}, {Sullivan}, {Sweatman}, {Suzuki}, {Tristram}, {Wada},
  {Yamai}, {Yock}, {Yonehara}, {MOA Collaboration}, {de Almeida}, {DePoy},
  {Dong}, {Jablonski}, {Jung}, {Kavka}, {Lee}, {Park}, {Pogge}, {}, {Yee},
  {{$\mu$}FUN Collaboration}, {Albrow}, {Bachelet}, {Batista}, {Brillant},
  {Caldwell}, {Cassan}, {Cole}, {Corrales}, {Coutures}, {Dieters}, {Dominis
  Prester}, {Donatowicz}, {Fouqu{\'e}}, {Greenhill}, {J{\o}rgensen}, {Kane},
  {Kubas}, {Marquette}, {Martin}, {Meintjes}, {Menzies}, {Pollard}, {Williams},
  {Wouters}, {PLANET Collaboration}, {Bramich}, {Dominik}, {Horne}, {Browne},
  {Hundertmark}, {Ipatov}, {Kains}, {Snodgrass}, {Steele}, {Street}, \&
  {RoboNet Collaboration}}]{MB12486}
{Hwang}, K.-H., {Choi}, J.-Y., {Bond}, I.~A., {et~al.} 2013, \apj, 778, 55,
  \dodoi{10.1088/0004-637X/778/1/55}

\bibitem[{{Hwang} {et~al.}(2018{\natexlab{a}}){Hwang}, {Udalski},
  {Shvartzvald}, {Ryu}, {Albrow}, {Chung}, {Gould}, {Han}, {Jung}, {}, {Yee},
  {Zhu}, {Cha}, {Kim}, {Kim}, {Kim}, {Lee}, {Lee}, {Lee}, {Park}, {Pogge},
  {KMTNet Collaboration}, {Skowron}, {Mr{\'o}z}, {Poleski}, {Koz{\l}owski},
  {Soszy{\'n}ski}, {Pietrukowicz}, {Szyma{\'n}ski}, {Ulaczyk}, {Pawlak}, {OGLE
  Collaboration}, {Bryden}, {Beichman}, {Calchi Novati}, {Gaudi}, {Henderson},
  {Jacklin}, {Penny}, \& {UKIRT Microlensing Team}}]{OB170173}
{Hwang}, K.-H., {Udalski}, A., {Shvartzvald}, Y., {et~al.} 2018{\natexlab{a}},
  \aj, 155, 20, \dodoi{10.3847/1538-3881/aa992f}

\bibitem[{{Hwang} {et~al.}(2018{\natexlab{b}}){Hwang}, {Kim}, {Kim}, {Gould},
  {Albrow}, {Chung}, {Han}, {Jung}, {Ryu}, {Shin}, {Shvartzvald}, {Yee},
  {Zang}, {Zhu}, {Cha}, {Kim}, {Lee}, {Lee}, {Lee}, {Park}, \&
  {Pogge}}]{KB160212}
{Hwang}, K.~H., {Kim}, H.~W., {Kim}, D.~J., {et~al.} 2018{\natexlab{b}},
  Journal of Korean Astronomical Society, 51, 197,
  \dodoi{10.5303/JKAS.2018.51.6.197}

\bibitem[{{Hwang} {et~al.}(2022){Hwang}, {Zang}, {Gould}, {Udalski}, {Bond},
  {Yang}, {Mao}, {Mao}, {Albrow}, {Chung}, {Han}, {Kil Jung}, {Ryu}, {Shin},
  {Shvartzvald}, {Yee}, {Cha}, {Kim}, {Kim}, {Kim}, {Lee}, {Lee}, {Lee},
  {Park}, {Pogge}, {Pogge}, {Mr{\'o}z}, {Poleski}, {Skowron}, {Szyma{\'n}ski},
  {Soszy{\'n}ski}, {Pietrukowicz}, {Koz{\l}owski}, {Ulaczyk}, {Rybicki},
  {Iwanek}, {Wrona}, {Gromadzki}, {Gromadzki}, {Abe}, {Barry}, {Bennett},
  {Bhattacharya}, {Fujii}, {Fukui}, {Hirao}, {Itow}, {Kirikawa}, {Kondo},
  {Koshimoto}, {Munford}, {Matsubara}, {Miyazaki}, {Muraki}, {Olmschenk},
  {Ranc}, {Rattenbury}, {Satoh}, {Shoji}, {Ishitani Silva}, {Sumi}, {Suzuki},
  {Tristram}, {Yonehara}, {Yonehara}, {Zhang}, {Zhu}, {Penny}, {Fouqu{\'e}}, \&
  {Fouqu{\'e}}}]{KB190253}
{Hwang}, K.-H., {Zang}, W., {Gould}, A., {et~al.} 2022, \aj, 163, 43,
  \dodoi{10.3847/1538-3881/ac38ad}

\bibitem[{{Ishitani Silva} {et~al.}(2022){Ishitani Silva}, {Ranc}, {Bennett},
  {Bond}, {Zang}, {Abe}, {Barry}, {Bhattacharya}, {Fujii}, {Fukui}, {Hirao},
  {Itow}, {Kirikawa}, {Kondo}, {Koshimoto}, {Matsubara}, {Matsumoto},
  {Miyazaki}, {Muraki}, {Olmschenk}, {Okamura}, {Rattenbury}, {Satoh}, {Sumi},
  {Suzuki}, {Toda}, {Tristram}, {Vandorou}, {Yama}, {Petric}, {Burdullis},
  {Fouqu{\'e}}, {Mao}, {Penny}, {Zhu}, {CFHT Microlensing Collaboration}, \&
  {Rau}}]{MB20135}
{Ishitani Silva}, S., {Ranc}, C., {Bennett}, D.~P., {et~al.} 2022, \aj, 164,
  118, \dodoi{10.3847/1538-3881/ac82b8}

\bibitem[{{Jiang} {et~al.}(2004){Jiang}, {DePoy}, {Gal-Yam}, {Gaudi}, {Gould},
  {Han}, {Lipkin}, {Maoz}, {Ofek}, {Park}, {Pogge}, {MuFun Collaboration},
  {Udalski}, {Kubiak}, {Szyma{\'n}ski}, {Szewczyk}, {{\.Z}ebru{\'n}},
  {Wyrzykowski}, {Soszy{\'n}ski}, {Pietrzy{\'n}ski}, {OGLE Collaboration},
  {Albrow}, {Beaulieu}, {Caldwell}, {Cassan}, {Coutures}, {Dominik},
  {Donatowicz}, {Fouqu{\'e}}, {Greenhill}, {Hill}, {Horne}, {J{\o}rgensen},
  {J{\o}rgensen}, {Kane}, {Kubas}, {Martin}, {Menzies}, {Pollard}, {Sahu},
  {Wambsganss}, {Watson}, {Williams}, \& {PLANET Collaboration}}]{Jiang2004}
{Jiang}, G., {DePoy}, D.~L., {Gal-Yam}, A., {et~al.} 2004, \apj, 617, 1307,
  \dodoi{10.1086/425678}

\bibitem[{{Jung} {et~al.}(2017){Jung}, {Udalski}, {Yee}, {Sumi}, {Gould},
  {Han}, {Albrow}, {Lee}, {Kim}, {Chung}, {Hwang}, {Ryu}, {}, {Zhu}, {Cha},
  {Kim}, {Lee}, {Park}, {Pogge}, {KMTNet Collaboration}, {Pietrukowicz},
  {Koz{\l}owski}, {Poleski}, {Skowron}, {Mr{\'o}z}, {Szyma{\'n}ski},
  {Soszy{\'n}ski}, {Pawlak}, {Ulaczyk}, {OGLE Collaboration}, {Abe}, {Bennett},
  {Barry}, {Bond}, {Asakura}, {Bhattacharya}, {Donachie}, {Freeman}, {Fukui},
  {Hirao}, {Itow}, {Koshimoto}, {Li}, {Ling}, {Masuda}, {Matsubara}, {Muraki},
  {Nagakane}, {Oyokawa}, {Rattenbury}, {Sharan}, {Sullivan}, {Suzuki},
  {Tristram}, {Yamada}, {Yamada}, {Yonehara}, \& {MOA
  Collaboration}}]{OB160733}
{Jung}, Y.~K., {Udalski}, A., {Yee}, J.~C., {et~al.} 2017, \aj, 153, 129,
  \dodoi{10.3847/1538-3881/aa5d07}

\bibitem[{{Jung} {et~al.}(2020){Jung}, {Udalski}, {Zang}, {Bond}, {Yee}, {Han},
  {Albrow}, {Chung}, {Gould}, {Hwang}, {Ryu}, {Shin}, {Shvartzvald}, {Cha},
  {Kim}, {Kim}, {Kim}, {Lee}, {Lee}, {Lee}, {Park}, {Pogge}, {KMTNet
  Collaboration}, {Mr{\'o}z}, {Szyma{\'n}ski}, {Skowron}, {Poleski},
  {Soszy{\'n}ski}, {Pietrukowicz}, {Koz{\l}owski}, {Ulaczyk}, {Rybicki},
  {Iwanek}, {Wrona}, {OGLE Collaboration}, {Abe}, {Barry}, {Bennett},
  {Bhattacharya}, {Donachie}, {Fujii}, {Fukui}, {Hirao}, {Itow}, {Kamei},
  {Kondo}, {Koshimoto}, {Li}, {Matsubara}, {Miyazaki}, {Muraki}, {Nagakane},
  {Ranc}, {Rattenbury}, {Satoh}, {Shoji}, {Suematsu}, {Sullivan}, {Sumi},
  {Suzuki}, {Tristram}, {Yamakawa}, {Yamamwaki}, {Yonehara}, \& {MOA
  Collaboration}}]{KB190842}
{Jung}, Y.~K., {Udalski}, A., {Zang}, W., {et~al.} 2020, \aj, 160, 255,
  \dodoi{10.3847/1538-3881/abbe93}

\bibitem[{{Jung} {et~al.}(2022){Jung}, {Zang}, {Han}, {Gould}, {Udalski},
  {Albrow}, {Chung}, {Hwang}, {Ryu}, {Shin}, {Shvartzvald}, {Yang}, {Yee},
  {Cha}, {Kim}, {Kim}, {Lee}, {Lee}, {Lee}, {Park}, {Pogge}, {KMTNet
  Collaboration}, {Mr{\'o}z}, {Szyma{\'n}ski}, {Skowron}, {Poleski},
  {Soszy{\'n}ski}, {Pietrukowicz}, {Koz{\l}owski}, {Ulaczyk}, {Rybicki},
  {Iwanek}, {Wrona}, \& {OGLE Collaboration}}]{2018_subprime}
{Jung}, Y.~K., {Zang}, W., {Han}, C., {et~al.} 2022, \aj, 164, 262,
  \dodoi{10.3847/1538-3881/ac9c5c}

\bibitem[{{Kennedy} \& {Kenyon}(2008)}]{snowline}
{Kennedy}, G.~M., \& {Kenyon}, S.~J. 2008, \apj, 673, 502,
  \dodoi{10.1086/524130}

\bibitem[{{Kim} {et~al.}(2018{\natexlab{a}}){Kim}, {Kim}, {Hwang}, {Albrow},
  {Chung}, {Gould}, {Han}, {Jung}, {Ryu}, {}, {Yee}, {Zhu}, {Cha}, {Kim},
  {Lee}, {Lee}, {Lee}, {Park}, {Pogge}, \& {The KMTNet
  Collaboration}}]{KMTeventfinder}
{Kim}, D.-J., {Kim}, H.-W., {Hwang}, K.-H., {et~al.} 2018{\natexlab{a}}, \aj,
  155, 76, \dodoi{10.3847/1538-3881/aaa47b}

\bibitem[{{Kim} {et~al.}(2018{\natexlab{b}}){Kim}, {Hwang}, {Shvartzvald},
  {Yee}, {Albrow}, {Cha}, {Chung}, {Gould}, {Han}, {Jung}, {Kim}, {Kim}, {Lee},
  {Lee}, {Lee}, {Park}, {Pogge}, {Ryu}, {Shin}, \& {Zang}}]{KMTAF}
{Kim}, H.-W., {Hwang}, K.-H., {Shvartzvald}, Y., {et~al.} 2018{\natexlab{b}},
  arXiv e-prints, arXiv:1806.07545.
\newblock \doarXiv{1806.07545}

\bibitem[{{Kim} {et~al.}(2016){Kim}, {Lee}, {Park}, {Kim}, {Cha}, {Lee}, {Han},
  {Chun}, \& {Yuk}}]{KMT2016}
{Kim}, S.-L., {Lee}, C.-U., {Park}, B.-G., {et~al.} 2016, Journal of Korean
  Astronomical Society, 49, 37, \dodoi{10.5303/JKAS.2016.49.1.037}

\bibitem[{{Kondo} {et~al.}(2021){Kondo}, {Yee}, {Bennett}, {Sumi}, {Koshimoto},
  {Bond}, {Gould}, {Udalski}, {Shvartzvald}, {Jung}, {Zang}, {Bozza},
  {Bachelet}, {Hundertmark}, {Rattenbury}, {Abe}, {Barry}, {Bhattacharya},
  {Donachie}, {Fukui}, {Fujii}, {Hirao}, {Silva}, {Itow}, {Kirikawa}, {Li},
  {Matsubara}, {Miyazaki}, {Muraki}, {Olmschenk}, {Ranc}, {Satoh}, {Shoji},
  {Suzuki}, {Tanaka}, {Tristram}, {Yamawaki}, {Yonehara}, {Mr{\'o}z},
  {Poleski}, {Skowron}, {Szyma{\'n}ski}, {Soszy{\'n}ski}, {Koz{\l}owski},
  {Pietrukowicz}, {Ulaczyk}, {Rybicki}, {Iwanek}, {Wrona}, {Albrow}, {Chung},
  {Han}, {Hwang}, {Kim}, {Shin}, {Cha}, {Kim}, {Kim}, {Lee}, {Lee}, {Lee},
  {Park}, {Pogge}, {Ryu}, {Beichman}, {Bryden}, {Novati}, {Carey}, {Gaudi},
  {Henderson}, {Zhu}, {Maoz}, {Penny}, {Dominik}, {J{\o}rgensen},
  {Longa-Pe{\~n}a}, {Peixinho}, {Sajadian}, {Skottfelt}, {Snodgrass},
  {Tregloan-Reed}, {Burgdorf}, {Campbell-White}, {Dib}, {Fujii}, {Hinse},
  {Khalouei}, {Rahvar}, {Rabus}, {Southworth}, {Tsapras}, {Street}, {Bramich},
  {Cassan}, {Horne}, {Wambsganss}, {Mao}, {Saha}, \& {ROME/REA Project
  Team}}]{OB181185}
{Kondo}, I., {Yee}, J.~C., {Bennett}, D.~P., {et~al.} 2021, \aj, 162, 77,
  \dodoi{10.3847/1538-3881/ac00ba}

\bibitem[{{Mao} \& {Paczynski}(1991)}]{Shude1991}
{Mao}, S., \& {Paczynski}, B. 1991, \apjl, 374, L37, \dodoi{10.1086/186066}

\bibitem[{{Min} {et~al.}(2011){Min}, {Dullemond}, {Kama}, \&
  {Dominik}}]{Min2011}
{Min}, M., {Dullemond}, C.~P., {Kama}, M., \& {Dominik}, C. 2011, \icarus, 212,
  416, \dodoi{10.1016/j.icarus.2010.12.002}

\bibitem[{{Muraki} {et~al.}(2011){Muraki}, {Han}, {Bennett}, {Suzuki},
  {Monard}, {Street}, {Jorgensen}, {Kundurthy}, {Skowron}, {Becker}, {Albrow},
  {Fouqu{\'e}}, {Heyrovsk{\'y}}, {Barry}, {Beaulieu}, {Wellnitz}, {Bond},
  {Sumi}, {Dong}, {Gaudi}, {Bramich}, {Dominik}, {Abe}, {Botzler}, {Freeman},
  {Fukui}, {Furusawa}, {Hayashi}, {Hearnshaw}, {Hosaka}, {Itow}, {Kamiya},
  {Korpela}, {Kilmartin}, {Lin}, {Ling}, {Makita}, {Masuda}, {Matsubara},
  {Miyake}, {Nishimoto}, {Ohnishi}, {Perrott}, {Rattenbury}, {Saito},
  {Skuljan}, {Sullivan}, {Sweatman}, {Tristram}, {Wada}, {Yock}, {MOA
  Collaboration}, {Christie}, {DePoy}, {Gorbikov}, {Gould}, {Kaspi}, {Lee},
  {Mallia}, {Maoz}, {McCormick}, {Moorhouse}, {Natusch}, {Park}, {Pogge},
  {Polishook}, {Shporer}, {Thornley}, {Yee}, {{\ensuremath{\mu}}FUN
  Collaboration}, {Allan}, {Browne}, {Horne}, {Kains}, {Snodgrass}, {Steele},
  {Tsapras}, {RoboNet Collaboration}, {Batista}, {Bennett}, {Brillant},
  {Caldwell}, {Cassan}, {Cole}, {Corrales}, {Coutures}, {Dieters}, {Dominis
  Prester}, {Donatowicz}, {Greenhill}, {Kubas}, {Marquette}, {Martin},
  {Menzies}, {Sahu}, {Waldman}, {Williams}, {Zub}, {PLANET Collaboration},
  {Bourhrous}, {Matsuoka}, {Nagayama}, {Oi}, {Randriamanakoto}, {IRSF
  Observers}, {Bozza}, {Burgdorf}, {Calchi Novati}, {Dreizler}, {Finet},
  {Glitrup}, {Harps{\o}e}, {Hinse}, {Hundertmark}, {Liebig}, {Maier},
  {Mancini}, {Mathiasen}, {Rahvar}, {Ricci}, {Scarpetta}, {Skottfelt},
  {Surdej}, {Southworth}, {Wambsganss}, {Zimmer}, {MiNDSTEp Consortium},
  {Udalski}, {Poleski}, {Wyrzykowski}, {Ulaczyk}, {Szyma{\'n}ski}, {Kubiak},
  {Pietrzy{\'n}ski}, {Soszy{\'n}ski}, \& {OGLE Collaboration}}]{MB09266}
{Muraki}, Y., {Han}, C., {Bennett}, D.~P., {et~al.} 2011, \apj, 741, 22,
  \dodoi{10.1088/0004-637X/741/1/22}

\bibitem[{{Nataf} {et~al.}(2013){Nataf}, {Gould}, {Fouqu{\'e}}, {Gonzalez},
  {Johnson}, {Skowron}, {}, {Szyma{\'n}ski}, {Kubiak}, {Pietrzy{\'n}ski},
  {Soszy{\'n}ski}, {Ulaczyk}, {Wyrzykowski}, \& {Poleski}}]{Nataf2013}
{Nataf}, D.~M., {Gould}, A., {Fouqu{\'e}}, P., {et~al.} 2013, \apj, 769, 88,
  \dodoi{10.1088/0004-637X/769/2/88}

\bibitem[{{Nataf} {et~al.}(2016){Nataf}, {Gonzalez}, {Casagrande}, {Zasowski},
  {Wegg}, {Wolf}, {Kunder}, {Alonso-Garcia}, {Minniti}, {Rejkuba}, {Saito},
  {Valenti}, {Zoccali}, {Poleski}, {Pietrzy{\'n}ski}, {Skowron},
  {Soszy{\'n}ski}, {Szyma{\'n}ski}, {Udalski}, {Ulaczyk}, \&
  {Wyrzykowski}}]{Nataf2016}
{Nataf}, D.~M., {Gonzalez}, O.~A., {Casagrande}, L., {et~al.} 2016, \mnras,
  456, 2692, \dodoi{10.1093/mnras/stv2843}

\bibitem[{{Paczy{\'n}ski}(1986)}]{Paczynski1986}
{Paczy{\'n}ski}, B. 1986, \apj, 304, 1, \dodoi{10.1086/164140}

\bibitem[{{Penny} {et~al.}(2019){Penny}, {Gaudi}, {Kerins}, {Rattenbury},
  {Mao}, {Robin}, \& {Calchi Novati}}]{MatthewWFIRSTI}
{Penny}, M.~T., {Gaudi}, B.~S., {Kerins}, E., {et~al.} 2019, \apjs, 241, 3,
  \dodoi{10.3847/1538-4365/aafb69}

\bibitem[{{Poindexter} {et~al.}(2005){Poindexter}, {Afonso}, {Bennett},
  {Glicenstein}, {Gould}, {Szyma{\'n}ski}, \& {Udalski}}]{Poindexter2005}
{Poindexter}, S., {Afonso}, C., {Bennett}, D.~P., {et~al.} 2005, \apj, 633,
  914, \dodoi{10.1086/468182}

\bibitem[{{Ranc} {et~al.}(2019){Ranc}, {Bennett}, {Hirao}, {Udalski}, {Han},
  {Bond}, {Yee}, {and}, {Albrow}, {Chung}, {Gould}, {Hwang}, {Jung}, {Ryu},
  {Shin}, {Shvartzvald}, {Zang}, {Zhu}, {Cha}, {Kim}, {Kim}, {Kim}, {Lee},
  {Lee}, {Lee}, {Park}, {Pogge}, {KMTNet Collaboration}, {Abe}, {Barry},
  {Bhattacharya}, {Donachie}, {Fukui}, {Itow}, {Kawasaki}, {Kondo},
  {Koshimoto}, {Li}, {Matsubara}, {Miyazaki}, {Muraki}, {Nagakane},
  {Rattenbury}, {Suematsu}, {Sullivan}, {Sumi}, {Suzuki}, {Tristram},
  {Yonehara}, {MOA Collaboration}, {Poleski}, {Mr{\'o}z}, {Skowron},
  {Szyma{\'n}ski}, {Soszy{\'n}ski}, {Koz{\l}owski}, {Pietrukowicz}, {Ulaczyk},
  \& {OGLE Collaboration}}]{OB151670}
{Ranc}, C., {Bennett}, D.~P., {Hirao}, Y., {et~al.} 2019, \aj, 157, 232,
  \dodoi{10.3847/1538-3881/ab141b}

\bibitem[{{Rota} {et~al.}(2021){Rota}, {Hirao}, {Bozza}, {Abe}, {Barry},
  {Bennett}, {Bhattacharya}, {Bond}, {Donachie}, {Fukui}, {Fujii}, {Silva},
  {Itow}, {Kirikawa}, {Koshimoto}, {Li}, {Matsubara}, {Miyazaki}, {Muraki},
  {Olmschenk}, {Ranc}, {Satoh}, {Sumi}, {Suzuki}, {Tristram}, \&
  {Yonehara}}]{MB06074}
{Rota}, P., {Hirao}, Y., {Bozza}, V., {et~al.} 2021, \aj, 162, 59,
  \dodoi{10.3847/1538-3881/ac0155}

\bibitem[{{Ryu} {et~al.}(2020){Ryu}, {Udalski}, {Yee}, {Penny}, {Zang},
  {Albrow}, {Chung}, {Gould}, {Han}, {Hwang}, {Jung}, {Shin}, {Shvartzvald},
  {Cha}, {Kim}, {Kim}, {Kim}, {Lee}, {Lee}, {Lee}, {Park}, {Pogge}, {KMTNet
  Collaboration}, {Mr{\'o}z}, {Szyma{\'n}ski}, {Skowron}, {Poleski},
  {Soszy{\'n}ski}, {Pietrukowicz}, {Koz{\l}owski}, {Ulaczyk}, {Rybicki},
  {Iwanek}, {Wrona}, {OGLE Collaboration}, {Mao}, {Fouque}, {Zhu}, {Wang}, \&
  {CFHT Microlensing Collaboration}}]{OB180532}
{Ryu}, Y.-H., {Udalski}, A., {Yee}, J.~C., {et~al.} 2020, \aj, 160, 183,
  \dodoi{10.3847/1538-3881/abaa3f}

\bibitem[{{Ryu} {et~al.}(2022){Ryu}, {Kil Jung}, {Yang}, {Gould}, {Albrow},
  {Chung}, {Han}, {Hwang}, {Shin}, {Shvartzvald}, {Yee}, {Zang}, {Cha}, {Kim},
  {Kim}, {Lee}, {Lee}, {Lee}, {Park}, \& {Pogge}}]{KMT2021_mass1}
{Ryu}, Y.-H., {Kil Jung}, Y., {Yang}, H., {et~al.} 2022, \aj, 164, 180,
  \dodoi{10.3847/1538-3881/ac8d6c}

\bibitem[{{Sako} {et~al.}(2008){Sako}, {Sekiguchi}, {Sasaki}, {Okajima}, {Abe},
  {Bond}, {Hearnshaw}, {Itow}, {Kamiya}, {Kilmartin}, {Masuda}, {Matsubara},
  {Muraki}, {Rattenbury}, {Sullivan}, {Sumi}, {Tristram}, {Yanagisawa}, \&
  {Yock}}]{Sako2008}
{Sako}, T., {Sekiguchi}, T., {Sasaki}, M., {et~al.} 2008, Experimental
  Astronomy, 22, 51, \dodoi{10.1007/s10686-007-9082-5}

\bibitem[{{Shvartzvald} {et~al.}(2017){Shvartzvald}, {Yee}, {Calchi Novati},
  {Gould}, {Lee}, {Beichman}, {Bryden}, {Carey}, {Gaudi}, {Henderson}, {Zhu},
  {Spitzer team}, {Albrow}, {Cha}, {Chung}, {Han}, {Hwang}, {Jung}, {Kim},
  {Kim}, {Kim}, {Lee}, {Park}, {Pogge}, {Ryu}, {}, \& {KMTNet
  group}}]{OB161195}
{Shvartzvald}, Y., {Yee}, J.~C., {Calchi Novati}, S., {et~al.} 2017, \apjl,
  840, L3, \dodoi{10.3847/2041-8213/aa6d09}

\bibitem[{{Skowron} {et~al.}(2011){Skowron}, {Udalski}, {Gould}, {Dong},
  {Monard}, {Han}, {Nelson}, {McCormick}, {Moorhouse}, {Thornley}, {Maury},
  {Bramich}, {Greenhill}, {Koz{\l}owski}, {Bond}, {Poleski}, {Wyrzykowski},
  {Ulaczyk}, {Kubiak}, {Szyma{\'n}ski}, {Pietrzy{\'n}ski}, {Soszy{\'n}ski},
  {OGLE Collaboration}, {Gaudi}, {Yee}, {Hung}, {Pogge}, {DePoy}, {Lee},
  {Park}, {Allen}, {Mallia}, {Drummond}, {Bolt}, {{$\mu$}FUN Collaboration},
  {Allan}, {Browne}, {Clay}, {Dominik}, {Fraser}, {Horne}, {Kains}, {Mottram},
  {Snodgrass}, {Steele}, {Street}, {Tsapras}, {RoboNet Collaboration}, {Abe},
  {Bennett}, {Botzler}, {Douchin}, {Freeman}, {Fukui}, {Furusawa}, {Hayashi},
  {Hearnshaw}, {Hosaka}, {Itow}, {Kamiya}, {Kilmartin}, {Korpela}, {Lin},
  {Ling}, {Makita}, {Masuda}, {Matsubara}, {Muraki}, {Nagayama}, {Miyake},
  {Nishimoto}, {Ohnishi}, {Perrott}, {Rattenbury}, {Saito}, {Skuljan},
  {Sullivan}, {Sumi}, {Suzuki}, {Sweatman}, {Tristram}, {Wada}, {Yock}, {MOA
  Collaboration}, {Beaulieu}, {Fouqu{\'e}}, {Albrow}, {Batista}, {Brillant},
  {Caldwell}, {Cassan}, {Cole}, {Cook}, {Coutures}, {Dieters}, {Dominis
  Prester}, {Donatowicz}, {Kane}, {Kubas}, {Marquette}, {Martin}, {Menzies},
  {Sahu}, {Wambsganss}, {Williams}, {Zub}, \& {PLANET Collaboration}}]{OB09020}
{Skowron}, J., {Udalski}, A., {Gould}, A., {et~al.} 2011, \apj, 738, 87,
  \dodoi{10.1088/0004-637X/738/1/87}

\bibitem[{{Sumi} {et~al.}(2010){Sumi}, {Bennett}, {Bond}, {Udalski}, {Batista},
  {Dominik}, {Fouqu{\'e}}, {Kubas}, {Gould}, {Macintosh}, {Cook}, {Dong},
  {Skuljan}, {Cassan}, {Abe}, {Botzler}, {Fukui}, {Furusawa}, {Hearnshaw},
  {Itow}, {Kamiya}, {Kilmartin}, {Korpela}, {Lin}, {Ling}, {Masuda},
  {Matsubara}, {Miyake}, {Muraki}, {Nagaya}, {Nagayama}, {Ohnishi}, {Okumura},
  {Perrott}, {Rattenbury}, {Saito}, {Sako}, {Sullivan}, {Sweatman}, {Tristram},
  {Yock}, {MOA Collaboration}, {Beaulieu}, {Cole}, {Coutures}, {Duran},
  {Greenhill}, {Jablonski}, {Marboeuf}, {Martioli}, {Pedretti}, {Pejcha},
  {Rojo}, {Albrow}, {Brillant}, {Bode}, {Bramich}, {Burgdorf}, {Caldwell},
  {Calitz}, {Corrales}, {Dieters}, {Dominis Prester}, {Donatowicz}, {Hill},
  {Hoffman}, {Horne}, {J{\o}rgensen}, {Kains}, {Kane}, {Marquette}, {Martin},
  {Meintjes}, {Menzies}, {Pollard}, {Sahu}, {Snodgrass}, {Steele}, {Street},
  {Tsapras}, {Wambsganss}, {Williams}, {Zub}, {PLANET Collaboration},
  {Szyma{\'n}ski}, {Kubiak}, {Pietrzy{\'n}ski}, {Soszy{\'n}ski}, {Szewczyk},
  {Wyrzykowski}, {Ulaczyk}, {OGLE Collaboration}, {Allen}, {Christie}, {DePoy},
  {Gaudi}, {Han}, {Janczak}, {Lee}, {McCormick}, {Mallia}, {Monard}, {Natusch},
  {Park}, {Pogge}, {Santallo}, \& {{\ensuremath{\mu}}FUN
  Collaboration}}]{OB07368}
{Sumi}, T., {Bennett}, D.~P., {Bond}, I.~A., {et~al.} 2010, \apj, 710, 1641,
  \dodoi{10.1088/0004-637X/710/2/1641}

\bibitem[{{Suzuki} {et~al.}(2016){Suzuki}, {Bennett}, {Sumi}, {Bond}, {Rogers},
  {Abe}, {Asakura}, {Bhattacharya}, {Donachie}, {Freeman}, {Fukui}, {Hirao},
  {Itow}, {Koshimoto}, {Li}, {Ling}, {Masuda}, {Matsubara}, {Muraki},
  {Nagakane}, {Onishi}, {Oyokawa}, {Rattenbury}, {Saito}, {Sharan}, {Shibai},
  {Sullivan}, {Tristram}, {Yonehara}, \& {MOA Collaboration}}]{Suzuki2016}
{Suzuki}, D., {Bennett}, D.~P., {Sumi}, T., {et~al.} 2016, \apj, 833, 145,
  \dodoi{10.3847/1538-4357/833/2/145}

\bibitem[{{Szyma{\'n}ski} {et~al.}(2011){Szyma{\'n}ski}, {Udalski},
  {Soszy{\'n}ski}, {Kubiak}, {Pietrzy{\'n}ski}, {Poleski}, {Wyrzykowski}, \&
  {Ulaczyk}}]{OGLEIII}
{Szyma{\'n}ski}, M.~K., {Udalski}, A., {Soszy{\'n}ski}, I., {et~al.} 2011,
  \actaa, 61, 83.
\newblock \doarXiv{1107.4008}

\bibitem[{{Tomaney} \& {Crotts}(1996)}]{Tomaney1996}
{Tomaney}, A.~B., \& {Crotts}, A. P.~S. 1996, \aj, 112, 2872,
  \dodoi{10.1086/118228}

\bibitem[{{Udalski}(2003)}]{Udalski2003}
{Udalski}, A. 2003, \actaa, 53, 291

\bibitem[{{Udalski} {et~al.}(1994){Udalski}, {Szymanski}, {Kaluzny}, {Kubiak},
  {Mateo}, {Krzeminski}, \& {Paczynski}}]{Udalski1994}
{Udalski}, A., {Szymanski}, M., {Kaluzny}, J., {et~al.} 1994, \actaa, 44, 227

\bibitem[{{Udalski} {et~al.}(2015){Udalski}, {Szyma{\'n}ski}, \&
  {Szyma{\'n}ski}}]{OGLEIV}
{Udalski}, A., {Szyma{\'n}ski}, M.~K., \& {Szyma{\'n}ski}, G. 2015, \actaa, 65,
  1.
\newblock \doarXiv{1504.05966}

\bibitem[{{Udalski} {et~al.}(2018){Udalski}, {Ryu}, {Sajadian}, {Gould},
  {Mr{\'o}z}, {Poleski}, {Szyma{\'n}ski}, {Skowron}, {Soszy{\'n}ski},
  {Koz{\l}owski}, {Pietrukowicz}, {Ulaczyk}, {Pawlak}, {Rybicki}, {Iwanek},
  {Albrow}, {Chung}, {Han}, {Hwang}, {Jung}, {}, {Shvartzvald}, {Yee}, {Zang},
  {Zhu}, {Cha}, {Kim}, {Kim}, {Kim}, {Lee}, {Lee}, {Lee}, {Park}, {Pogge},
  {Bozza}, {Dominik}, {Helling}, {Hundertmark}, {J{\o}rgensen},
  {Longa-Pe{\~n}a}, {Lowry}, {Burgdorf}, {Campbell-White}, {Ciceri}, {Evans},
  {Figuera Jaimes}, {Fujii}, {Haikala}, {Henning}, {Hinse}, {Mancini},
  {Peixinho}, {Rahvar}, {Rabus}, {Skottfelt}, {Snodgrass}, {Southworth}, \&
  {von Essen}}]{OB171434}
{Udalski}, A., {Ryu}, Y.-H., {Sajadian}, S., {et~al.} 2018, \actaa, 68, 1.
\newblock \doarXiv{1802.02582}

\bibitem[{{Wang} {et~al.}(2022){Wang}, {Zang}, {Zhu}, {Hwang}, {Udalski},
  {Gould}, {Han}, {Albrow}, {Chung}, {Jung}, {Kim}, {Ryu}, {Shin},
  {Shvartzvald}, {Yee}, {Cha}, {Kim}, {Kim}, {Kim}, {Lee}, {Lee}, {Lee},
  {Park}, {Pogge}, {Poleski}, {Mr{\'o}z}, {Skowron}, {Szyma{\'n}ski},
  {Soszy{\'n}ski}, {Pietrukowicz}, {Koz{\l}owski}, {Ulaczyk}, {Rybicki},
  {Iwanek}, {Wrona}, {Gromadzki}, {Yang}, {Mao}, \& {Zhang}}]{OB180383}
{Wang}, H., {Zang}, W., {Zhu}, W., {et~al.} 2022, \mnras, 510, 1778,
  \dodoi{10.1093/mnras/stab3581}

\bibitem[{{Wang} {et~al.}(2018){Wang}, {Calchi Novati}, {Udalski}, {Gould},
  {Mao}, {Zang}, {Beichman}, {Bryden}, {Carey}, {Gaudi}, {Henderson},
  {Shvartzvald}, {Yee}, {Spitzer Team}, {Mr{\'o}z}, {Poleski}, {Skowron},
  {Szyma{\'n}ski}, {Soszy{\'n}ski}, {Koz{\l}owski}, {Pietrukowicz}, {Ulaczyk},
  {Pawlak}, {OGLE Collaboration}, {Albrow}, {Chung}, {Han}, {Hwang}, {Jung},
  {Ryu}, {Shin}, {Zhu}, {Cha}, {Kim}, {Kim}, {Kim}, {Lee}, {Lee}, {Lee},
  {Park}, {Pogge}, \& {KMTNet Collaboration}}]{OB171130}
{Wang}, T., {Calchi Novati}, S., {Udalski}, A., {et~al.} 2018, \apj, 860, 25,
  \dodoi{10.3847/1538-4357/aabcd2}

\bibitem[{{Wozniak}(2000)}]{Wozniak2000}
{Wozniak}, P.~R. 2000, \actaa, 50, 421

\bibitem[{{Yan} \& {Zhu}(2022)}]{CSST_Wei}
{Yan}, S., \& {Zhu}, W. 2022, Research in Astronomy and Astrophysics, 22,
  025006, \dodoi{10.1088/1674-4527/ac3c44}

\bibitem[{{Yang} {et~al.}(2022){Yang}, {Zang}, {Gould}, {Yee}, {Hwang},
  {Christie}, {Sumi}, {Zhang}, {Mao}, {Albrow}, {Chung}, {Han}, {Jung}, {Ryu},
  {Shin}, {Shvartzvald}, {Cha}, {Kim}, {Kim}, {Kim}, {Lee}, {Lee}, {Lee},
  {Park}, {Pogge}, {Drummond}, {Maoz}, {McCormick}, {Natusch}, {Penny}, {Zhu},
  {Bond}, {Abe}, {Barry}, {Bennett}, {Bhattacharya}, {Donachie}, {Fujii},
  {Fukui}, {Hirao}, {Itow}, {Kirikawa}, {Kondo}, {Koshimoto}, {Li},
  {Matsubara}, {Muraki}, {Miyazaki}, {Olmschenk}, {Ranc}, {Rattenbury},
  {Satoh}, {Shoji}, {Silva}, {Suzuki}, {Tanaka}, {Tristram}, {Yamawaki},
  {Yonehara}, \& {MOA Collaboration}}]{KB210171}
{Yang}, H., {Zang}, W., {Gould}, A., {et~al.} 2022, \mnras, 516, 1894,
  \dodoi{10.1093/mnras/stac2023}

\bibitem[{{Yee} {et~al.}(2021){Yee}, {Zang}, {Udalski}, {Ryu}, {Green},
  {Hennerley}, {Marmont}, {Sumi}, {Mao}, {Gromadzki}, {Mr{\'o}z}, {Skowron},
  {Poleski}, {Szyma{\'n}ski}, {Soszy{\'n}ski}, {Pietrukowicz}, {Koz{\l}owski},
  {Ulaczyk}, {Rybicki}, {Iwanek}, {Wrona}, {Albrow}, {Chung}, {Gould}, {Han},
  {Hwang}, {Jung}, {Kim}, {Shin}, {Shvartzvald}, {Cha}, {Kim}, {Kim}, {Lee},
  {Lee}, {Lee}, {Park}, {Pogge}, {Bachelet}, {Christie}, {Hundertmark}, {Maoz},
  {McCormick}, {Natusch}, {Penny}, {Street}, {Tsapras}, {Beichman}, {Bryden},
  {Novati}, {Carey}, {Gaudi}, {Henderson}, {Johnson}, {Zhu}, {Bond}, {Abe},
  {Barry}, {Bennett}, {Bhattacharya}, {Donachie}, {Fujii}, {Fukui}, {Hirao},
  {Silva}, {Itow}, {Kirikawa}, {Kondo}, {Koshimoto}, {Alex Li}, {Matsubara},
  {Muraki}, {Miyazaki}, {Olmschenk}, {Ranc}, {Rattenbury}, {Satoh}, {Shoji},
  {Suzuki}, {Tanaka}, {Tristram}, {Yamawaki}, {Yonehara}, \& {MOA
  Collaboration}}]{OB190960}
{Yee}, J.~C., {Zang}, W., {Udalski}, A., {et~al.} 2021, \aj, 162, 180,
  \dodoi{10.3847/1538-3881/ac1582}

\bibitem[{{Yoo} {et~al.}(2004){Yoo}, {DePoy}, {Gal-Yam}, {Gaudi}, {Gould},
  {Han}, {Lipkin}, {Maoz}, {Ofek}, {Park}, {Pogge}, {Mu-Fun Collaboration},
  {Udalski}, {Soszy{\'n}ski}, {Wyrzykowski}, {Kubiak}, {Szyma{\'n}ski},
  {Pietrzy{\'n}ski}, {Szewczyk}, {{\.Z}ebru{\'n}}, \& {OGLE
  Collaboration}}]{Yoo2004}
{Yoo}, J., {DePoy}, D.~L., {Gal-Yam}, A., {et~al.} 2004, \apj, 603, 139,
  \dodoi{10.1086/381241}

\bibitem[{{Zang} {et~al.}(2021{\natexlab{a}}){Zang}, {Han}, {Kondo}, {Yee},
  {Lee}, {Gould}, {Mao}, {de Almeida}, {Shvartzvald}, {Zhang}, {Albrow},
  {Chung}, {Hwang}, {Jung}, {Ryu}, {Shin}, {Cha}, {Kim}, {Kim}, {Kim}, {Lee},
  {Lee}, {Park}, {Pogge}, {Drummond}, {Tan}, {Nascimento J{\'u}nior}, {Maoz},
  {Penny}, {Zhu}, {Bond}, {Abe}, {Barry}, {Bennett}, {Bhattacharya},
  {Donachie}, {Fujii}, {Fukui}, {Hirao}, {Itow}, {Kirikawa}, {Koshimoto}, {Alex
  Li}, {Matsubara}, {Muraki}, {Miyazaki}, {Olmschenk}, {Ranc}, {Rattenbury},
  {Satoh}, {Shoji}, {Silva}, {Sumi}, {Suzuki}, {Tanaka}, {Tristram},
  {Yamawaki}, {Yonehara}, {Petric}, {Burdullis}, \& {Fouqu{\'e}}}]{KB200414}
{Zang}, W., {Han}, C., {Kondo}, I., {et~al.} 2021{\natexlab{a}}, Research in
  Astronomy and Astrophysics, 21, 239, \dodoi{10.1088/1674-4527/21/9/239}

\bibitem[{{Zang} {et~al.}(2021{\natexlab{b}}){Zang}, {Hwang}, {Udalski},
  {Wang}, {Zhu}, {Sumi}, {Yee}, {Gould}, {Mao}, {Zhang}, {Albrow}, {Chung},
  {Han}, {Jung}, {Ryu}, {Shin}, {Shvartzvald}, {Cha}, {Kim}, {Kim}, {Kim},
  {Lee}, {Lee}, {Lee}, {Park}, {Pogge}, {Mr{\'o}z}, {Skowron}, {Poleski},
  {Szyma{\'n}ski}, {Soszy{\'n}ski}, {Pietrukowicz}, {Koz{\l}owski}, {Ulaczyk},
  {Rybicki}, {Iwanek}, {Wrona}, {Gromadzki}, {Bond}, {Abe}, {Barry}, {Bennett},
  {Bhattacharya}, {Donachie}, {Fujii}, {Fukui}, {Hirao}, {Itow}, {Kirikawa},
  {Kondo}, {Koshimoto}, {Li}, {Matsubara}, {Muraki}, {Miyazaki}, {Olmschenk},
  {Ranc}, {Rattenbury}, {Satoh}, {Shoji}, {Ishitani Silva}, {Suzuki}, {Tanaka},
  {Tristram}, {Yamawaki}, {Yonehara}, {Beichman}, {Bryden}, {Calchi Novati},
  {Carey}, {Gaudi}, {Henderson}, {Johnson}, \& {Spitzer Team}}]{OB191053}
{Zang}, W., {Hwang}, K.-H., {Udalski}, A., {et~al.} 2021{\natexlab{b}}, \aj,
  162, 163, \dodoi{10.3847/1538-3881/ac12d4}

\bibitem[{{Zang} {et~al.}(2022{\natexlab{a}}){Zang}, {Yang}, {Han}, {Lee},
  {Udalski}, {Gould}, {Mao}, {Zhang}, {Zhu}, {Albrow}, {Chung}, {Hwang},
  {Jung}, {Ryu}, {Shin}, {Shvartzvald}, {Yee}, {Cha}, {Kim}, {Kim}, {Kim},
  {Lee}, {Lee}, {Park}, {Pogge}, {Mr{\'o}z}, {Skowron}, {Poleski},
  {Szyma{\'n}ski}, {Soszy{\'n}ski}, {Pietrukowicz}, {Koz{\l}owski}, {Ulaczyk},
  {Rybicki}, {Iwanek}, {Wrona}, \& {Gromadzki}}]{2019_prime}
{Zang}, W., {Yang}, H., {Han}, C., {et~al.} 2022{\natexlab{a}}, \mnras, 515,
  928, \dodoi{10.1093/mnras/stac1883}

\bibitem[{{Zang} {et~al.}(2022{\natexlab{b}}){Zang}, {Shvartzvald}, {Udalski},
  {Yee}, {Lee}, {Sumi}, {Zhang}, {Yang}, {Mao}, {Novati}, {Gould}, {Zhu},
  {Beichman}, {Bryden}, {Carey}, {Gaudi}, {Henderson}, {Mr{\'o}z}, {Skowron},
  {Poleski}, {Szyma{\'n}ski}, {Soszy{\'n}ski}, {Pietrukowicz}, {Koz{\l}owski},
  {Ulaczyk}, {Rybicki}, {Iwanek}, {Wrona}, {Albrow}, {Chung}, {Han}, {Hwang},
  {Jung}, {Ryu}, {Shin}, {Cha}, {Kim}, {Kim}, {Kim}, {Lee}, {Lee}, {Park},
  {Pogge}, {Bond}, {Abe}, {Barry}, {Bennett}, {Bhattacharya}, {Donachie},
  {Fujii}, {Fukui}, {Hirao}, {Itow}, {Kirikawa}, {Kondo}, {Koshimoto}, {Li},
  {Matsubara}, {Muraki}, {Miyazaki}, {Ranc}, {Rattenbury}, {Satoh}, {Shoji},
  {Suzuki}, {Tanaka}, {Tristram}, {Yamawaki}, {Yonehara}, {Bachelet},
  {Hundertmark}, {Jaimes}, {Maoz}, {Penny}, {Street}, \& {Tsapras}}]{OB180799}
{Zang}, W., {Shvartzvald}, Y., {Udalski}, A., {et~al.} 2022{\natexlab{b}},
  \mnras, 514, 5952, \dodoi{10.1093/mnras/stac1631}

\bibitem[{{Zhang} \& {Gaudi}(2022)}]{ZhangGaudi2022}
{Zhang}, K., \& {Gaudi}, B.~S. 2022, \apjl, 936, L22,
  \dodoi{10.3847/2041-8213/ac8c2b}

\bibitem[{{Zhang} {et~al.}(2022){Zhang}, {Gaudi}, \& {Bloom}}]{Zhang2022}
{Zhang}, K., {Gaudi}, B.~S., \& {Bloom}, J.~S. 2022, Nature Astronomy, 6, 782,
  \dodoi{10.1038/s41550-022-01671-6}

\bibitem[{{Zhu} {et~al.}(2014){Zhu}, {Penny}, {Mao}, {Gould}, \&
  {Gendron}}]{Zhu2014ApJ}
{Zhu}, W., {Penny}, M., {Mao}, S., {Gould}, A., \& {Gendron}, R. 2014, \apj,
  788, 73, \dodoi{10.1088/0004-637X/788/1/73}

\bibitem[{{Zhu} {et~al.}(2017){Zhu}, {Udalski}, {Calchi Novati}, {Chung},
  {Jung}, {Ryu}, {}, {Gould}, {Lee}, {Albrow}, {Yee}, {Han}, {Hwang}, {Cha},
  {Kim}, {Kim}, {Kim}, {Kim}, {Lee}, {Park}, {Pogge}, {KMTNet Collaboration},
  {Poleski}, {Mr{\'o}z}, {Pietrukowicz}, {Skowron}, {Szyma{\'n}ski},
  {Koz{\l}owski}, {Ulaczyk}, {Pawlak}, {OGLE Collaboration}, {Beichman},
  {Bryden}, {Carey}, {Fausnaugh}, {Gaudi}, {Henderson}, {Shvartzvald},
  {Wibking}, \& {Spitzer Team}}]{Zhu2017spitzer}
{Zhu}, W., {Udalski}, A., {Calchi Novati}, S., {et~al.} 2017, \aj, 154, 210,
  \dodoi{10.3847/1538-3881/aa8ef1}

\end{thebibliography}

\begin{center}  
\begin{longtable*}{c c c c c c c c c c c}
    \caption{Information of 2016--2019 KMTNet AnomalyFinder planetary sample with $q < 10^{-4}$ solutions}
    \label{tab:-4planet}\\
    \hline
    \hline
    Event Name & KMTNet Name & $\log q$ & $s$ & $|u_0|$ & Method & $\Delta\chi^2$ & Caustic-Crossing & Anomaly Type & Field \\ 
    \hline
    KB161105$^1$ & KB161105 & $-5.194 \pm 0.248$ & $1.143 \pm 0.009$ & 0.171 & Discovery & 0.0 & yes & bump & sub-prime \\
     &  & $-4.423 \pm 0.197$ & $1.136 \pm 0.011$ & 0.153 & & 2.3 & yes & & \\
     &  & $-4.184 \pm 0.206$ & $1.106 \pm 0.013$ & 0.154 & & 2.7 & no & & \\
     &  & $-5.027 \pm 0.080$ & $0.888 \pm 0.007$ & 0.148 & & 3.5 & no & & \\
     &  & $-4.069 \pm 0.182$ & $0.892 \pm 0.005$ & 0.154 & & 4.4 & yes & & \\
    \hline
    OB160007$^2$ & KB161991 & $-5.168 \pm 0.131$ & $2.829 \pm 0.009$ & 1.253 & Discovery &  & yes & bump & prime \\
    \hline
    OB191053$^3$ & KB191504 & $-4.885 \pm 0.035$ & $1.406 \pm 0.011$ & 0.373 & Discovery & & yes & bump & prime \\
    \hline
    OB190960$^4$ & KB191591 & $-4.830 \pm 0.041$ & $1.029 \pm 0.001$ & 0.0061 & Recovery & 0.0 & yes & bump & sub-prime \\
     &  & $-4.896 \pm 0.024$ & $0.997 \pm 0.001$ & 0.0060 & & 1.0 & yes & & \\
     &  & $-4.896 \pm 0.024$ & $0.996 \pm 0.001$ & 0.0059 & & 1.9 & yes & & \\
     &  & $-4.845 \pm 0.043$ & $1.028 \pm 0.001$ & 0.0061 & & 2.1 & yes & & \\
    \hline
    KB180029$^5$ & KB180029 & $-4.737 \pm 0.047$ & $0.999 \pm 0.002$ & 0.027 & Recovery & 0.0 & yes & bump & sub-prime \\
    & & $-4.746 \pm 0.050$ & $1.028 \pm 0.002$ & 0.027 & & 0.2 & yes & & \\ 
    & & $-4.740 \pm 0.045$ & $0.999 \pm 0.002$ & 0.027 & & 2.1 & yes & & \\ 
    & & $-4.736 \pm 0.050$ & $1.028 \pm 0.002$ & 0.027 & & 2.2 & yes & & \\ 
    \hline
    KB191806$^1$ & KB191806 & $-4.714 \pm 0.116$ & $1.035 \pm 0.009$ & 0.0255 & Discovery & 0.0 & no & dip & sub-prime  \\
     &  & $-4.717 \pm 0.117$ & $1.034 \pm 0.009$ & 0.0257 & & 0.4 & no & & \\
     &  & $-4.724 \pm 0.117$ & $0.938 \pm 0.007$ & 0.0260 & & 0.7 & no & & \\
     &  & $-4.734 \pm 0.109$ & $0.938 \pm 0.007$ & 0.0251 & & 1.1 & no & & \\
    \hline
    OB170173$^6$ & KB171707 & $-4.606 \pm 0.042$ & $1.540 \pm 0.031$ & 0.867 & Recovery & 0.0 & yes & bump & prime \\
     &  & $-4.195 \pm 0.068$ & $1.532 \pm 0.025$ & 0.844 & & 3.5 & yes & & \\
    \hline
    KB171194$^1$ & KB171194 & $-4.582 \pm 0.058$ & $0.806 \pm 0.010$ & 0.256 & Discovery & & no & dip & sub-prime \\  
    \hline
    KB181988$^7$ & KB181988 & $-4.544_{-0.168}^{+0.300}$ & $0.97 \pm 0.03$ & 0.014 & Recovery & 0.0 & no & dip & sub-prime \\
    & & $-4.759_{-0.618}^{+0.698}$ & $1.01 \pm 0.05$ & 0.014 & & 0.1 & no & & \\
    \hline
    KB190842$^8$ & KB190842 & $-4.389 \pm 0.031$ & $0.983 \pm 0.013$ & 0.0066 & Recovery & & no & bump & prime \\
    \hline
    KB190253$^9$ & KB190253 & $-4.387 \pm 0.076$ & $1.009 \pm 0.009$ & 0.0559 & Discovery & 0.0 & no & dip & prime \\
    & & $-4.390 \pm 0.080$ & $0.929 \pm 0.007$ & 0.0555 & & 0.3 & no & & \\
    \hline
    OB180977$^9$ & KB180728 & $-4.382 \pm 0.045$ & $0.897 \pm 0.007$ & 0.147 & Discovery & & yes & dip & prime \\
    \hline   
    KB171003$^1$ & KB171003 &  $-4.373 \pm 0.144$ & $0.910 \pm 0.005$ & 0.179 & Discovery & 0.0 & no & dip & sub-prime \\
    & & $-4.260 \pm 0.152$ & $0.889 \pm 0.004$ & 0.179 & & 0.2 & no & & \\ 
    \hline  
    OB171806$^1$ & KB171021 & $-4.352 \pm 0.171$ & $0.857 \pm 0.008$ & 0.026 & Discovery & 0.0 & yes & bump & sub-prime \\
    &  & $-4.392 \pm 0.180$ & $0.861 \pm 0.007$ & 0.025 & & 0.2 & yes & &  \\
    &  & $-4.441 \pm 0.168$ & $1.181 \pm 0.011$ & 0.026 & & 8.3 & yes & &  \\
    &  & $-4.317 \pm 0.126$ & $1.190 \pm 0.012$ & 0.027 & & 8.4 & yes & &  \\
    \hline
    OB161195$^{10}$ & KB160372 & $-4.325 \pm 0.037$ & $0.989 \pm 0.004$ & 0.0526 & Recovery & 0.0 & no & bump & prime \\
    & & $-4.318 \pm 0.038$ & $1.079 \pm 0.004$ & 0.0526 & & 0.1 & no & & \\ 
    \hline   
    OB170448$^2$ & KB170090 & $-4.296 \pm 0.149$ & $3.157 \pm 0.022$ & 1.482 & Discovery & & yes & bump & prime \\
    & & $-2.705 \pm 0.045$ & $0.431 \pm 0.004$ & 1.486 & & 5.8 & yes & & \\
    & & $-3.969 \pm 0.086$ & $3.593 \pm 0.045$ & 1.611 & & 9.7 & yes & & \\ 
    \hline
    KB191367$^1$ & KB191367 & $-4.303 \pm 0.118$ & $0.939 \pm 0.007$ & 0.083 & Discovery & 0.0 & no & dip & sub-prime \\
     & & $-4.298 \pm 0.103$ & $0.976 \pm 0.007$ & 0.082 & & 0.2 & no & & \\ 
    \hline   
    KB170428$^1$ & KB170428 &  $-4.295 \pm 0.072$ & $0.882 \pm 0.004$ & 0.205 & Discovery & 0.0 & no & dip & prime \\
     & & $-4.302 \pm 0.075$ & $0.915 \pm 0.005$ & 0.205 & & 0.1 & no & & \\
    \hline   
    OB171434$^{11}$ & KB170016 & $-4.242 \pm 0.011$ & $0.979 \pm 0.001$ & 0.043 & Recovery & 0.0 & yes & dip & prime \\
    & & $-4.251 \pm 0.012$ & $0.979 \pm 0.001$ & 0.043 & & 4.0 & yes & & \\ 
    \hline
    OB181185$^{12}$ & KB181024 & $-4.163 \pm 0.014$ & $0.963 \pm 0.001$ & 0.0069 & Recovery & & no & bump & prime \\
    \hline
    OB181126$^{13}$ & KB182064 & $-4.130 \pm 0.280$ & $0.852 \pm 0.040$ & 0.0083 & Discovery & 0.0 & no & dip & prime \\
    & & $-4.260 \pm 0.290$ & $1.154 \pm 0.052$ & 0.0082 & & 2.1 & no & & \\
    \hline
    OB180506$^9$ & KB180835 & $-4.117 \pm 0.133$ & $1.059 \pm 0.021$ & 0.0884 & Discovery & 0.0 & no & dip & prime \\
    & & $-4.109 \pm 0.126$ & $0.861 \pm 0.018$ & 0.0884 & & 0.4 & no & & \\
    \hline
    KB181025$^{14}$ & KB181025 & $-4.081 \pm 0.141$ & $0.937 \pm 0.021$ & 0.0071 & Recovery & 0.0 & no & bump & prime \\
     &  & $-3.789 \pm 0.133$ & $0.883 \pm 0.025$ & 0.0086 & & 8.4 & no & & \\
    \hline
    OB171691$^{15}$ & KB170752 & $-4.013 \pm 0.152$ & $1.003 \pm 0.014$ & 0.0495 & Recovery & 0.0 & yes & bump & sub-prime \\
    & & $-4.150 \pm 0.141$ & $1.058 \pm 0.011$ & 0.0483 & & 0.4 & yes & & \\
    \hline
    OB180532$^{16}$ & KB181161 & $-4.011 \pm 0.053$ & $1.013 \pm 0.001$ & 0.0082 & Recovery & 0.0 & yes & dip & prime \\
     &  & $-4.033 \pm 0.047$ & $1.011 \pm 0.001$ & 0.0071 & & 2.0 & yes & & \\
     &  & $-3.926 \pm 0.049$ & $1.013 \pm 0.001$ & 0.0089 & & 4.6 & yes & & \\
     &  & $-4.016 \pm 0.076$ & $1.011 \pm 0.001$ & 0.0074 & & 5.4 & yes & & \\
    \hline
    KB160625$^2$ & KB160625 & $-3.628 \pm 0.226$ & $0.741 \pm 0.009$ & 0.073 & Discovery & 0.0 & yes & bump & prime \\
    &  & $-4.138 \pm 0.159$ & $1.367 \pm 0.018$ & 0.075 & & 1.0 & yes & & \\
    &  & $-3.746 \pm 0.291$ & $0.741 \pm 0.009$ & 0.072 & & 1.0 & yes & & \\
    &  & $-4.499 \pm 0.266$ & $1.358 \pm 0.015$ & 0.076 & & 3.3 & yes & & \\
    \hline   
    KB160212$^{17}$ & KB160212 & $-1.434 \pm 0.072$ & $0.829 \pm 0.007$ & 0.328 & Recovery & 0.0 & yes & bump & prime \\
    &  & $-4.310 \pm 0.070$ & $1.427 \pm 0.014$ & 0.615 & & 6.6 & yes & & \\
    &  & $-4.315 \pm 0.099$ & $1.434 \pm 0.012$ & 0.619 & & 8.0 & yes & & \\
    &  & $-4.082 \pm 0.080$ & $1.430 \pm 0.015$ & 0.617 & & 8.7 & yes & & \\
    \hline
    \hline
    \multicolumn{11}{c}{NOTE: For each planet, we only consider the models that have $\Delta\chi^2 < 10$ compared to the best-fit model. ``Discovery'' represents that the planet was}\\
 
    \multicolumn{11}{c}{discovered using AnomlyFinder, and ``Recovery'' means that the planet was first discovered from by-eye searches and then recovered by AnomlyFinder.}\\
   
    \multicolumn{11}{c}{Reference: 1. This work; 2. in prep; 3. \cite{OB191053}; 4. \cite{OB190960}; 5. \cite{KB180029}, Zhang et al. in prep;}\\
    
    \multicolumn{11}{c}{6. \cite{OB170173}; 7. \cite{KB181988}; 8. \cite{KB190842}; 9. \cite{KB190253}; 10. \cite{OB161195}, \cite{OB161195_MOA},}\\
    
    \multicolumn{11}{c}{Zhang et al. in prep; 11. \cite{OB171434}; 12. \cite{OB181185}; 13. \cite{2018_prime}; 14. \cite{KB181025}; 15. \cite{OB171691};} \\
    
    \multicolumn{11}{c}{16. \cite{OB180532}; 17. \cite{KB160212}.} \\
    
\end{longtable*}
\end{center}

\end{document}